\documentclass[epsf]{emulateapj}
\shorttitle{M87 Globular Cluster Kinematics}
\shortauthors{Strader et al.}
\def\etal{et~al.}
\def\kms{\,km~s$^{-1}$}

\def\rh{$r_{\rm h}$}
\def\gsim{\;\rlap{\lower 2.5pt
 \hbox{$\sim$}}\raise 1.5pt\hbox{$>$}\;}
\def\lsim{\;\rlap{\lower 2.5pt
   \hbox{$\sim$}}\raise 1.5pt\hbox{$<$}\;}

\usepackage{subfigure}

\begin{document}

\title{Wide-Field Precision Kinematics of the M87 Globular Cluster System}

\author{Jay Strader\altaffilmark{1,2}, Aaron J.~Romanowsky\altaffilmark{3}, Jean P. Brodie\altaffilmark{3},  Lee R. Spitler\altaffilmark{4}, Michael A. Beasley\altaffilmark{3,5}, \\
Jacob A. Arnold\altaffilmark{3}, Naoyuki Tamura\altaffilmark{6}, Ray M. Sharples\altaffilmark{7}, Nobuo Arimoto\altaffilmark{8}}
\email{jstrader@cfa.harvard.edu}

\altaffiltext{1}{Harvard-Smithsonian Center for Astrophysics, Cambridge, MA 02138}
\altaffiltext{2}{Hubble Fellow, now Menzel Fellow}
\altaffiltext{3}{UCO/Lick Observatory, University of California, Santa Cruz, CA, 95064}
\altaffiltext{4}{Center for Astrophysics \& Supercomputing, Swinburne University, Hawthorn, VIC 3122, Australia}
\altaffiltext{5}{Instituto de Astrofisica de Canarias, Via Lactea s/n, 38200 La Laguna, Tenerife, Spain}
\altaffiltext{6}{Subaru Telescope, National Astronomical Observatory of Japan, Hilo, HI 96720}
\altaffiltext{7}{Department of Physics, University of Durham, South Road, Durham, United Kingdom}   
\altaffiltext{8}{National Astronomical Observatory of Japan, Mitaka, Tokyo 181-8588, Japan}

\begin{abstract}

We present the most extensive combined photometric and spectroscopic study to date of the enormous globular cluster (GC) system around M87, the central giant elliptical galaxy in the nearby Virgo cluster. Using observations from DEIMOS and LRIS at Keck, and Hectospec on the MMT, we derive new, precise radial velocities for 451 GCs around M87, with projected radii from $\sim 5$ to $185$ kpc. We combine these measurements with literature data for a total sample of 737 objects, which we use for a re-examination of the kinematics of the GC system of M87. The velocities are analyzed in the context of archival wide-field photometry and a novel \emph{Hubble Space Telescope} catalog of half-light radii, which includes sizes for 344 spectroscopically confirmed clusters. We use this unique catalog to identify 18 new candidate ultra-compact dwarfs, and to help clarify the relationship between these objects and true GCs. 

We find much lower values for the outer velocity dispersion and rotation of the GC system than in earlier papers, and also differ from previous work in seeing no evidence for a transition in the inner halo to a potential dominated by the Virgo cluster, nor for a truncation of the stellar halo. We find little kinematical evidence for an intergalactic GC population. Aided by the precision of the new velocity measurements, we see significant evidence for kinematical substructure over a wide range of radii, indicating that M87 is in active assembly. A simple, scale-free analysis finds less dark matter within $\sim$~85~kpc than in other recent work, reducing the tension between X-ray and optical results. In general, out to a projected radius of $\sim 150$ kpc, our data are consistent with the notion that M87 is not dynamically coupled to the Virgo cluster; the core of Virgo may be in the earliest stages of assembly.

\end{abstract}


\section{Introduction}\label{sec:intro}

Giant elliptical galaxies at the centers of galaxy clusters, often called brightest cluster galaxies (BCGs) or cluster-central galaxies, are extreme systems that provide stringent tests for theories of galaxy formation. Current models of BCG formation predict dualistic histories where most of the stellar mass is formed very early in the highest-$\sigma$ peaks of the dark matter distribution, while much of the {\it assembly} of mass occurs later through the merging of smaller galaxies that are already old (e.g., \citealt{2007MNRAS.375....2D}).

Observations support and indeed motivate the old stellar ages in this scenario (e.g., \citealt{2009MNRAS.398..133L}). However, the assembly situation is less clear. There are observational reports of high-$z$ BCGs having similar sizes and stellar masses to those in the local Universe \citep{2009Natur.458..603C,2011MNRAS.414..445S}. Such findings could contradict the theoretical expectations for late-epoch growth, although the evidence is still in dispute (e.g., \citealt{2011ApJ...726...69A}).

The late assembly of BCGs can be considered an extreme example of the general theoretical picture of two-phase assembly for massive galaxies, where a seed galaxy grows gradually through the accretion of an extended outer envelope, primarily through minor mergers 
\citep{2006MNRAS.365..747A,2009ApJ...702.1058Z,2010ApJ...725.2312O,2011MNRAS.413.1373W,2011MNRAS.413.3022D,2011MNRAS.416.2802F}. 
Direct evidence for this process has been found via number density, chemical, and kinematical transitions in the halos of various nearby galaxies \citep{1978ApJ...225..357S,2007Natur.450.1020C,2009MNRAS.394.1249C,2010MNRAS.407L..26C,2009MNRAS.398...91P,2009MNRAS.395L..34A,2011MNRAS.413.2943F,2011MNRAS.414..770H,2011ApJ...736L..26A}, but BCGs provide arguably the clearest opportunity to study such transitions---especially in cases with massive, extended cD envelopes.

A few cD galaxies are known to show dramatically increasing stellar velocity dispersions with radius, implying cD envelopes that are associated with the surrounding cluster and may originate in disrupted or tidally stripped galaxies (e.g., \citealt{1979ApJ...231..659D,1999MNRAS.307..131C,2010A&A...520L...9V,2011ApJ...728L..39N}). Such transitions can be mapped out to larger radii using the kinematics of halo globular clusters (GCs; e.g., \citealt{2011A&A...531A.119R}). The remarkable abundance of GCs around BCGs, of up to 50,000 per system, not only makes kinematic studies feasible but has also long implied an unusual formational pathway for these galaxies' outer regions (e.g., \citealt{1982AJ.....87.1465F}), compared to the halos of most other galaxies which have relatively fewer GCs.

M87 (= NGC~4486) is the nearest example of a massive elliptical galaxy at the center of a cluster (Virgo).  It  hosts a population of $\sim 10^4$ GCs along with a mild cD envelope, and is a natural target for exploring the build-up of BCG halos in detail. Deep imaging of the core regions of Virgo has revealed a host of faint tidal streamers around M87 that presumably originate from ongoing accretion events \citep{2005ApJ...631L..41M,2010ApJ...715..972J}. However, despite this system's proximity, its kinematics and dynamics are not well constrained.

M87 was in fact the first galaxy outside the Local Group to be successfully studied with GC kinematics, followed by further extensive studies \citep{1987AJ.....93..779H,1987AJ.....93...53M,1990AJ.....99.1823M,1997ApJ...486..230C,2000AJ....119..162C,1998AJ....116.2237K,2001ApJ...559..812H, 2001ApJ...559..828C}. However, in recent years the observational focus has turned to other systems, leaving the current data set for M87 relatively obsolete. The typical velocity precision is much worse than that possible with modern instrumentation, and there are minimal data outside of $\sim$~35~kpc, which is just where the cD envelope begins to emerge \citep{2009ApJS..182..216K}.

One of the most notable results from these GC studies was the discovery of rapid rotation in the outer regions of M87, suggesting either spin-up from a major merger or shear from large-scale motions of the Virgo core. Another finding was a rapid outward increase in velocity dispersion that implied a transition to dynamics governed by the large-scale cluster potential. More recently, a handful of planetary nebula (PN) spectra have been used to argue for just the opposite: a truncated stellar envelope and little sign of rotation \citep{2009A&A...502..771D}.

The goal of the present study is to revisit the kinematics of M87's GCs by exploiting the new generation of spectrographs on large telescopes. The combined depth, accuracy, and field of view of the observations presented here are the best yet obtained for any galaxy.  We aimed to map the kinematics over a large baseline in radius, covering the possible transition region between galaxy, cD envelope, and surrounding cluster. As we will discuss, many of the previous findings about the M87 halo are overturned in the light of this extensive data set.

Given the richness of the data, there are numerous topics of interest that could be examined but are not covered in this paper. In particular, the themes of GC sizes, metallicities, and phase-space substructure are considered in depth in separate papers  (\citealt{2011arXiv1109.5696B};  Strader et al., in preparation; \citealt{Roman11}).

The distance we adopt for M87 is 16.5~Mpc, which is consistent with analyses of surface brightness fluctuations \citep{2009ApJ...694..556B},  GC sizes \citep{2009ApJ...705..237M}, and the tip of the red giant branch \citep{2010A&A...524A..71B}. One arcmin of angular distance corresponds to 4.80~kpc.

The effective radius $R_{\rm e}$ and luminosity of a galaxy like M87 are highly uncertain and perhaps ill-posed quantities. For reference, \citet{2009ApJS..182..216K} found $R_{\rm e}\simeq$~3.2\arcmin\ (16~kpc) and $M_V \simeq -22.86$.

This paper is outlined as follows. \S\ref{sec:data} presents the new GC observations, while \S\ref{sec:reduce} discusses the spectroscopic data reduction. \S\ref{sec:compile} compares and combines old and new data, and provides careful classification of GCs and ``contaminants'' such as foreground stars. The photometric and kinematical properties of the GC system are analyzed in \S\ref{sec:photprop} and \S\ref{sec:kin}, respectively. \S\ref{sec:dyn} provides a brief dynamical analysis. \S\ref{sec:disc} discusses several implications of our results, including the classification of GCs and the formation of M87 and BCGs in general. \S\ref{sec:summ} summarizes the details of our findings.

\section{Data}\label{sec:data}

Here we outline the acquisition of the new GC data sets presented in this paper. Spectroscopic data from Keck/DEIMOS, MMT/Hectospec, and Keck/LRIS are covered by \S\ref{sec:deimos}, \ref{sec:hecto}, and \ref{sec:lris}, respectively.  A log for all spectroscopic observing runs can be found in Table~\ref{tab:obsrun}.
Those interested in an overview of the spatial positions of the GC spectra can glance forward to Figure~\ref{fig:twod}. The archival CFHT imaging is discussed in \S\ref{sec:photom}.

This paper does {\it not} incorporate the small data set of velocities obtained with DEIMOS around the outer ``stream A'', which is presented in \citet{Roman11}.

Note that we will use the usual term ``radial velocity", referring to the observer's line-of-sight, which should not be confused with other quantities involving the internal (projected) radius of M87.

\subsection{Keck/DEIMOS Data}\label{sec:deimos}

We selected GC candidates from the Subaru/Suprime-Cam imaging study of \citet{2006MNRAS.373..588T}, who identified  potential clusters using $BVI$ imaging. Preference was given to objects with $I < 23$ and to blue candidates, since most of the clusters are large radii are expected to be metal-poor.

Spatially, the \citet{2006MNRAS.373..588T} data extend directly east from M87 to an approximate projected galactocentric radius of $\sim 2$ deg, encompassing also the elliptical galaxy NGC 4552 at a radius of $\sim 1.2$ deg. We designed four Keck/DEIMOS slitmasks, each covering an area of $\sim$~16\arcmin$\times$~5\arcmin. These were arranged with minimal overlap in a $2 \times 2$ grid covering a total area of $\sim$~32\arcmin$\times$~10\arcmin, starting $\sim$~8\arcmin\ from the center of M87.

The two inner masks had 70--80 candidates each and the outer masks 50--60 GC candidates each, for a total of 254 possible clusters (two objects were duplicated on another mask, giving 256 slitlets). Not all of these were high-quality candidates; objects outside of the nominal range of color expected for true M87 GCs, and objects likely too faint to produce a usable spectrum, were used to fill unused mask area. All slits were cut with a width of 1\arcsec\ and a minimum length of 5\arcsec.

The four masks were observed using DEIMOS on 2007 March 20--21, with an exposure time of 3--3.5 hours per mask, divided into individual exposures of 30 min. The weather was clear and seeing ranged from 0.8--0.9\arcsec. We used the 1200 l/mm grating centered at 7500 \AA, yielding a resolution of $\sim 1.5$ \AA. This setup gives coverage of the full Ca~{\small II} triplet (CaT, with rest frame wavelengths of 8498, 8542, and 8662 \AA) for most clusters with radial velocities typical of objects associated with M87. At the blue end, we also cover H$\alpha$ for many clusters, which can serve as a valuable velocity check for low signal-to-noise ($S/N$) spectra. This is our standard setup for GC observations with DEIMOS, used also in \citet{2009AJ....137.4956R}, \citet{2011MNRAS.415.3393F}, and \citet{2011ApJ...736L..26A}, but here with longer exposure times to compensate for the fainter median magnitude of our targets due to the low densities of GCs in the far outer halo of M87.

\subsubsection{Additional DEIMOS Data}

As a check on some of the unusually high radial velocities reported in the literature, we observed one additional mask in March 2010, centered $\sim$~7\arcmin\ south of M87 at an approximate position angle of 120 degrees. Candidates were split among those with known velocities and previously unobserved objects. The total exposure time was 36 min, split into two exposures, with approximately 1.0\arcsec\ seeing. The setup was identical to that for the earlier DEIMOS data.

\subsection{MMT/Hectospec Data}\label{sec:hecto}

The initial results of our DEIMOS observations motivated us to obtain follow-up data with complete azimuthal coverage.  Hectospec is a fiber-fed instrument on MMT with 300 fibers covering a 1 degree field of view \citep{2005PASP..117.1411F}. GC candidates were selected from the CFHT imaging described in \S\ref{sec:photom}.  Preference was given to candidates with $g < 22$, but with no color restrictions beyond the basic GC candidate cuts. Objects were primarily selected at large radii, since local sky subtraction must be performed with individual sky fibers and sky subtraction in the inner regions of the galaxy (well interior to any of our fiber locations) is more challenging. 244 fibers were assigned to GC candidates, with the remainder to guide stars or sky. Each fiber subtends a diameter of 1.5\arcsec.

A single Hectospec field, centered on M87, was observed for 2 hours (divided into 20 min exposures) in February 2010. The conditions were clear and the seeing was 0.6--0.7\arcsec. Observations were made with the 270 l/mm grating, which yields a spectral coverage of 3700--9100 \AA\ and a resolution of $\sim 5$ \AA.

\subsection{Keck/LRIS Data}\label{sec:lris}

Most spectroscopic studies of GCs in M87 have targeted very luminous GCs, with typical $i$-band luminosities of $\sim$~$1.2\times10^6 L_\odot$, surpassing all but the brightest clusters in the Milky Way.  With a goal of determining the metallicity distribution of somewhat fainter GCs in M87, in April 2010 we observed four slitmasks in the central region of the galaxy  ($\sim$~1.5--5.5\arcmin) with Keck/LRIS. The stellar populations of these GCs will be discussed in a separate paper; here we consider only the radial velocities. As with the Hectospec observations, candidates were selected from CFHT imaging (\S\ref{sec:photom}). The median targeted object had $g = 22.5$, and no preference was given to objects of a certain color (beyond the basic GC selection criteria), so the objects observed should be an accurate reflection of the intrinsic color distribution of GCs at the galactocentric distances surveyed.

Each mask had $\sim 50-55$ candidates, with 209 targets total. All slits had a width of 1\arcsec\ and a minimum length of 4\arcsec. At the time of the observations, the red side of LRIS was malfunctioning, so observations were made only on the blue side. All data were taken binned 2x2, yielding a scale of 0.27\arcsec\ per pixel. We used the 600 l/mm grating, which gives $\sim 2600$ \AA\ of spectral coverage. The exact wavelength coverage for each object depends on the spatial position in the mask, but a typical range was from $\sim 3300-5600$ \AA\, with the upper limit set by the dichroic cutoff. With a 1\arcsec\ slit, the resolution was $\sim$~3.6--4.1 \AA\, depending on wavelength, with a binned dispersion of $\sim 1.26$ \AA\ per pixel.

Total exposure times for each mask ranged from 2.5--3.5 hr, with individual exposures of 30 min. Seeing was variable between 0.6--0.9\arcsec.

\subsection{Photometry}\label{sec:photom}

The \citet{2006MNRAS.373..588T} imaging extends to extremely large radii, but with non-uniform azimuthal coverage. For this reason, we chose to use for our default GC photometry the new $gri$ CFHT/Megacam imaging of M87 published in \citet{2009ApJ...703..939H}\footnote{As Harris mentions, these data will soon be superseded by the Next Generation Virgo Cluster Survey (NGVS), which should allow superior selection of GC candidates.}. These data cover $\sim$~1 square degree and so offer complete radial coverage to a projected radius of $\sim$~30\arcmin\ (144 kpc). 

We downloaded the stacked CFHT images from the CADC archive and reduced and analyzed the data in a manner very similar to that described in \citet{2009ApJ...703..939H}; all of our measurements are consistent with that study, although we differ in the interpretation of a few points discussed below. The $gri$ magnitudes were roughly calibrated to the SDSS system by matching objects with the SDSS DR7 catalog and calculating a constant offset for each filter.
The matches were done to the DR7 psfMag values (since the objects are unresolved) and over the magnitude range 18--22 for each filter. We cannot rule out the possibility of small variations in these zeropoints over the Megacam field of view, but the resulting accuracy of photometric calibration is sufficient for nearly all of our scientific goals. The one exception is our investigation of possible GC color gradients (see further discussion in \S\ref{sec:beyond}).

We corrected the photometry for foreground reddening based on \citet{2010ApJ...719..415P}, which uses passively evolving galaxies to correct the \citet{1998ApJ...500..525S} reddening map over the SDSS DR7 spectroscopic area. Note that we have applied spatial variations from the \citet{1998ApJ...500..525S} map (which has a resolution of 6\arcmin), rather than taking a single mean reddening value. This exercise is important when studying the data on large angular scales, as there are highly non-uniform patches of Galactic cirrus in the Virgo direction, and we found variations in the $E(g-i)$ correction of up to 0.02~mag.

\section{Reduction and Radial Velocities}\label{sec:reduce}

In this section we discuss the data reduction and determination of radial velocities for all spectra. Keck/DEIMOS, MMT/Hectospec, and Keck/LRIS data are covered in \S 3.1, 3.2, and 3.3 respectively. \S 3.4 compares velocities for common objects among the three instruments as a consistency check.

\subsection{Original DEIMOS Data}

The spectra were reduced in a standard manner by using the IDL spec2d pipeline\footnote{http://astro.berkeley.edu/\textasciitilde cooper/deep/spec2d/}. Two-dimensional images were  flatfielded and wavelength calibrated using internal lamps, followed by sky subtraction and an optimal  extraction of one-dimensional spectra.

We derived heliocentric radial velocities as described in \citet{2009AJ....137.4956R}. In brief, GC spectra around the CaT were cross-correlated with a library of DEIMOS stellar templates from mid-F to late-K. Confirmation of the velocity required visual identification of at least two lines; these were usually CaT lines, but in several cases a single Ca line and H$\alpha$ were used. 

Radial velocity uncertainties were estimated from the width of the cross-correlation peak and by the variation in velocity obtained from different templates. As discussed in \citet{2009AJ....137.4956R}, cross-checks from multiple DEIMOS observations of the same GCs in previous work and from different instruments all suggest that our DEIMOS uncertainty estimates are accurate in a random sense. However, these uncertainties do not reflect the possibility of catastrophic misidentifications of the principal cross-correlation peak. This is a particular problem for low signal-to-noise spectra in the CaT region, which is prone to sky-line residuals. Hence we require visual confirmation of multiple lines; nonetheless, such catastrophic failures have evidently occurred in the literature (\S\ref{sec:merge}), and we cannot exclude their existence from the present data.

We classify our 254 candidate GCs as follows, based on the spectroscopic results. 63 appear to be bona fide M87 GCs, of which 60 are newly confirmed. A further 44 are Galactic stars (see the \S\ref{sec:stars} for details of star--GC separation). 26 are background galaxies, mostly objects with multiple narrow emission lines. For the remaining 122 candidates, no reliable velocity could be obtained, generally because of low $S/N$. A subset of these objects had cross-correlation velocities consistent with M87 GCs, but had at most one identifiable line; many could be true clusters.

These numbers suggest a nominal success rate of $\sim$~50\% for identifying GCs, but the true percentage is higher, since many marginal objects were included as mask filler. This suggests that three-band imaging in decent conditions is an efficient method for finding GCs, even in the far outskirts of galaxies where the surface density of clusters is low.

The basic data for all of these observed DEIMOS objects are given in Table~\ref{tab:allcand}. The median velocity uncertainty for confirmed GCs is 9~\kms.

\subsubsection{Supplementary DEIMOS Data}

As discussed above, we observed one additional mask, closer to the center of M87, for 36 min in March 2010. Data reduction and analysis were identical to that for the earlier DEIMOS data. Owing to the short exposure time, reliable velocities could only be determined for 29 objects of the 112 initial targets. 27 of these are GCs and two are stars. Data for these 29 objects are given in Table~\ref{tab:keck2}.

\subsection{MMT/Hectospec}

The Hectospec data were pipeline reduced in a standard manner as described in \citet{2007ASPC..376..249M}. While this pipeline also produces radial velocities through cross-correlation with a template library, we chose to manually derive velocities for our objects. Heliocentric radial velocities for the Hectospec data were estimated through cross-correlation with a normalized Hectospec template of an M31 globular cluster over the wavelength range 4000--5450 \AA. Regions with significant sky emission were excluded.

We estimated velocity uncertainties by Monte Carlo simulations using the error spectra. To these values, we add an error of 8~\kms\ in quadrature due to uncertainties in the wavelength calibration. The median total uncertainty for all of the spectra is 14~\kms.

Of the 244 candidates, 240 produced an identifiable radial velocity. Of these, 172 appear to be true GCs and the other 68 foreground stars (see \S\ref{sec:class}). Since this sample of candidates is brighter than in the DEIMOS observations, a higher fraction of foreground stars and a lower fraction of background galaxies is expected, as observed.

The basic data for all observed Hectospec objects are given in Table~\ref{tab:allcand2}.

\subsection{Keck/LRIS}

The bulk of the LRIS data reduction was done with the XIDL/LowRedux package\footnote{http://www.ucolick.org/\textasciitilde xavier/LowRedux/}. This pipeline performs the standard steps of bias subtraction, flat fielding, illumination correction, wavelength calibration, and sky subtraction before an optimal extraction. As the LRIS calibration lamps cannot be used for flat fields in the blue, the pipeline uses pixel flats taken with the grism in place during twilight.

For approximately 15\% of the slits, the pipeline extraction did not produce an optimal spectrum; in most cases this was due to very short slits for which the sky subtraction failed. These slits were reduced manually in IRAF using standard tasks.

Unlike Keck/DEIMOS and MMT/Hectospec, LRIS has significant flexure---shifts of 1--2 pixels over two hours are quite common. Such shifts correspond to radial velocity shifts $> 100$~\kms\ and so must be addressed.\footnote{LRIS flexure issues may be responsible for the apparent velocity zero-point offset between the studies of \citet{1997ApJ...486..230C} and \citet{2000AJ....119..162C}.} Each individual exposure was corrected for flexure by cross-correlation of the sky spectrum with a template before coaddition of the exposures for each object. The proper location of the bright \ion{O}{1} line at 5577.34 \AA\ indicates the overall fidelity of the process.

Heliocentric radial velocities of the coadded spectra were derived through cross-correlation with a library of appropriate templates over the wavelength range 3900--5300 \AA. The intrinsic velocity uncertainties were estimated with Monte Carlo simulations using the error spectra. Total velocity uncertainties were computed by adding these values in quadrature with the wavelength calibration uncertainty and that due to the flexure correction, which varied depending on the mask. The median uncertainty for all of the spectra is 26~\kms. While we believe that these estimated uncertainties are reasonable, few of our objects have existing velocities in the literature because of their faintness (see \S\ref{sec:compile}), and so only minimal external checks can be performed. Given the potential flexure issues, future study of a subset of these GCs would be valuable.

The four masks contained 209 designed slits, and the data reduction yielded 204 usable spectra. 200 are probable M87 GCs, with only 2 stars, one background galaxy, and one ambiguous object (see \S\ref{sec:clean}), for an excellent targeting efficiency of $> 95$\% in the abundant inner regions of M87.

The basic data for all observed LRIS objects are given in Table~\ref{tab:lriscand}.

\subsection{Self-Consistency of the New Radial Velocities}\label{sec:self}

The new radial velocities were derived from spectra taken with three different instruments on four separate observing runs, and so checking the consistency of these velocities is desirable. As these new measurements all have much smaller uncertainties than in previous studies, we first consider only new data; in \S\ref{sec:compile} we compare our new velocities to those in the literature.

Each of our data sets has differing coverage in galactocentric radius, azimuth, and target magnitude range. As such, the sample overlap is limited. Those objects observed with LRIS have no counterparts in our other data. There are five GCs in common between Hectospec and the original DEIMOS sample, and a further six between Hectospec and the supplementary DEIMOS data. 

The median difference between the combined set of DEIMOS radial velocities and those from Hectospec is $15\pm5$~\kms\ (all such errors quoted are standard errors). These are  If the first and second DEIMOS runs are taken individually, the respective differences are $4\pm10$~\kms\ and $18\pm6$~\kms. The distribution of the observed differences between the MMT and DEIMOS radial velocities is consistent with our estimated velocity uncertainties once these offsets are applied (see also Figure~\ref{fig:vel_comp} below). Given the small values of the offsets and the limited number of comparison GCs, we do not apply a common offset to the final set of velocities. However, we cannot exclude the possibility of systematic offsets of order $\sim$~10~\kms\ among these data.

\begin{figure}
%\epsscale{0.85}
\epsscale{1.2}
%\plotone{spec.eps}
\plotone{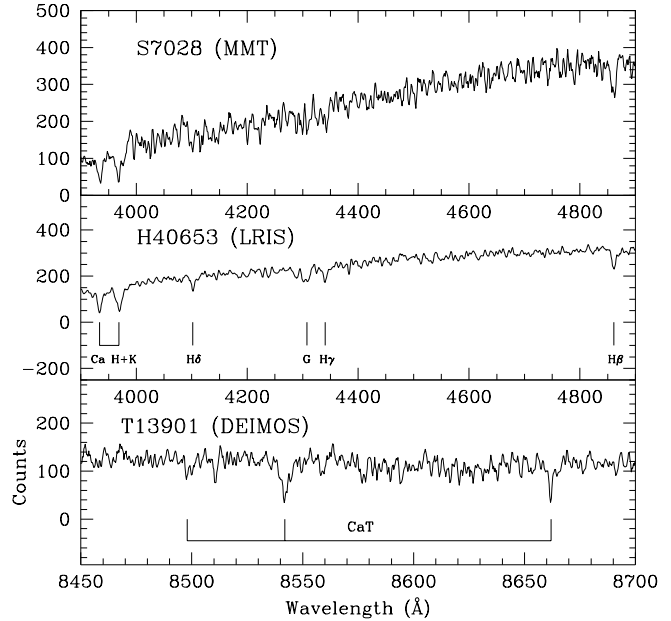}
\figcaption[spec]{\label{fig:medspec}
Sample spectra from each of the three instruments employed. Each GC was chosen to have a median $i_0$ magnitude close to that of all confirmed GCs for that dataset. Representative spectral regions are displayed: 3900--4900 \AA\ for MMT and LRIS, and the Ca~{\small II} triplet region for DEIMOS. Prominent spectral lines are marked. The spectra have each been shifted to zero redshift and smoothed with a 3-pixel boxcar for display.}
\end{figure}

In Figure~\ref{fig:medspec} we show sample spectra, over a restricted spectral range, from each of the instruments employed. We deliberately show spectra for each dataset that have $i_{0}$ magnitudes near that of the median for that sample, so the plotted spectra can be considered representative of our data.

\section{A Photometric and Spectroscopic Compilation}\label{sec:compile}

This section describes our comprehensive catalog of photometry and radial velocities for candidate star clusters associated with M87. \S 4.1 compiles velocities from the literature. \S 4.2 compares our new velocities to those from the literature to establish a common velocity scale. In \S 4.3 we derive new half-light radii for many clusters using archival \emph{Hubble Space Telescope} imaging. \S 4.4 discusses the classification of objects, including those from the literature. \S 4.5 summarizes the catalog and \S 4.6 discusses the separation of  M87 GCs into metallicity subpopulations.

\subsection{Literature Data}

To form a complete dataset for dynamical analysis, we combine our 451 GC velocities with data from the literature. Not all of these velocities are unique---there are 24 duplicates with the literature (with 5 additional stellar repeats), as well as some internal duplicates. As our starting point, we take the collected catalog of radial velocities from \citet[hereafter H+01]{2001ApJ...559..812H} (see also \citealt{2001ApJ...559..828C}). Other than the new CFHT/MOS velocities published in that  paper, the previous data in the compilation were taken from \citet{1987AJ.....93..779H,1987AJ.....93...53M,1990AJ.....99.1823M,1997ApJ...486..230C}; 
and \citet{2000AJ....119..162C}.

To this compilation, we have added radial velocities for some objects observed as part of studies of ``ultra-compact dwarfs" (UCDs) or ``dwarf-globular transition objects" (DGTOs) in the Virgo cluster. \citet{2006AJ....131..312J}, \citet{2008MNRAS.389.1539F}, and \citet{2010ApJ...724L..64P} present low-resolution spectra\footnote{There are also 6 velocities reported in \citet{2009MNRAS.394.1801F}. However, these are all duplicate measurements from the literature and from our new data, and inter-comparisons of the velocities suggest that these authors' measurement uncertainties may be underestimated. Therefore we do not use them in this paper, pending further  investigation.}; we also include higher resolution studies of a small number of objects, using Keck/ESI, undertaken by \citet{2005ApJ...627..203H}, \citet{2007PhDT.........4H}, and \citet{2007AJ....133.1722E}. Together, these papers report radial velocities for 42 unique objects associated with M87 or the Virgo cluster itself. In \S\ref{sec:class} we address the classification of these objects.

H+01 used both linear fits and simple offsets to correct all of the radial velocities to a common system. We adopt slightly different corrections than did H+01, preferring constant offsets if possible. There are 48 objects in common between CFHT/MOS and \citet{1997ApJ...486..230C} with reliable radial velocities. The median difference is $33\pm19$~\kms, which we adopt as a common offset. While there are only eight GCs in the overlap between CFHT/MOS and the later study of \citet{2000AJ....119..162C},  the evidence for an offset is even stronger (median difference $77\pm33$~\kms), and we use this value.

\citet{1990AJ.....99.1823M} published original velocities as well as previously published ones from \citet{1987AJ.....93...53M} and \citet{1987AJ.....93..779H}. The reported velocities for the five clusters in common between \citet{1987AJ.....93..779H} and their work are straight averages. Considering the sixteen objects in common between the Mould~\etal\ studies and the CFHT/MOS data, there is no statistically significant evidence for an offset in the velocity scale. There is marginal evidence for an offset between the velocities in \citet{1987AJ.....93..779H} and those of both H+01 and Mould~\etal\ However, the sign of the offset differs between the comparison samples, and so we choose to make no correction.

Given the substantial set of new radial velocities and the large velocity dispersion of the GC system, these offsets actually have very little effect on the first-order dynamical analysis. However, they are relevant for the examination of outlying velocities and higher order moments of the velocity distribution, as well as overall consistency.

In addition to these zero-point corrections, we follow \citet{2001ApJ...553..722R} in renormalizing the published uncertainties of the studies prior to H+01 by small factors to enforce consistency among the uncertainties of different  papers (or, in some cases, providing uncertainties where none were published). In particular, we adopt a common velocity uncertainty of 106~\kms\ for all objects from \citet{1987AJ.....93...53M,1990AJ.....99.1823M,1987AJ.....93..779H,1997ApJ...486..230C}; and a value of 77~\kms\ for \citet{2000AJ....119..162C}. When combining velocities the weighted average was used. A few of the oldest (1990 or previous) velocities were excluded from combined values if they deviated significantly from more than one recent measurement for the same object.

There are two exceedingly popular objects in the vicinity of M87, S314\footnote{We refer to  objects from the H+01 compilation with an ``S" prefix since they are on the revised numbering system of \citet{1981ApJ...245..416S}. Rather than extend that system further, all new GCs with imaging from \citet{2009ApJ...703..939H} have IDs beginning with ``H";  those with imaging from \citet{2006MNRAS.373..588T} are given a prefix of ``T", even though a subset of these objects may be in the Strom \etal\ catalog.} and S1280, each of which has five published radial velocities. Regrettably, the latter is a foreground star.

\subsection{Merging the Datasets}\label{sec:merge}

In \S\ref{sec:self} we established---at least to the degree possible with the current observations---that our new radial velocities appear to be self-consistent and have reasonable uncertainty estimates. Now we discuss the integration of the current results with velocities from the literature.

There are eighteen objects in common between the CFHT/MOS radial velocities reported in H+01 and the present study. These matches are not all happenstance; the supplementary DEIMOS mask was designed to re-target a number of existing GCs with unusual reported velocities. Of the eighteen matches, fifteen show good agreement. The median difference between the new and old velocities is $7 \pm 23$~\kms\ and the uncertainty estimates appear to be accurate.

However, three of the objects are ``catastrophic" outliers, such that the new velocity differs by $> 5\sigma$ from the literature velocity (for other examples in the literature, see \citealt{2008ApJ...674..857L} and \citealt{2010ApJ...709..377P}). These objects are listed, together with the other matches, in Table~\ref{tab:glob_match}. The area around each of the GCs was checked for close companions to rule out target misidentification, but no viable interlopers were identified. The likely culprit is the incorrect choice of the primary peak in the velocity cross-correlation in the previous work. A general method to identify further outliers is unclear. While one of these objects (S923) is among the faintest in the H+01 MOS sample and is also peculiar as we shall see later, the other two (S878 and S1074) do not have unusual magnitudes or colors. It should be noted that we specifically targeted extreme velocities in our supplementary DEIMOS mask; our new observations do not provide strong evidence that the fraction of unreliable velocities in the literature is high (see also \S\ref{sec:clean}).

The other principal sources for the H+01 compilation are \citet{1997ApJ...486..230C} and \citet{2000AJ....119..162C}. Since (at least nominally) these two papers have already been normalized to a common scale, we consider them together. For thirteen objects in common between the present study and these two papers, the median difference is $8 \pm 24$~\kms, and the distribution of the normalized differences supports the accuracy of the adopted uncertainty estimates.

\begin{figure}
\epsscale{1.2}
\plotone{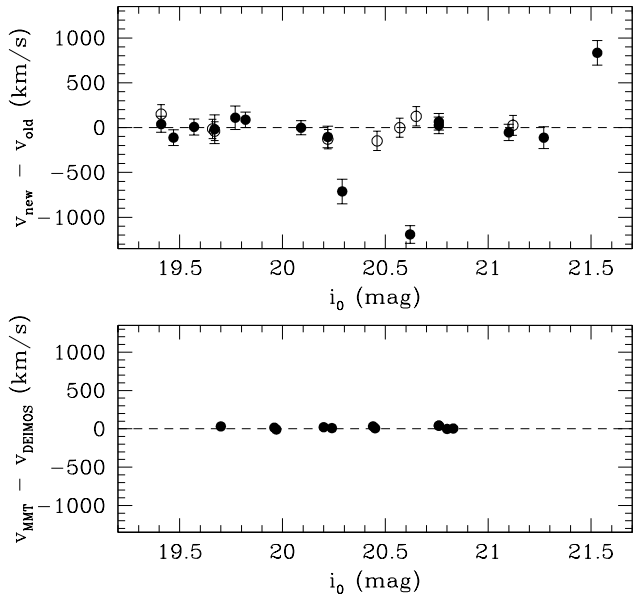}
\figcaption[spec]{\label{fig:vel_comp}
The top panel shows the difference between our new radial velocities and the literature values as a function of $i_{0}$ magnitude, restricted to H+01 (filled circles) and \citet[open circles]{1997ApJ...486..230C} for simplicity. The three ``catastrophic" outliers discussed in \S 4.2 are apparent. The lower panel shows the same quantities, but comparing our MMT and DEIMOS repeats, discussed in \S 3.4.}
\end{figure}

In Figure~\ref{fig:vel_comp} we summarize these results. The top panel plots the radial velocity difference between our new data and the dominant literature studies (H+01 and \citealt{1997ApJ...486..230C}) as a function of $i_0$. The scatter is  consistent with the reported error bars, with the exception of the three catastrophic outliers. The bottom panel, with identical axis limits, plots these same quantities but for the repeat measurements between our MMT and DEIMOS spectra (\S 3.4). The higher precision of the new radial velocities is evident.

\subsection{Cluster Sizes}\label{sec:size}

For this study, there are two reasons why sizes\footnote{This term should be taken as a synonym for projected half-light radii \rh\ for the remainder of the paper, unless otherwise stated.} are useful. First, they can be used to distinguish unresolved foreground stars from actual GCs when the radial velocity is ambiguous. In addition, a subset of our objects are probable UCDs with potentially different kinematics from the GCs; these objects can be partially identified through their large sizes.

For as many GCs as possible, we collected sizes from the literature. The relevant papers included extensive ACS studies \citep{2009ApJS..180...54J,2009ApJ...705..237M} as well as the older WFPC2-based paper from \citet{2001AJ....121.2974L}. In addition, we obtained sizes for UCDs (or candidates) from \citet{2008AJ....136..461E}, \citet{2005ApJ...627..203H}, and \citet[but see below]{2007PhDT.........4H}. We took sizes for a total of 147 objects from these papers, but this is only a fraction of the total spectroscopic catalog. All sizes were scaled to a common distance of 16.5 Mpc.

For the remainder of the objects, we searched the \emph{Hubble Space Telescope} ({\it HST}) archive for images suitable for estimating GC sizes. We considered images taken with any of ACS, WFPC2, or STIS, and in any broadband optical filter. Owing to the large number of cycles in which WFPC2 was the default parallel instrument, the majority of useful images are from this camera.

Our image analysis varied somewhat depending on the instrument used. For ACS data, we used the default drizzled images from the Hubble Legacy Archive, and estimated the point spread function (PSF) directly from a number of bright stars on the drizzled image. For WFPC2 data, we instead downloaded the basic calibrated images from the MAST archive. After cosmic ray rejection with L.A.~Cosmic \citep{2001PASP..113.1420V}, the individual chip images were averaged together. In order to preserve the PSF in severely undersampled WFPC2 images, we only coadded images taken at the same position; we made no image shifts. TinyTim\footnote{http://www.stsci.edu/software/tinytim/tinytim.html} was used to calculate position and filter-dependent PSFs for each WFPC2 cluster. These theoretical PSFs were subsequently convolved with the standard diffusion kernel that accounts for charge diffusion in the CCD. Charge diffusion is likely to be wavelength dependent and is one of the dominant sources of error in estimating sizes for barely resolved objects in WFPC2 images. Finally, the few GCs with STIS data were analyzed in a similar manner to the WFPC2 images.

Using these images and PSFs, we estimated cluster effective (half-light) radii using \emph{ishape} \citep{1999A&AS..139..393L}. This program convolves models of star clusters with a supplied PSF and determines best-fit parameters by maximizing the likelihood function. Given the moderate signal-to-noise of most of our clusters on the images, we fit elliptical King models with a fixed concentration ($r_t$/$r_0$) of 30 to each of the GCs. For nearly all objects, a fitting radius of 5 pixels was used; we adopted a larger radius for a few unusually large clusters. The effective radii reported are circularized and assume a distance of 16.5 Mpc.
Besides the new size measurements, we have also remeasured the sizes of a handful of the fainter objects from the Ha{\c s}egan catalogs, now using fixed concentrations (see \citealt{2011arXiv1109.5696B} for further details).

Determining uncertainties for cluster sizes is challenging even with homogeneous data; the dominant uncertainties are systematic. With our range of instruments, filters, and exposure times, supplying an accurate error estimate for each cluster is implausible. To assess the typical uncertainty, we analyzed WFPC2 images in F555W and F606W of the central regions of M87 that overlap with the published ACS GC size catalog of \citet{2009ApJS..180...54J}. F555W is the filter for which the charge diffusion is best characterized; F606W is the dominant filter in our catalog. These WFPC2 images have exposure times close to those of the ACS images, and the comparison objects have a distribution of magnitudes similar to those of our catalog.

Considering the \citet{2009ApJS..180...54J} sizes themselves: there are measurements in two different bands (F475W and F850LP), with a systematic difference of $\sim 0.25$ pc between the filters (F850LP sizes are larger; \citealt{2005ApJ...634.1002J}). The typical reported uncertainty among the bright ACS GCs is 0.15--0.20 pc. We use a straight average of the size estimates in both filters for comparison to the WFPC2 sizes. Unlike for our sizes, \citet{2009ApJS..180...54J} do not assume a fixed King model concentration, but (i) moderate variations in the concentration affect the half-light radius very little, and (ii) the $S/N$ of the majority of GCs is insufficient to constrain this parameter in any case.

The median difference between our F555W sizes and those in Jordan \etal\ is $-0.70$ pc, with a standard deviation of $\sim 0.5$ pc. In F606W, the median difference is $+0.17$ pc, but with a nearly identical dispersion. Given the small uncertainties of the ACS sizes themselves, this suggests that the typical random errors for our WFPC2 sizes are 0.4--0.5 pc (about 15\% for a normal GC). \citet{2001AJ....121.2974L} suggested approximate uncertainties of 1 pc for WFPC2 data of GCs in Virgo cluster galaxies, but the bulk of their objects were fainter than ours.

More troubling is the large systematic difference between the WFPC2 F555W and F606W half-light radii. While the ACS sizes are not free from error---and indeed the filter-to-filter size variations indicate the presence of systematic effects (see also \citealt{2006AJ....132.1593S})---the smaller pixels and better-sampled PSF suggest naively that ACS sizes are probably more accurate than those from WFPC2. Thus it is surprising that the difference from ``truth" is largest in the F555W filter, for which the effects of charge diffusion are expected to be best understood\footnote{Convolution with an older, more diffuse kernel, which was at one time recommended by the WFPC2 Data Handbook, produces an even larger discrepancy with the ACS sizes.}. These results suggest that WFPC2 charge diffusion may be poorly characterized, even in F555W. To our knowledge no other direct comparisons between WFPC2 and ACS sizes exist. In any case, the bulk of our sizes are measured in F606W, which seems to agree best with the ACS effective radii, and we do not make any corrections to our measured sizes.

Table~\ref{tab:sizes} lists our size estimates, along with the instrument and filter, for the interested reader. In total, we estimated half-light radii for 197 objects that are spectroscopic members of M87\footnote{A single object, H38352, is unresolved in its WFPC2 images. Its radial velocity of $\sim$~1100~\kms\ confirms M87 membership; we estimate an upper limit of $\sim 1$ pc for its effective radius.}. Thus 344 clusters---nearly half of the spectroscopic catalog---have sizes. We have also verified that an additional 39 objects with low radial velocities are unresolved foreground stars; these are marked as such in Table~\ref{tab:stars}. This is by far the largest publicly-available data set of GCs around any galaxy with both measured sizes and spectroscopic confirmation, and enables us to clarify the relations between UCDs and GCs (\S\ref{sec:ucds} and \citealt{2011arXiv1109.5696B}).

\subsection{Object Classification}\label{sec:class}

One of the key ingredients in our kinematical and dynamical analyses is to have a clean sample of very-high-probability GCs. To derive this sample, we use a combination of colors, luminosities,  sizes, and velocities to distinguish GCs from other ``contaminating'' objects. The two obvious classes of interlopers are background galaxies and Galactic foreground stars; a third, more subtle group are the UCDs, which inhabit a similar color-magnitude space to the GCs. 

\subsubsection{Ultra-compact dwarfs}\label{sec:ucds}

Considering the UCDs first: the largest objects, with sizes approaching \rh~$\sim$~100~pc, are almost surely a different class of objects than ``normal'' GCs, and can be considered bona fide galaxies or UCDs. However, there is a substantial gray area of smaller objects that could include both the most compact UCDs and the most extended GCs. True UCDs are unlikely to share the kinematical properties of normal GCs and so should be excluded from the pure GC sample.

\begin{figure}
%\epsscale{0.85}
\epsscale{1.2}
\plotone{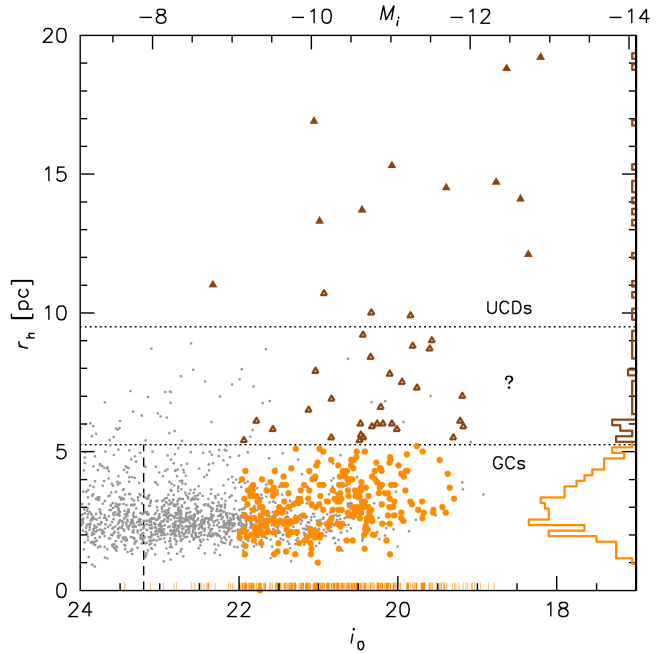}
\figcaption[M87GCsize2cx.ps]{\label{fig:sizes}
Half-light radius versus $i$-band magnitude for compact stellar systems.
Small gray points show GC candidates in the central {\it HST}/ACS field 
\citep{2009ApJS..180...54J}, while larger symbols show objects in the spectroscopic sample.
Dotted lines show our diagnostic boundaries for GCs, UCDs, and uncertain/transition
objects, with accordingly different symbols for the spectroscopic objects in these regions.
Spectroscopically-confirmed objects with no size measurements are plotted 
as tick-marks near \rh~$=0$. The vertical dashed line shows the approximate GCLF turnover magnitude.
A histogram on the right-hand side shows the number of objects in size bins.
There are another 20 UCDs with \rh~$>$~20~pc that are not visible on this plot.}
\end{figure}

Distinguishing between UCDs and GCs is not a problem we will solve in this paper, but we will try to at least provide some clear boundaries for the case of M87. In Figure~\ref{fig:sizes} we show a diagnostic diagram of size versus magnitude. Considering both the sample of objects imaged with {\it HST}/ACS in the central regions of M87 (Jord\'an et al.) and our spectroscopic sample with sizes over a wider field (\S\ref{sec:size}), there is a clear ridge-line of GCs centered at \rh~$\sim$~2.5 pc over most of the luminosity range, with a gradual upturn to $\sim$~3.0--3.5~pc beginning at magnitudes brighter than $i_0 \sim 21$. At these bright magnitudes, there is also a broad tail of objects extending to very large sizes. These general observations parallel those of \citet{2009ApJ...699..254H} for GCs in a combined sample of {\it HST}/ACS BCGs, although there appear to be fewer large objects in the Harris sample than in M87; whether this difference is real (perhaps because such objects are preferentially found at large radii) or due to selection effects is unclear.

\begin{figure*}
%\epsscale{0.85}
\epsscale{0.9}
\plotone{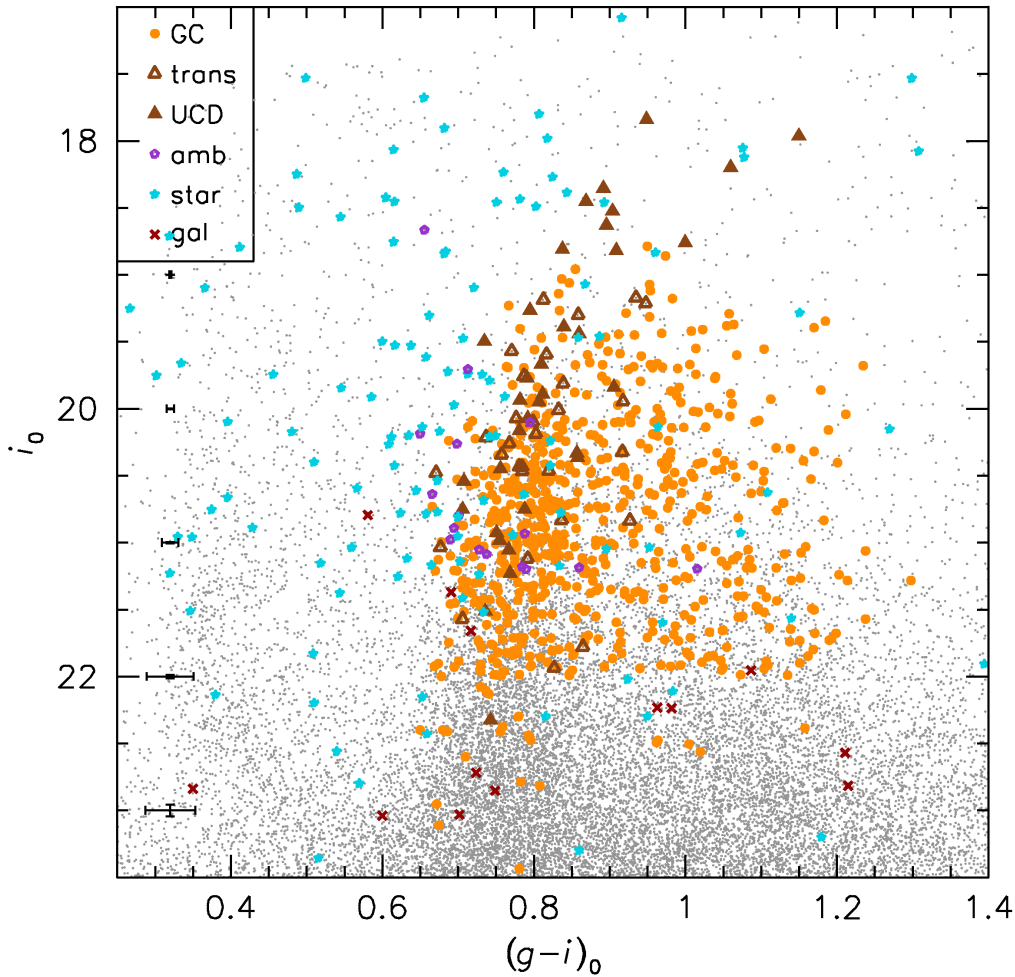}
\figcaption[M87GCcol5av.ps]{\label{fig:CMD}
Color-magnitude diagram of objects around M87.
Gray points are all point sources in the CFHT field-of-view,
while the other symbols are spectroscopic
objects identified as GCs, UCDs, transition objects, foreground stars, 
background galaxies, and ambiguous objects (in the $\sim$~150--350~\kms\ range),
as identified in the legend at upper left.
All objects with spectroscopic data, whether old or new, are included here.
Note that a few of the GCs were outside the CFHT field-of-view, so their photometry
shown here is based on approximate conversions from the Subaru $BVI$ photometry;
such conversions for background galaxies are less accurate, and we only show
cases with direct $gri$ photometry.
Typical photometric uncertainties are shown by error bars near the
left-hand axis.
}
\end{figure*}

Following the envelope of GC sizes from faint to bright magnitudes, we identify \rh~$\sim$~5.25~pc as an approximate upper boundary to the ``normal'' GC distribution. To classify objects as UCDs, we use a boundary of \rh~$>$~9.5~pc, which is somewhat arbitrary but consistent with findings in other environments from the literature (e.g., \citealt{2008MNRAS.389.1924F,2010ApJ...712.1191T}). Using this criterion, there are $34$ UCDs that are confirmed members of M87,
including $18$ that we have newly idenfitied. Objects in the intermediate size range of \rh~$\sim$~5--10~pc may include a mix of GCs and UCDs, and for now we designate them as transition objects\footnote{A notable Local Group object in the transition region is $\omega$~Cen (\rh~$\sim$~7--8~pc), while M54 and G1 (with \rh~$\sim$~3--4~pc)  would be considered ``normal'' GCs. NGC~2419 would also be classified as a UCD with our criteria.}.

We illustrate our adopted size boundaries in Figure~\ref{fig:sizes}. We emphasize that although many previous studies of UCDs have included a minimum
luminosity as a classification criterion (e.g., $M_i \la -11$), the low luminosities of UCDs that we have discovered here demonstrate that such a boundary is
an artifact of observational selection effects (see \citealt{2011arXiv1109.5696B} for a fuller discussion). At very faint magnitudes, $M_i \ga -9$, M87 may harbor a population of extended clusters as found in the Milky Way. This could include some bona fide GCs that have undergone relaxation-driven expansion, but our spectroscopic sample does not go faint enough for this complication to be a worry.

Even the ``pure'' samples may include some misclassified objects (because there may not be clean boundaries in principle, and in practice our sizes are only accurate at the $\sim$~10--15\% level anyway), but these should be very few and would not compromise our analyses in general. Slightly more problematic is the large sample of spectroscopic objects {\it without} measured sizes, as these will include a substantial population of of unrecognized UCDs and transition objects. We estimate this ``contamination'' fraction will be $\sim$~50\% at the bright end  ($i_0 \la 20$), and $\sim$~10\% for fainter objects, and so we will generally excluded the brighter objects from our kinematical analyses.

\begin{figure}
%\epsscale{0.85}
\epsscale{1.2}
\plotone{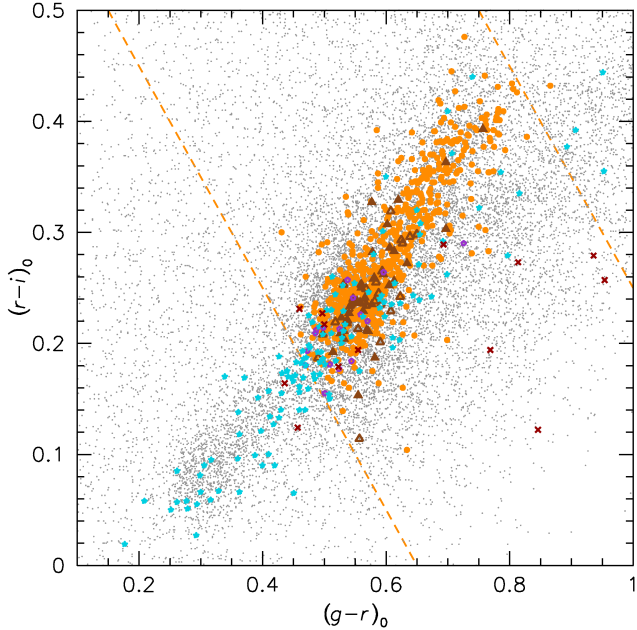}
\figcaption[M87GCcol4ab.ps]{\label{fig:colcol}
Color-color diagnostic diagram of objects around M87.
Symbols are as in Figure~\ref{fig:CMD}.
Dashed lines show GC color boundaries of $(g-i)_0=$~0.65 and 1.25.
}
\end{figure}

Color and magnitude diagnostic diagrams are shown in Figures~\ref{fig:CMD} and  \ref{fig:colcol}. An important result from these diagrams is the color distribution of the UCDs and transition objects.  These are mostly drawn from the blue side of the color distributions, and they appear to follow a coherent color-magnitude relation that is tilted and distinct from the general blue GC subpopulation. \citet{2008AJ....136..461E} show that the nuclei of Virgo dwarf elliptical galaxies follow a similar offset relation, with the brightest Virgo UCDs falling at the upper end of said relation, as possible evidence of the ``threshing" hypothesis for UCDs. Our observation that UCDs and transition objects have a similar color--luminosity relation provides evidence that most of the transition objects are in fact an extension of the UCD population rather than being extended GCs, although this conclusion is necessarily preliminary.

A more in-depth discussion of the M87 UCDs and transition objects is covered in another paper \citep{2011arXiv1109.5696B}. One additional comment here is that Figure~\ref{fig:sizes} suggests that the size-luminosity correlation for UCDs is not as strong as conventionally thought (e.g., \citealt{2008AJ....136..461E}).

\subsubsection{Foreground Stars}\label{sec:stars}

\citet{2001ApJ...559..828C} used a combination of $C-T_{1}$ color and radial velocity to separate Galactic foreground stars from GCs. All objects with $v > 200$~\kms\ and a color in the range $0.8 \le C-T_1 \le 2.35$ were classified as GCs, while objects with $v < 200$~\kms\ (regardless of color) were classified as stars. In this paper we also use a combination of photometry and radial velocity to select GCs, with the addition of high-resolution {\it HST} imaging where available. We supplement the Subaru/Suprime-Cam photometry from \citet{2006MNRAS.373..588T} with $gri$ CFHT/Megacam photometry (see \citealt{2009ApJ...703..939H}; \S\ref{sec:photom}) and $gz$ {\it HST}/ACS data \citep{2009ApJS..180...54J} when possible; the original H+01 $CT_{1}$ photometry is also used.

Since M87 has a systemic velocity of $1307 \pm 8$~\kms\ \citep{2000MNRAS.313..469S} and the GCs have a large velocity dispersion ($\sim 350$--400~\kms; \citealt{2001ApJ...559..828C}) it is likely that some true M87 GCs will have radial velocities comparable to foreground stars. If we assume that the Galactic halo is purely pressure supported, the stellar velocity distribution should peak at a heliocentric radial velocity of 40~\kms\ (assuming $v_{\rm LSR} =$~244~\kms; \citealt{2009ApJ...704.1704B}) with an approximate dispersion of 120~\kms. Therefore, foreground stars with velocities larger than $\sim $~300~\kms\ should be uncommon, and those $>$~500~\kms\ nearly nonexistent. Expectations for M87 are less clear: the velocity dispersion of the GC system is large and varies with radius, substantial rotation could be present, and, as we discuss below, there is compelling evidence for kinematical substructure. Especially for the latter reason, we do not perform a full likelihood analysis to assess M87 membership probability that would rely on the assumption of a Gaussian velocity distribution. Nonetheless, objects with very low radial velocities ($<$~150~\kms) are much more likely to be foreground stars than GCs.

We initially consider all objects with $v > 350$~\kms\ to be GCs and all objects with $v < 150$~\kms\ to be stars, unless there is strong evidence to the contrary. Unfortunately this (unavoidable) selection makes it more difficult to identify some of the most obvious intergalactic GCs, as we will discuss later. There are three objects (from the older spectroscopic data sets) apparently below this lower boundary that are {\it HST} confirmed GCs: S7017 ($v = 94\pm101$~\kms),  S7006 ($v = 104\pm58$~\kms), and S7007 ($v = 138\pm94$~\kms). {\it HST} imaging is among the most trustworthy methods for separating stars from GCs, since nearly all luminous M87 GCs are reliably resolved with WFPC2 or ACS images of typical duration. A single {\it HST}-confirmed foreground star (S161) lies above the upper velocity limit of 350~\kms, but it has a very large velocity uncertainty ($v = 446\pm246$~\kms). 

There are 50 objects in the combined catalog in the most ambiguous range, $150 < v < 350$~\kms. Of these, 11 have certain classifications from {\it HST} imaging: three are true GCs, and the other eight stars. Of the remaining 39 objects, the highest radial velocity is 289~\kms\ (even though the sample is bounded at 350~\kms), suggesting that the bulk of these objects are indeed foreground stars. The color-color diagram (Figure~\ref{fig:colcol}) gives supporting evidence. The locus of spectroscopically confirmed stars begins blueward of the GC sequence, crosses this sequence in the region of metal-poor GCs, and extends far to the red. Many ambiguous objects which lie far from the GC sequence can be conclusively identified as foreground stars (and are marked as such in the diagrams). However, it is evident that halo dwarfs in the approximate color range $0.5 \la (g-r)_0 \la 0.6$ cannot be cleanly separated from GCs with the existing photometry. For this paper, we classify all of these uncertain objects (marked as open purple pentagons in Figures~\ref{fig:CMD} and \ref{fig:colcol}) as stars, but note that a subset may be true GCs.

Two objects in the velocity range $350 < v < 500$~\kms do not have {\it HST} imaging. S1644 ($v = 496\pm68$~\kms) is in the region in $gri$ color-color space occupied by metal-rich GCs and we classify it as a GC. H58443 ($v = 441\pm24$~\kms) has colors consistent with either a very metal-poor GC or a foreground star and is likely to be a GC.

The data show some unusual features outside of the ambiguous velocity range discussed above. Two of the reddest objects in the Hectospec sample (H66419 and H20573) also have low radial velocities for M87 members: $596\pm11$~\kms\ and $683\pm10$~\kms, respectively. Both objects have high signal-to-noise spectra and the velocities are not in doubt. However, these radial velocities are close to or exceeding Galactic escape velocity at their likely (but unmeasured) distances. Therefore, it is unlikely that they are foreground stars.

As an aid to dynamical analyses and future observations, in Table~\ref{tab:stars} we have compiled a list of the probable spectroscopic Galactic foreground stars. This table does not include objects from \citet{1997ApJ...486..230C} and \citet{2000AJ....119..162C} that were identified as stars but without published radial velocities (S178, S221, S246, S354, S681, S996, S1610). These objects all have colors very dissimilar from those of GCs and are unlikely to be re-observed in the future.\footnote{Some other surveys, such as \citet{2008MNRAS.389.1539F}, also unfortunately did not publish their lists of foreground stars and so are likewise excluded.}

\subsubsection{Cleaning the Catalog}\label{sec:clean}

A small number of objects from H+01 have velocities consistent with being GCs but with photometry or other information that contradicts this classification. We discuss these objects in turn.

The object S923 (mentioned in \S\ref{sec:merge}) was previously observed by H+01 with a reported $v = 1943\pm134$~\kms. Our new LRIS velocity is catastrophically different at $2777\pm26$~\kms, which would place it among the highest reported for an M87 GC. The spectrum is of high quality and the object is clearly very metal-poor, consistent with its blue color ($g-r = 0.49$). An archival WFPC2 image suggests an unusual, asymmetric morphology, with a resolved core (of size $\sim 2$ pc) and two small protuberances. Given the morphology and the outlying velocity, we exclude this object from our analysis; better {\it HST} imaging is needed for this very interesting candidate.

S7023 presents an unusual case. The object closest to the position reported by H+01 is clearly resolved in the CFHT image; it is also present in an archival ACS image and has a tight core with a diffuse envelope. This morphology and the outlying radial velocity ($415\pm134$~\kms) are generally consistent with the idea that it is an ultra-compact dwarf galaxy in the Virgo cluster with a half-light radius of tens of pc. However, its $g-r$ and $g-i$ colors are too red for a normal old stellar population---and, in addition, are inconsistent with the reported $C-T_{1}$ color. The nearest potential interloper is located 4\arcsec\ from the listed position and is a background galaxy. We exclude this object from our analysis but further investigation is warranted. Both S923 and S7023 are listed at the end of Table~\ref{tab:allcand3}.

S1602 has $v \sim 1010$~\kms, but a $C-T_1$ color of 3.42. The CFHT $g$-band image of S1602 shows that it is a close pair; both components have colors too red for a GC, and morphologies consistent with those of background galaxies, so we exclude it. In addition, S7008 and S7012, which have H+01 velocities nominally consistent with M87 membership, are also apparent background galaxies. We do not know if their identifications are incorrect or if these are additional catastrophic radial velocity failures; in any case, they are removed from the sample of GCs. S508, with a large reported velocity in H+01 ($v \sim$~5800~\kms), is another clear background galaxy.

Finally, we have resolved a subset of the ambiguous object identifications in H+01. For example, in the close pair S8052, one of the objects is a background galaxy, so the targeted GC is not uncertain. In the pair S8051, only one of the objects has $gri$ colors consistent with that of a GC. S9001 and S9002 are both close pairs for which one (ambiguous) member has a published radial velocity. We have obtained new velocities for one member of each pair; each disagrees with the published velocity. In the absence of other information, we assume that the new observations cover novel objects, and so we can now correctly identify S9001 and S9002.

\subsubsection{GCs in Dwarf Galaxies?}\label{sec:dwarf}

In our quest for a pure catalog of M87 GCs, we also flag any objects that have a reasonable chance of being bound to one of the dwarf galaxies near to M87 (NGC~4476, NGC~4478, NGC~4486A, NGC~4486B, IC~3443). To do this, we follow a similar approach to \citet{2007MNRAS.382.1342F} and define a tidal radius $R_{\rm t}$ for each dwarf:
\begin{equation}
R_{\rm t} \simeq R_{\rm p} \left(\frac{M_{\rm d}}{3 M_r}\right)^{1/3} ,
\end{equation}
where $R_{\rm p}$ is its distance from M87, $M_{\rm d}$ is its mass, and $M_r$ is the mass of M87 enclosed within $r=R_{\rm p}$. Both $R_{\rm t}$ and $R_{\rm p}$ should in principle be 3D radii, and the pericenter rather than instantaneous radius should be used, but we will assume the observed projected radii work well enough. If we do identify an object within a dwarf tidal radius, we examine its velocity relative to the dwarf, and compare this to the escape velocity, which we define as $v_{\rm esc} \simeq \sqrt{2} v_{\rm c}$, where $v_{\rm c}$ is the local circular velocity of the dwarf. If the object is still consistent with being dwarf-bound, we remove it from the overall kinematics sample.

To estimate the $M_{\rm d}$ values, we construct simplified models for the dwarfs based on spectroscopic and photometric data from the literature, and for $M_r$, we use the dynamical results for M87 from \citet{2011ApJ...729..129M}. In the end, we identify one object which may be bound to NGC~4478: H35970.

\subsection{The Final Catalog}

Accounting for these classifications, and combining the new velocities with those from the literature, gives a total sample of 737 objects associated with M87. 707 fall into our pure GC or transition samples, while the remainder are UCDs or other extended objects. The final catalog, including all potential M87 members, is given in Table~\ref{tab:allcand3}. Where relevant, the listed radial velocity is a weighted average of the available values. The $gri$ photometry is listed in this table; all other photometry has been included in a supplementary table (Table~\ref{tab:photsupp}), with the exception of the removal of some $CT_{1}$ measurements that strongly disagreed with other data. Except for the H+01 $CT_{1}$ data, the photometry is corrected for foreground reddening, as described here or in the relevant papers.\footnote{For the reader who wishes to make the correction, \citet{2001ApJ...559..828C} report a mean reddening of $E(C-T_{1}) = 0.045$ mag in the direction of M87. Note also that there were some complications in the photometric calibration of the Suprime-Cam $BVI$ data (see \citealt{2006MNRAS.373..588T}). Comparing to the CFHT $gri$ data, we find chip-level zeropoint variations of typically 0.1~mag for a subset of the Suprime-Cam chips; one should keep this caveat in mind when making use of the $BVI$ data.}.

Cluster astrometry in this catalog was recalibrated using the CFHT $g$-band image and USNO coordinates for all objects within the field of view. These coordinates have proven sufficiently accurate for reliable multi-object spectroscopy. A few sources from the literature could not be uniquely identified or fell on image defects and so retain their previous positions. In addition, a number of objects from the \citet{2006MNRAS.373..588T} catalog lie outside of the field of view of the CFHT image. 

The profusion of papers on GCs and UCDs in M87 has led to a corresponding proliferation in object IDs. We have attempted to keep our naming transparent while generally retaining principal IDs used in refereed publications with spectra. All objects within the field of view of the \citet{2009ApJ...703..939H} CFHT imaging have been given ``H" IDs and these are listed as alternate IDs where applicable.

\subsection{Globular Cluster Subpopulation Classification}\label{sec:class2}

Although one of the goals if this paper is to examine afresh the classification of M87 GCs into independent subpopulations, as a starting point we first adopt a classical bimodal separation into two components: a blue (metal-poor) subpopulation and a red (metal-rich) subpopulation. A detailed analysis of stellar populations from a subset of our spectra will be the subject of a follow-up paper, but in an initial inspection of the spectra, we see no sign of GCs younger than $\sim$~2~Gyr. This further motivates the interpretation of the color variations as tracing metallicity differences for old populations.

We classify objects into subpopulations using as many independent photometric estimates as possible. For $C-T_{1}$ and $g-z$ a simple color cut is used: $C-T_{1} = 1.42$ and $g-z = 1.20$, respectively. For GCs with $BVI$ and $gri$, we empirically define the GC sequence in color-color space and separate clusters using a line orthogonal to the GC sequence and passing through $V-I = 1.10$ and $g-i=0.93$, respectively. These boundary values are suggested by mixture modeling of the color distributions and are identical to those used in \citet{2006MNRAS.373..601T} and \citet{2009ApJ...703..939H}, respectively.

For 92\% of the GCs, the colors agree uniformly on the classification, or there is only a single color. Another 6\% (typically GCs close to the middle of the color distribution) have at least three independent sets of photometry that do not all agree; we choose the classification favored by two or more sources. Fourteen GCs of the total sample of 707 have equivocal evidence for classification. In these cases we use, in order of priority, $C-T_{1}$, $g-z$, $gri$, and $BVI$. 

Using this scheme, 486 (69\%) of the objects are classified as blue metal-poor GCs, with the remainder as red metal-rich clusters. For completeness, the UCD candidates have also been classified, but are not included in the above statistics.

\section{Photometric Properties of the Globular Cluster System}\label{sec:photprop}

Although \citet{2009ApJ...703..939H} has already analyzed the CFHT imaging data in detail, for consistency we repeat some of his analyses, and include various extensions of relevance to our overall photometric and spectroscopic study. In \S 5.1 we fit the radial surface density profiles of GCs using S\'{e}rsic functions. \S 5.2 discusses contamination, and \S 5.3 the ellipticity of the GC system. In \S 5.4, aided by spectroscopy, we examine trends of the GC colors and subpopulations.

\subsection{Radial Density Profiles}\label{sec:raddist}

The number density distribution of GCs with radius is important for a number of reasons, including as input to dynamical modeling of the GC system. The three-dimensional distribution is inferred from deprojection of the surface density, which for an extended system like M87's demands wide-field imaging. The metal-poor and metal-rich GC subsystems can be analyzed independently, and require derivations of their density profiles separately.

Several previous studies have derived surface density profiles for the separate GC subpopulations in M87, including \citet{2001ApJ...559..828C}, \citet{2006MNRAS.373..601T}, \citet{2007MNRAS.382.1947F}, and \citet{2009ApJ...703..939H}.

Nearly 700 spectroscopically confirmed M87 GCs are present in the CFHT photometric catalog. As shown in Figure~\ref{fig:colcol}, these empirically define the GC sequence in color-color space. The GCs scatter around a mean locus extending approximately from ($g-r$, $r-i$) = (0.48, 0.18) to (0.78, 0.43). We experimented with using rather stringent criteria in color-color space to select GCs, but found that the slight advantage in rejection of contaminants did not overcome the loss of many true clusters. Instead, we used the relatively loose color cuts in $g-r$ and $g-i$ from \citet{2009ApJ...703..939H}. Objects were selected over the magnitude range $19 < i < 22.5$, with the faint limit chosen to avoid incompleteness issues relatively close to the center of the galaxy. Extended galaxies were excluded using a cut on $g$-band FWHM. Metal-poor and metal-rich GCs were separated using a line perpendicular to the mean color locus  and passing through $g-i = 0.93$. 

We chose to fit the GC surface density profiles with \citet{1968adga.book.....S} functions for several reasons: (i) these have deprojected analytic forms suitable for use in dynamical modeling; (ii) they extrapolate to finite total GC numbers; and (iii) both the observed stellar components of galaxies and their simulated dark matter halos are reproduced well by S\'ersic-like functions, so it is only natural to suppose that the GC systems are as well.

The GC counts were binned in projected radius over the range 1\arcmin--29\arcmin. Corrections were made in the binned profile for the chip gaps present in the CFHT images. We assumed Poissonian errors both from the GC counts and, to be conservative, from a constant contaminant level of 0.5~arcmin$^{-2}$. We also assumed a systematic uncertainty in the contaminant level, set to 0.15~arcmin$^{-2}$ for the full sample and 0.10~arcmin$^{-2}$ for the individual blue and red subpopulations. 

We fit a S\'{e}rsic function of the form: $N_0 \exp[-(R/R_{s})^{1/m}]$, with normalization $N_0$, scale radius $R_{s}$, and index $m$. The contamination level was left to vary freely. Note that ``contaminant"  refers to any objects not part of the GC system of M87---these may be foreground stars, background galaxies, or potentially even intergalactic GCs with a much more extended spatial distribution than that of M87. Fits were done for the full sample and blue and red subpopulations separately. 

The best-fit contaminant levels are $0.51\pm0.16$, $0.18\pm0.17$, and $0.36\pm0.09$~arcmin$^{-2}$ for the full, blue, and red samples respectively. These are consistent with previous estimates: \citet{2009ApJ...703..939H} found background densities of $0.26$ and $0.36$~arcmin$^{-2}$ for the blue and red objects, respectively, and \citet{2006MNRAS.373..601T} reported a value of $\sim 0.2$~arcmin$^{-2}$ for blue ``intergalactic" GCs (IGCs). Note, however, that the very different luminosity functions of GCs and contaminants imply that the inferred density of interlopers will strongly depend on the selection function of candidates.

\begin{figure}
%\epsscale{0.85}
\epsscale{1.2}
\plotone{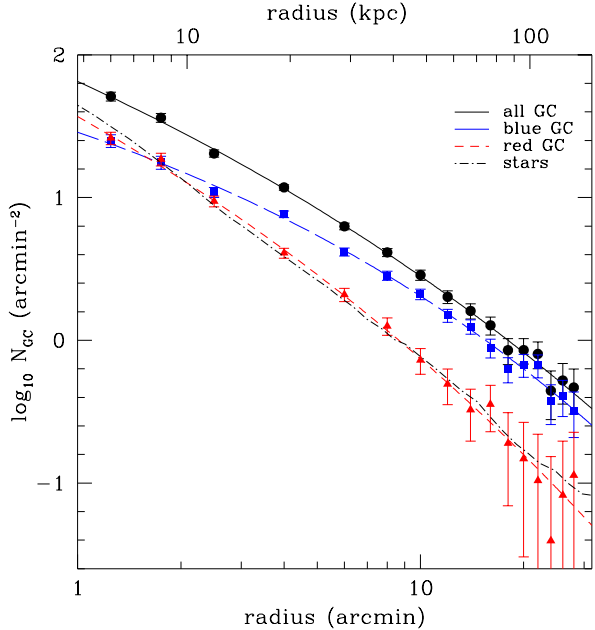}
\figcaption[m87sd2.eps]{\label{fig:sd}
Surface density profiles for GCs in M87. The full sample of GCs is given by the black circles, with the blue and red subpopulations plotted as blue squares and red triangles respectively. The S\'{e}rsic fits are overplotted in the appropriate colors as solid, long-dashed, and short-dashed lines (for full, blue, and red samples). The integrated light profile from \citet{2009ApJS..182..216K} is plotted as a dot-dashed line, and is a good match to the red GC profile over an extended radial range. The GC profiles are constructed in circular bins, and the galaxy light is plotted versus circular-equivalent radius: $r_{\rm SMA} (1-\epsilon)^{1/2}$.}
\end{figure}

The S\'{e}rsic fits are plotted, together with the surface density estimates, in Figure~\ref{fig:sd}. As is typical for GC systems (and well-established for M87; \citealt{2001ApJ...559..828C}), the metal-poor GCs are more extended than the metal-rich GCs, with the crossover point at a projected radius of $\sim 8$--10 kpc.  Out to the edge of our photometric data at $\sim 30\arcmin$, we see no evidence for either a sharp truncation of the outer density profiles, nor for a flattening that could mark the transition to an intergalactic or an obviously accreted component (cf~\citealt{2011MNRAS.414..770H}).  \S\ref{sec:transit} contains further discussion of this issue.

Figure~\ref{fig:sd} also plots the galaxy light distribution from \citet{2009ApJS..182..216K}. While the normalization of this M87 profile is arbitrary, it is clearly a good match to the surface density of the metal-rich GCs over a wide radial range (see also \citealt{2006MNRAS.373..601T}). The deviations occur in the central few kpc, where the GCs have a flat core-like distribution, and in the very outer halo ($\ga 100$ kpc). This lends credence to the generic idea that the metal-rich GCs formed along with the bulk of the field star population in early-type galaxies (see the review in \citealt{2006ARA&A..44..193B}). The normalization factor is such that the inferred mass in metal-rich GCs is $\sim 0.28$\% of M87 over the matching range in radius.

The derived parameters of the surface density fits are given in Table~\ref{tab:fits} (except for the background levels, which are given in Table~\ref{tab:interlop}). Table~\ref{tab:fits} also contains the results of fits for which the background level is fixed to the spectroscopic estimate determined in \S\ref{sec:specest} and listed in Table~\ref{tab:interlop}. It is challenging to assess the uncertainties on these parameters, since they are strongly coupled; the difference between the fixed and varying background fits gives a general sense of the constraints. Nonetheless, a range of values would likely provide reasonable fits. For example, modulo normalization factors, the $R^{1/4}$ fits presented in  \citet{2006MNRAS.373..601T} and \citet{2009ApJ...703..939H} (equivalent to $m=4$ S\'{e}rsic functions) generally match the data well. 

For reference, the implied globular cluster system (GCS) parameters for the fits with free background levels are as follows.  The GCS effective radius is $\simeq$~6.8\arcmin, 22\arcmin, 23\arcmin\ ($\simeq$~32, 106, 110~kpc) for the red subpopulation, blue subpopulation, and total system, respectively, with corresponding S\'{e}rsic indices of $\sim 5.3$, 3.7, and 5.5. The respective total number of GCs is $\simeq$~2000, 5500, 8900, although nearly half of the total GC population inferred from these fits is beyond the edge of our photometric field of view, and these values exclude the faint end of the GC luminosity function (with $i > 22.5$).

\subsection{Spectroscopic Estimates of the Contamination}\label{sec:specest}

Our S\'{e}rsic fits left the contaminant level as a free parameter. Here we combine the original, large radius DEIMOS spectroscopy and CFHT imaging to give an independent estimate of the density of interlopers.

The radial overlap between the DEIMOS and CFHT data begins at $\sim$~8\arcmin\ from the center of M87 and extends to the edge of the CFHT field at $\sim$~30\arcmin. Of the objects classified with DEIMOS and with CFHT photometry, there are 54 confirmed GCs and 44 confirmed contaminants. We then applied the selection function used for the surface density estimates in \S\ref{sec:raddist} to both the GC and contaminant samples, yielding 46 GCs and 20 contaminants. These samples were binned in radius as for the surface density estimates. This procedure allows multiple independent estimates of the contamination fraction, albeit with large individual uncertainties because of the small numbers of objects. The final surface density of contaminants in each radial bin is calculated by multiplying the local contaminant fraction by the overall surface density in that bin. The same method can be used on the blue and red GC subpopulations separately, though with necessarily larger uncertainties.

Table~\ref{tab:interlop} compares the spectroscopic and photometric estimates of the contaminant level for both the full sample of GCs and for the blue and red GCs. The listed spectroscopic values are a weighted average of the independent radial bins. There is excellent agreement between the photometric and spectroscopic estimates (and, indeed, even better agreement between the spectroscopic estimates and the photometric values of \citealt{2009ApJ...703..939H}). These comparisons suggest that our values are accurate estimates of the background level for the assumed selection function.

\subsection{Ellipticity}

Most kinematical and dynamical analyses using discrete tracers assume spherical symmetry for simplicity. This assumption could lead to  systematic errors in the mass modeling of real early-type galaxies, which are rarely spherical. M87 is notably aspherical, with its stellar isophotal ellipticity increasing from $\sim$~0.2 at 3\arcmin\ to $\sim$~0.45 at 13\arcmin\ \citep{2009ApJS..182..216K}. \citet{1994ApJ...422..486M} estimated the ellipticity of the GC system in the inner regions of M87, finding it roughly consistent with that of the galaxy light.

In this subsection we estimate the ellipticity of the GC system of M87, both as a whole and for the blue and red subpopulations. Such a subpopulation analysis has been done only rarely in the literature on early-type galaxies (e.g., \citealt{1997A&A...327..503K}). A number of methods have been used in the literature to determine ellipticities for sparsely sampled two-dimensional data; here we use the method of moments \citep{1980MNRAS.191..325C,1994ApJ...422..486M}. In general, maximum likelihood is a preferable approach, but we found that maximum likelihood estimates using the formalism of \citet{1995AJ....109.1033M}, when applied to radially binned data, were frequently unstable.

\begin{figure}
%\epsscale{0.85}
\epsscale{1.2}
\plotone{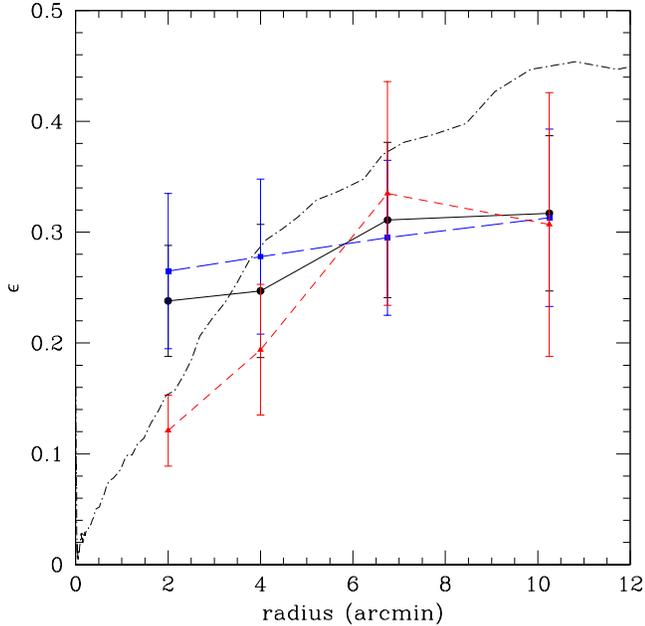}
\figcaption[ell.eps]{\label{fig:ellip}
Ellipticity profiles for the M87 GC system. Symbols are as in Figure~\ref{fig:sd}, including the integrated light profile of M87 from \citet{2009ApJS..182..216K}.
Circular-equivalent radii are plotted.}
\end{figure}

\begin{figure*}
%\epsscale{0.85}
%\epsscale{1.2}
\epsscale{0.9}
\plotone{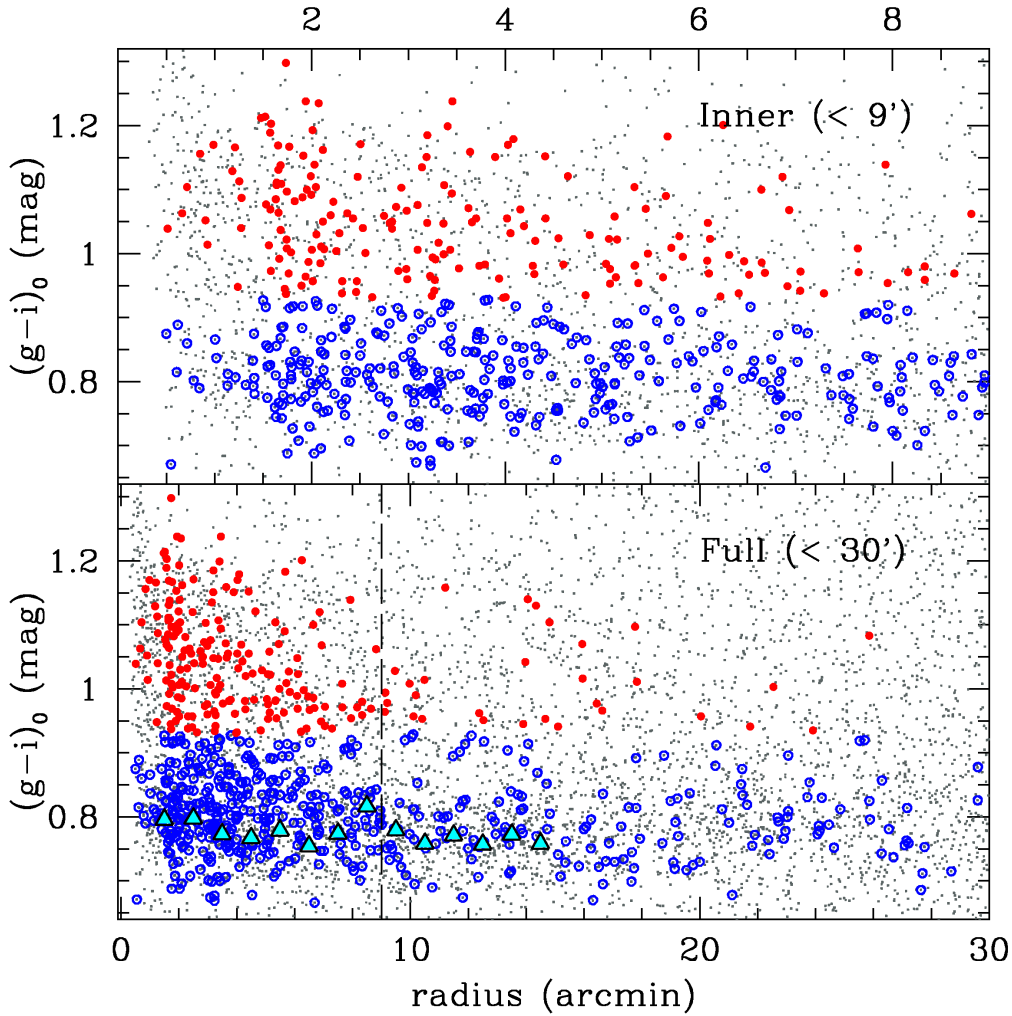}
\figcaption[c]{\label{fig:colgrad}
M87 GC colors versus galactocentric radius.  Small black points show the photometric
sample, while colored points show the spectroscopically-confirmed sample, with blue
and red colors denoting our primary color classification [using a division
at $(g-i)_0=0.93$]. The cyan triangles show our
estimates of the blue GCs peak location in radial bins using a photometric sample
(see main text for details).
The lower panel shows nearly the full radial range of confirmed GCs (out to 30\arcmin);
the upper panel zooms in on the inner part of the cluster system within 9\arcmin. 
}
\end{figure*}

We bin the data into four unequal radial bins in the range 1\arcmin--13\arcmin, with the outer limit set by incomplete spatial coverage due to the chip gaps. Both the ellipticity and position angle are allowed to vary. The derived ellipticity values are plotted in Figure~\ref{fig:ellip}, along with the circularized galaxy light from \citet{2009ApJS..182..216K}. The ellipticity profile of the GC system as a whole appears somewhat more constant than that of the galaxy light. The red GCs have a steeper ellipticity profile, out to the last measurement where their surface density becomes comparable to that of the background and thus very uncertain. The ellipticity profile for red GCs has a similar slope to that of the galaxy light. By contrast, the blue GCs have a nearly constant ellipticity profile, consistent with $\epsilon \simeq 0.29$ at all radii. The ellipticity values for the full GC sample in the inner regions agree with those obtained by \citet{1994ApJ...422..486M}.

The ellipticity and position angle estimates are given in Table~\ref{tab:elliptic}.

\subsection{Beyond Bimodality}\label{sec:beyond}

The preceding analyses have operated under the simplifying assumption that the M87 GCs fall into two standard subpopulations (red and blue) whose identities are invariant with respect to GC luminosity and galactocentric distance.  However, the richness of our data set allows us to explore the extent to which this simple scenario is valid.
Below we examine color trends with radius (\S\ref{sec:colgrad}), color substructures (\S\ref{sec:colsub}), luminosity trends (\S\ref{sec:lumtrend}), and connections with field stars (\S\ref{sec:cstars}).

\subsubsection{Color Gradients}\label{sec:colgrad}

\begin{figure*}
%\epsscale{0.85}
\epsscale{1.15}
\plotone{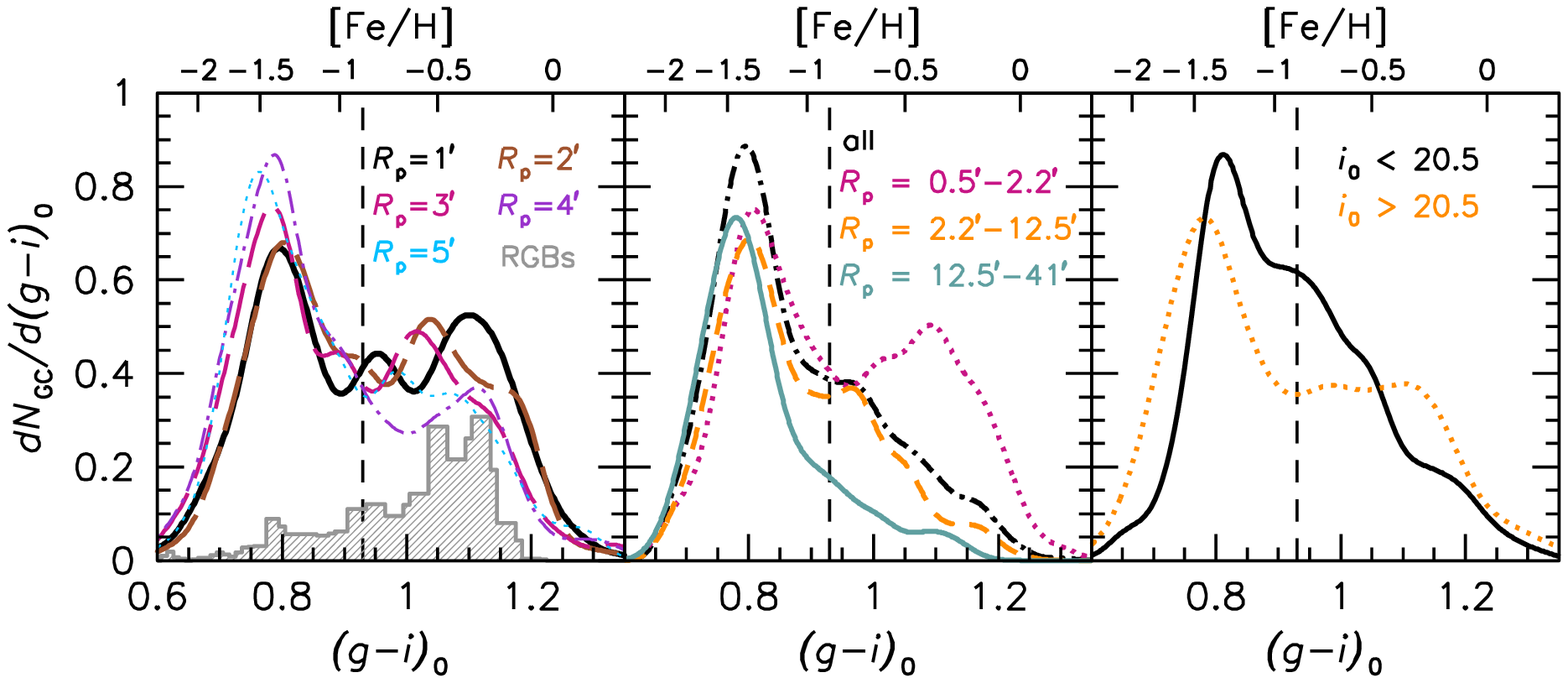}
%\epsscale{0.35} \plotone{dens.eps} \plotone{M87GCcol9b.ps} \plotone{bright.ps}
\figcaption[M87GCcol9be.eps]{\label{fig:coldist}
Distributions of colors for M87 GCs, with a small amount of smoothing applied,
and arbitrary normalizations. The left panel shows the central photometric sample in several radial bins 
as indicated in the panel legend. The middle panel shows the GCs with spectroscopic confirmation, out to larger
distances, in several radial bins. The right panel shows the central photometric GCs binned by bright and faint
magnitudes. The uncertainties in individual color measurements are generally $\lsim$~0.02~mag.
The gray shaded histogram in the left panel shows RGB data from \citet{2010A&A...524A..71B},
converting from metallicity to color using a relation from \citet{2010AJ....140.2101S};
note that the RGB data are not reliable for $(g-i)_0 \gsim 1.12$ ([Fe/H]~$\gsim -0.3$).
}
\end{figure*}

We first examine the color distribution of GCs versus distance $R_{\rm p}$ in Figure~\ref{fig:colgrad}, which shows the $(g-i)_0$ colors for the full photometric sample of GCs and for the spectroscopic objects which pass the photometric selection criteria (95\% of the total sample). As the bulk of the spectroscopic objects were selected without regard to color (especially our new LRIS and Hectospec samples), the bivariate distribution of spectroscopic objects should be a reasonable representation of the full set of GCs, as it generally appears to be. The reader can look ahead to Figure~\ref{fig:coldist} to see density plots of $(g-i)_0$
for both photometric and spectroscopic GC samples broken down by projected radius and magnitude.

The plot of the full radial range shows several remarkable features. Although the stated division between blue and red GCs appears reasonable in the inner region, in the radial range $\sim$~7--14\arcmin\ ($\sim$~35--75~kpc) the distribution of {\it red} GC colors appears to change: the peak either shifts blueward (as assumed in the power-law models of \citealt{2009ApJ...703..939H}---see his section 6) or is replaced by a population of intermediate-color objects. Intermediate-color GCs were in fact reported in the central regions using {\it HST}/ACS (\citealt{2006AJ....132.2333S}; also visible in \citealt{2006ApJ...639...95P} as residuals from their bimodal model), and the new component at larger radius may be an extension of this inner ``subpopulation''. An inspection of the inner radial plot shows that the overall red GC color transition may start at a projected radius of $\sim$~4\arcmin\ ($\sim$~20~kpc), outside of which the principal red $(g-i)_0$ edge of the metal-rich GCs begins to shift. 

At large radii, the blue GCs continue to have a well-defined and roughly constant-color peak, but with noticeable radial variations. \citet{2009ApJ...703..939H} reported a color gradient in the blue GCs within a radius of $\sim$~12\arcmin, and with a slope of $-0.025$ versus log radius. We can approximately reproduce this result through a linear least-squares fit to the photometric sample of blue GCs. However, we can now see that the fit is clearly affected by the presence of intermediate-color objects at small radii, and the evidence for a slope at radii $>$~3\arcmin\ is marginal. 

To more accurately trace the mode of the blue GC distribution, we defined a sample of fainter GCs ($21 < i_{0} < 23$) to eliminate a portion of the intermediate-color population (which we find below to be a stronger component at brighter magnitudes), while retaining good statistics. Density estimates with a fixed Gaussian kernel of 0.03 mag\footnote{This value is close to the optimal \citet{1986desd.book.....S} kernel bandwidth for typical values of $\sigma$ and $N$ for bins in the blue peak.} were constructed in 1\arcmin\ radial bins. 

The blue mode of each bin is plotted in Figure~\ref{fig:colgrad}, out to a radius of  $\sim$~15\arcmin\, after which contamination of the photometric sample becomes significant.  The ``bump'' in the color at $\sim$~8\arcmin\ appears real in that it reflects the actual color distribution of GCs (rather than just small number statistics). However, outside of a radius of $\sim$~3--5\arcmin\ ($\sim$~15--25~kpc), there is no evidence for a monotonic radial gradient among the blue GCs. This transition from an inner radial gradient to an essentially constant color has also been seen in the GC systems of the Milky Way \citep{2001stcl.conf..223H} and of the giant elliptical NGC 1407 \citep{2011MNRAS.413.2943F}, with implications that we will discuss in \S\ref{sec:disc}.

We also notice another color ``bump'' in the blue GCs at $R_{\rm p}\sim$~21--27\arcmin\ ($\sim$~100--130~kpc), which is confirmed by the spectroscopic sample. These objects are spread out nearly uniformly in azimuth, so it is unlikely (but still possible) that Galactic cirrus on a smaller resolution than the \citet{1998ApJ...500..525S} dust maps is contributing to this feature. Internal reddening gradients could also play a role in the color changes in principle, but would require invoking {\it ad hoc} dust distributions\footnote{Reddening from dust (whether internal or external) would produce identical color shifts for both blue and red subpopulations, but this test is difficult to apply because the red GCs do not maintain a clear, contiguous color peak over a large range in radius.}.

As previously discussed in \S\ref{sec:photom}, we emphasize that possible photometric gradients are dependent on accurate large scale photometric calibration, which we cannot guarantee. Therefore, these conclusions are still tentative, pending the forthcoming photometry from NGVS.

\subsubsection{Color Subpopulations and Substructure}\label{sec:colsub}

An impression garnered from both panels is the presence of substructure in the GC color distribution, especially in the red GCs. No firm conclusions can be drawn from the spectroscopic data alone, but one has the impression of clumping in the metal-rich subpopulation, both at small radii and in the outer regions.

To explore the red GC variations further, we constructed density plots of  the $(g-i)_0$ color of all photometric (not just spectroscopic) GCs. A fixed kernel size of 0.03 mag was used. Since our focus was on the red GCs, we binned in ellipses, starting with a semi-major axis of 1\arcmin\ and with the ellipticity equal to that of the galaxy light at that radius. The results for a range of 1--5\arcmin\ in semi-major axis are shown in the left panel of Figure~\ref{fig:coldist}.  The blue GC subpopulation looks similar in each bin, with some evidence for a mild negative color gradient as already discussed. However, the distribution of red GCs varies significantly among the bins. There is no clear monotonic trend, and some of the bins (especially the innermost one) show evidence for multiple subpopulations within the ``red'' peak.

Small number statistics are very unlikely to be the cause of these variations. There are more than 100 red GCs in each bin (excepting the last one, which has 92). If we assume conservatively that the red GCs have an intrinsic color distribution with $\sigma = 0.1$ in $(g-i)_0$, then the standard error of the mean is $\leq$~0.01~mag per bin. However, the scaled median absolute deviation\footnote{We use this robust estimator of the scale (which asymptotically approaches $\sigma$ for a normal distribution) since the underlying distribution of the red mode is not necessarily Gaussian.} of the red peaks in these five bins is 0.048 mag. By contrast, the blue GCs show much more uniformity, and the median absolute deviation of the blue peak color is 0.012 mag.

To extend this analysis to larger radii, we consider next the spectroscopic sample, which is virtually free of contamination. Note that later in this paper we will elect to use only our new data set because of the large impact on the kinematical and dynamical analyses of occasional catastrophic radial velocity errors in the older data, but such errors do not affect our photometric analyses and here we use the full combined sample.

The middle panel of Figure~\ref{fig:coldist} shows the color distribution of the spectroscopic sample in several radial bins. At the small radii corresponding roughly to the ACS field-of-view ($R_{\rm p} \la$~2.2\arcmin\ or $\la$~10~kpc), the classic blue/red color bimodality is evident. At larger radii, the red subpopulation becomes less dominant as expected based on the relative radial distributions of blues and reds (see \S\ref{sec:raddist}). However, the average colors of the red distribution also become dramatically bluer, and the data are more suggestive of a shift between separate red and intermediate-color peaks than of a simple gradient of the red peak. This impression is supported by kinematic evidence in \S\S\ref{sec:param} and \ref{sec:subpop}.

This curious intermediate-color peak does not seem to be a manifestation of a very extended redwards tail of the blue subpopulation; if so, it would be strong at all radii (see also \citealt{2009MNRAS.397.1003F}). The third ``peak'' clearly has a spatial distribution that is more extended than the ``classic'' redder peak but less extended than the blue peak. For the sake of examining the third peak for unusual kinematical behavior later in this paper, we carry out a trimodal fit to the photometric data over the radial range  $R\sim$~1.5\arcmin--4\arcmin, using alternatively circular or elliptical annuli. The three peaks inferred make roughly equal contributions to the GC numbers and we adopt color boundaries of $(g-i)_0\sim$~0.86 and $\sim$~1.01.\footnote{For example, for the circular fit the modes of the subpopulations are $(g-i)_0$ = 0.76, 0.94, and 1.08.}

\subsubsection{Bright GCs}\label{sec:lumtrend}

We next consider luminosity dependencies of the GC subpopulations. Separating out the photometric GCs with magnitudes in the range $i_0$=19--20.5, we show the color distribution for the small-to-intermediate-radius regions (0.5\arcmin--9\arcmin\ or 2--40~kpc) in the right panel of Figure~\ref{fig:coldist}, along with the fainter GCs for comparison. The colors of both the blue and the red subpopulations are clearly different between the bright and faint GCs. The blue shift is the well-known blue-tilt.  The red shift was previously noticed by \citet{2009ApJ...703..939H} as a tentative and surprising ``negative mass-metallicity relation'', but we regard the trend as secure, and indicative of a third, intermediate-color subpopulation rather than of a red-tilt.

Given the small number statistics for the bright objects, we are not able to determine where exactly any magnitude boundary is between the bright and faint objects (we have adopted $i_0=20.5$ as a rough value). Nor are we able to determine much about the spatial distributions of the bright objects except to note that the intermediate-color objects are somewhat more extended in radius than the redder objects. We do see some peculiar indications that the locations of the color peaks for both bright and faint objects shift dramatically inside radii of $\sim$~2\arcmin\ ($\sim$~10~kpc).  Our present data set is too limited to pursue this issue in detail, but we will find in \S\ref{sec:param} that the GC kinematics are also unusual inside this radius.

For additional insights, we turn to the spectroscopic sample, where we find that the intermediate-color objects, with $(g-i)_0 \sim$~0.85--0.95,
include a larger fraction of bright ($i < 19.5$) and large ($r_{\rm h} \ga$~10~pc) objects than are found at other colors, particularly at large radii. This result supports the idea that there is a third subpopulation with an unusual luminosity distribution, and further implies that this subpopulation
is more strongly ``contaminated'' with UCDs---although this fraction is still low ($<$~10\%), with the fainter and smaller  intermediate-color objects showing no unusual photometric properties.

The possible convergence of the GC color-magnitude distribution to a unimodal distribution at the bright end has been previously discussed in other galaxies by \citet{1998AJ....116.2854O}, \citet{2003AJ....125.1908D}, \citet{2006ApJ...636...90H}, \citet{2009AJ....137.4956R}, and \citet{2011MNRAS.413.2943F}. Also relevant is the growing number of early-type galaxies with centrally-concentrated, intermediate-color GC subpopulations without strongly unusual luminosity functions. Examples include NGC 4365 (\citealt{2005A&A...443..413L} and references therein) and NGC 5846 \citep{2006A&A...455..453C}.

We confirm this pattern for M87, but in addition, the high-quality, spatially extended and spectroscopically-confirmed photometric data set allows us to find more detailed evidence for substructures in color and magnitude space. Our next step will be to consider the new dimension of {\it kinematics} to disentangle the subpopulations.

\subsubsection{The Connection to Field Stars}\label{sec:cstars}

Before continuing with the kinematics analysis, we conclude this section with a look at the results for resolved red-giant branch (RGB) stars from deep ACS imaging of the central regions of M87 ($\sim$~2\arcmin\ or $\sim$~10~kpc) by \citet{2010A&A...524A..71B}. These authors inferred a stellar metallicity distribution which we superimpose on our central GC color distributions (Figure~\ref{fig:coldist}, left panel) after using the results of \citet{2010AJ....140.2101S} to transform $(g-i)_0$ color to metallicity. Although such comparisons are ultimately critical for establishing links between GC and field star subpopulations, unfortunately it is not yet possible to reach strong conclusions. The high-metallicity RGB peak that nominally seems to match up well with the far-red GC peak is actually truncated above $(g-i)_0 \sim 1.12$ ([Fe/H]~$\sim -0.3$) because of observational limitations. We already know that the field stars in M87 are redder overall than the corresponding red GC peak (e.g., \citealt{2007MNRAS.382.1947F,2010MNRAS.406.1125S}).

The apparent low- and intermediate-metallicity RGB peaks may be artifacts of the metallicity modeling functions as well. Still, it remains a remarkable fact that the rich complement of metal-poor GCs is accompanied by only an extremely meager metal-poor field star population (as is already known in other cases such as the Milky Way and M31, and discussed in detail by \citealt{2001stcl.conf..223H}). The origin of this high ``specific frequency'' is a great mystery,  but it does allow metal-poor stellar halos in distant systems to be probed indirectly using GCs. 

\section{Kinematics}\label{sec:kin}

Here we provide various characterizations of the kinematics of the M87 GC system, including its rotation, velocity dispersion, and velocity kurtosis. A key point throughout is to compare the results using the previous versus the new velocity data. Also, unless otherwise stated, our ``GC'' samples used for kinematical and dynamical analysis have been pruned of UCD-candidates (any with sizes of $r_{\rm h} \ga$~5~pc or magnitudes $i_0 < 20$), owing to the likelihood that these have systematically different orbits from the bulk of the GC system.

One complication is that as discussed in \citet{Roman11}, analysis of position-velocity phase space with M87 GCs reveals a large unrelaxed substructure in the $\sim$~50--100~kpc region. For the moment will we ignore this fact and proceed with conventional approaches to kinematical and dynamical analysis that assume a well mixed and relaxed system. In most galaxies studied to date with GC kinematics, the velocity data were either not extensive or high resolution enough to detect substructures, while even in M87 there may well be additional lurking substructures that we have not yet clearly identified. Therefore the conventional approaches may have widespread liabilities which are beyond the scope of this paper to consider in detail (cf.~\citealt{2006ApJ...643..154Y}).

We begin in \S\ref{sec:gen} with an overview of the individual GC positions and velocities, then present our ``kinemetric'' techniques along with some initial analyses in \S\ref{sec:kinem}. We carry out more detailed kinematical analyses of subpopulations in \S\S\ref{sec:param} and \ref{sec:subpop}. Looking further forward, \S\ref{sec:transit} will examine potential halo transitions, with \S\ref{sec:merger}  discussing possibilities for a recent merger.

\subsection{General Overview}\label{sec:gen}

\begin{figure}
%\epsscale{0.85}
\epsscale{1.2}
\plotone{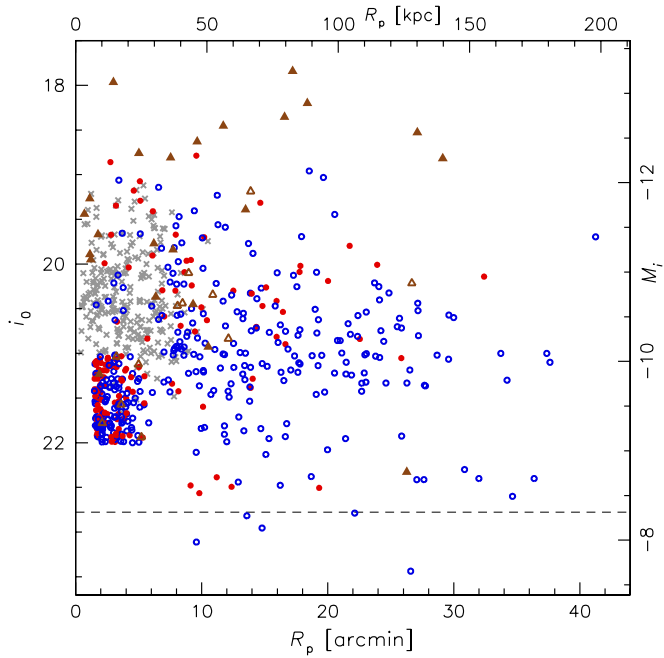}
\figcaption[M87GCpos9l.ps]{\label{fig:magrad}
Magnitude versus galactocentric radius for spectroscopic objects associated with M87.
The $i_0$-band magnitude is shown on the left axis, and absolute magnitude on
the right, with a dashed line indicating the approximate GCLF turnover magnitude.
Gray $\times$ symbols show the previous data compilation from \citet{2001ApJ...559..812H},
while circles show the new data, with open blue and filled red circles indicating
``blue'' and ``red'' GCs, respectively.
The new LRIS data are found at small radii between $i=21$ to $22$;
the DEIMOS data are at large radii down to $i \sim23$, and the Hectospec data are
at a range of radii down to $i \sim 21$.
Filled and open triangles show UCDs and transition objects
(with gray and brown colors for old and new data, respectively).
}
\end{figure}

A succinct way to illustrate the novel regions of GC parameter space in M87 covered by the new data is the galactocentric radius versus magnitude (Figure~\ref{fig:magrad}). The radial range is now extended from $\sim$~8\arcmin\ ($\sim$~40~kpc) previously to almost 40\arcmin\ ($\sim$~200 kpc), and the magnitude limit deepened from $i_0\sim$~21 to $i_0\sim$~22.5. The one area where the new data do not surpass the previous is a narrow radial range around $\sim$~6\arcmin ($\sim$~30~kpc). It should also be noted that the central regions where the old and new datasets have the most overlap are completely disjoint in the magnitude ranges probed. Therefore dependencies of the kinematics on magnitude should be kept in mind as possible explanations for any discrepancies found between the old and new results.

\begin{figure*}
\epsscale{1.17}
\plottwo{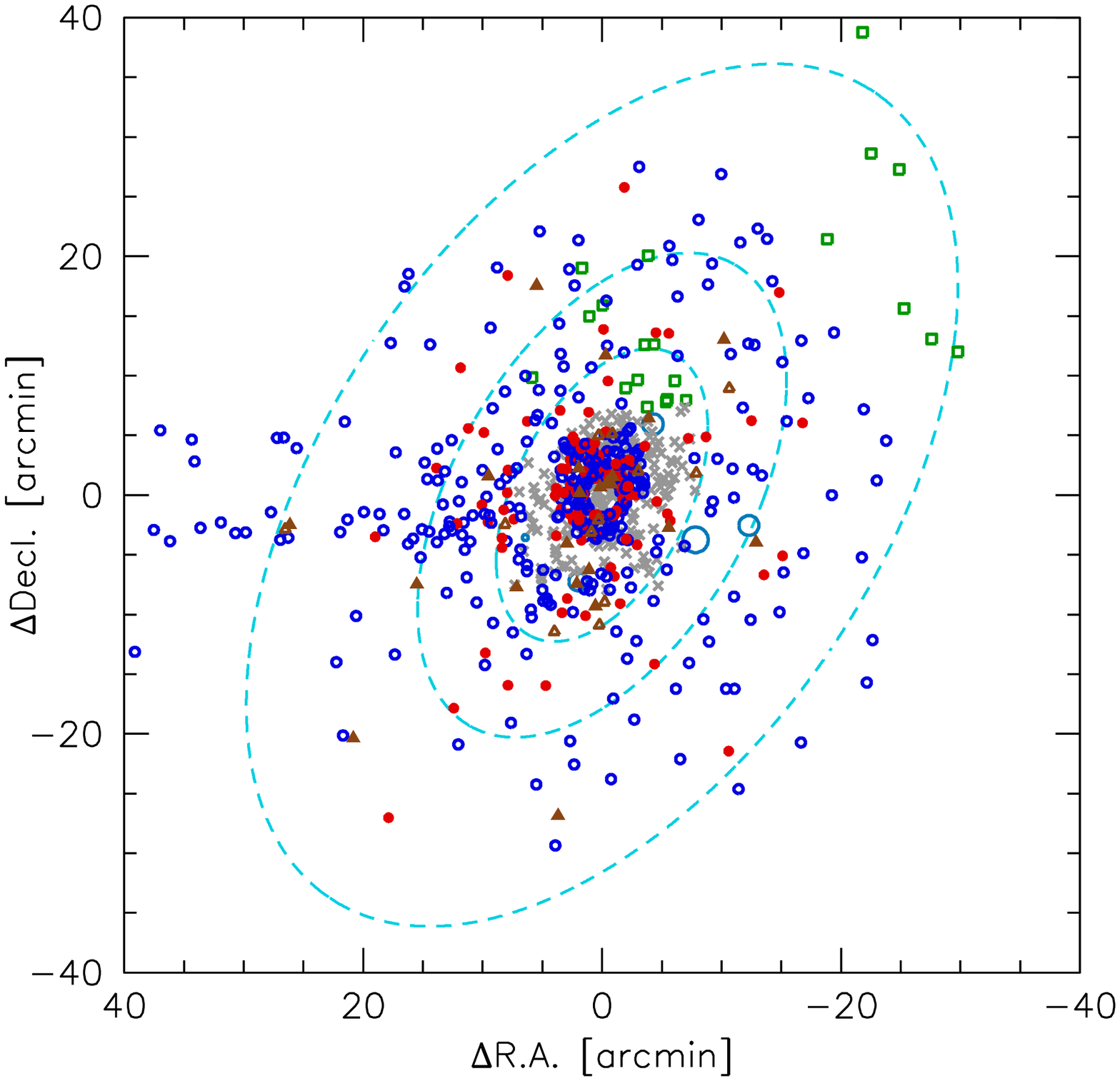}{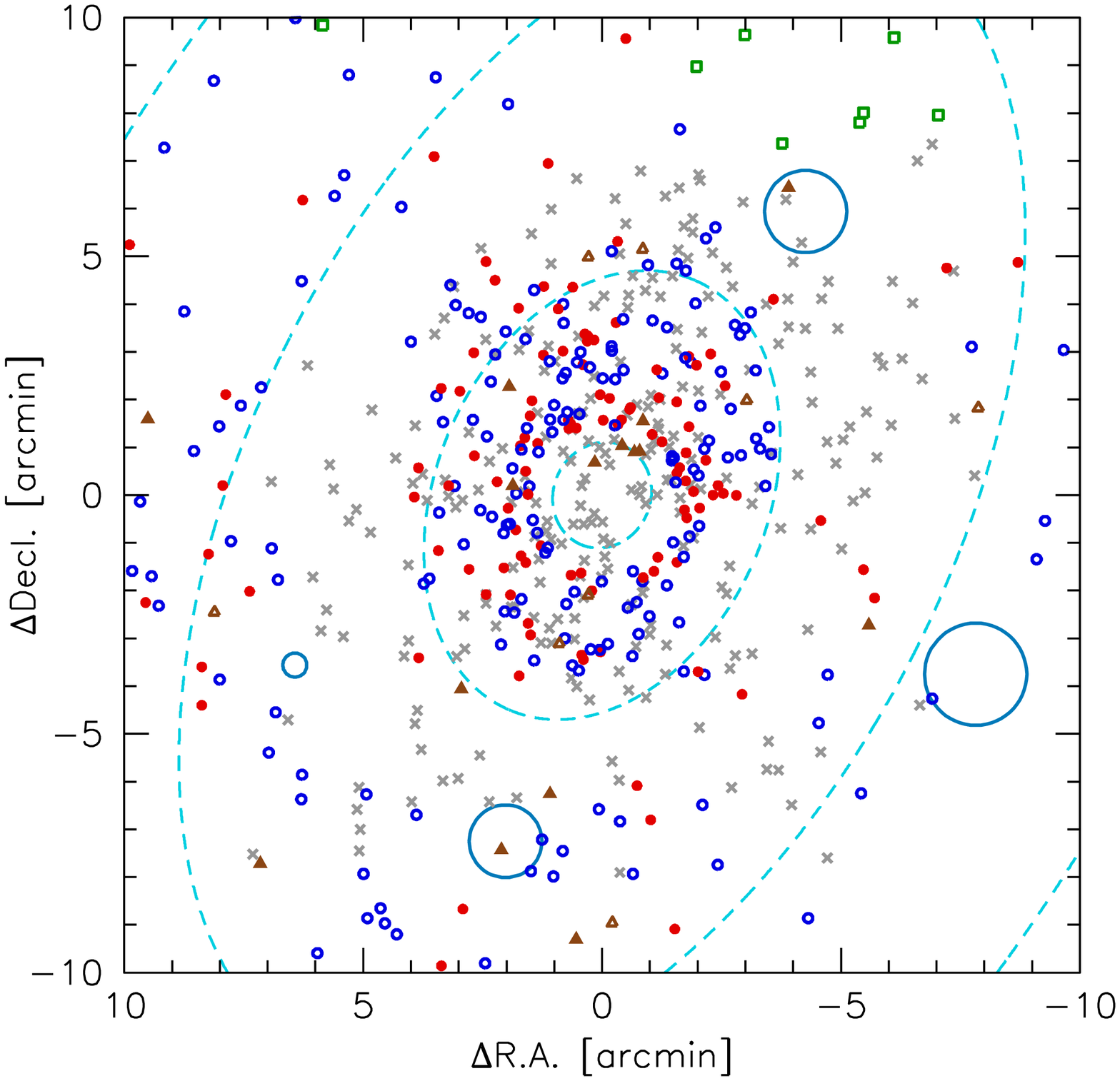}
\figcaption[M87GCpos8aj+8ak]{\label{fig:twod}
Two-dimensional distribution of GCs with velocity measurements;
the left panel shows an overall view and the right panel is zoomed in.
Symbols are as in Figure~\ref{fig:magrad}, with
green squares showing PNe from \citet{2009A&A...502..771D}.
Dashed blue ellipses show sample idealized stellar isophotes from \citet{2009ApJS..182..216K},
at $R_m \sim$~1\arcmin, 4\arcmin, 10\arcmin, 17\arcmin, 31\arcmin.
Larger blue circles show five dE galaxies, with the sizes marking their
estimated tidal radii.
}
\end{figure*}

\begin{figure*}
%\epsscale{0.85}
\epsscale{1.15}
\plotone{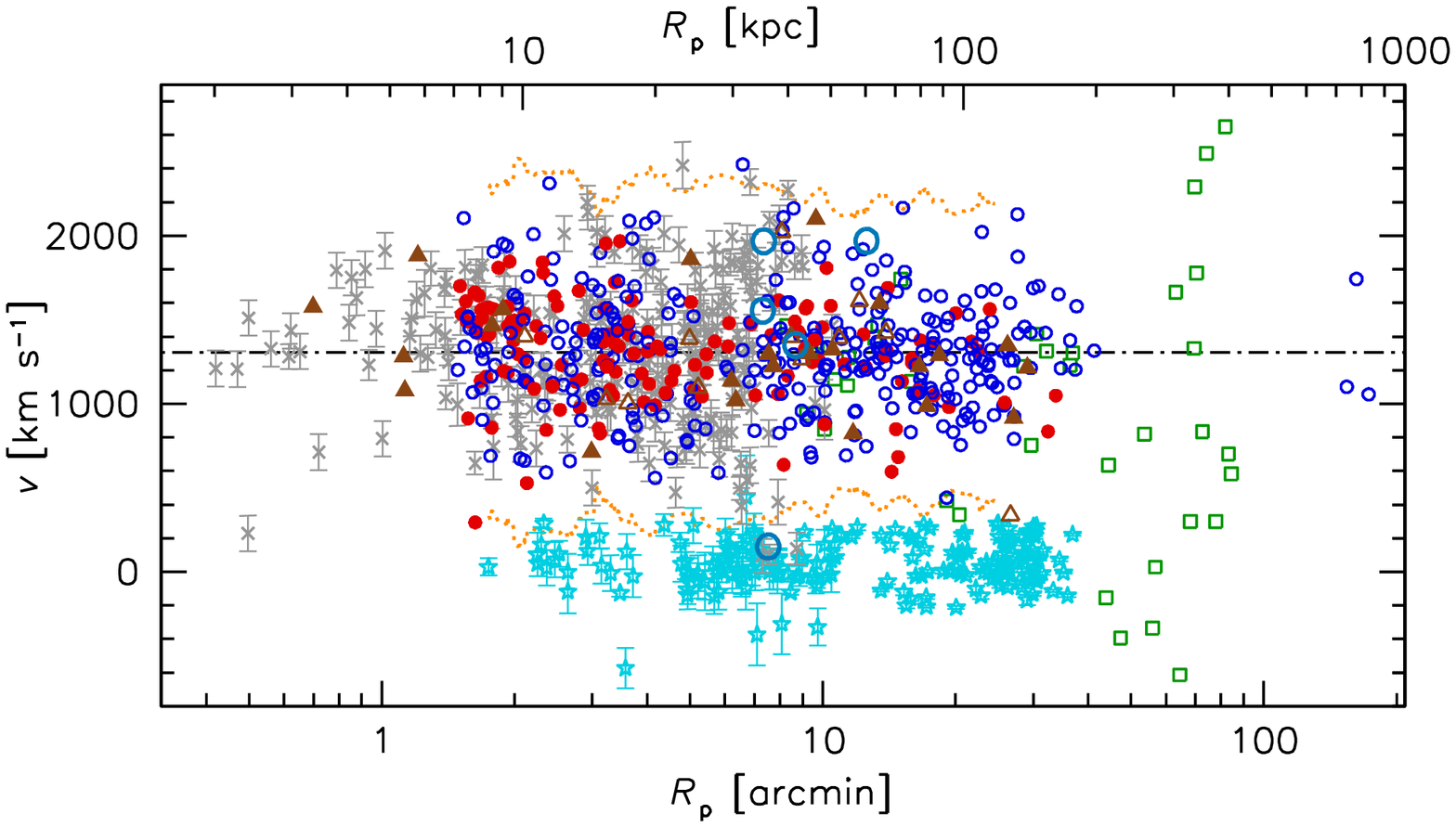}
\figcaption[velR6eh.eps]{\label{fig:velR}
Radial velocities versus the galactocentric radius.
Symbols are as in Figure~\ref{fig:twod},
with light blue star symbols showing objects identified as stars.
One of the dEs is hard to see at $\sim$~9\arcmin, $1350$~\kms.
Error bars show GC velocity measurement uncertainties in the case of the old data;
for the new data, these
are not shown since they would be in almost all cases smaller than the data points.
The orange dotted curves show the $\pm$~3~$\sigma$ envelopes for the
new GC data set.
}
\end{figure*}

The two-dimensional positions of the velocity data are illustrated by Figure~\ref{fig:twod}. The new LRIS and Hectospec data provide fairly uniform azimuthal coverage at small and large radii, respectively. The median velocity of the combined GC data from these two regions is 1300~\kms, with the medians from the two data subsamples consistent within the uncertainties. We adopt $v_{\rm sys}=1307$~\kms\ as the systemic velocity \citep{2000MNRAS.313..469S}. The DEIMOS masks provide additional data points in the south and east regions, including an eastwards extension to very large radii. As we found above, the GCS appears to be significantly flattened, and it may be appropriate to consider GC kinematics stratified on ellipses. Since the red GCS is consistent with the flattening $q$ and position angle of the stellar isophotes, we will adopt these better-determined quantities as our default for the elliptical circular-equivalent radius $R_m \equiv R_a \sqrt{q}$.

We next plot the GC velocities versus radius in Figure~\ref{fig:velR}, including both old and new data highlighted by different color schemes. A number of interesting features are visible by eye, where in particular, the scatter in velocities around $v_{\rm sys}$ is related to the velocity dispersion. In the $R \sim 1.5'$--$5'$ range, the old and new data behave similarly, but at $\sim 5'$--$8'$, the new data appear to have a much lower dispersion. At even larger radii ($\ga 8'$), the data show strong ``clumping'' near $v_{\rm sys}$. The latter feature is probably related to intrinsic unrelaxed substructure in the outer GCS, which we analyze in detail in \citet{Roman11}. In the meantime, we caution that our results outside $\sim 8'$ ($\sim 40$~kpc), in the region of suspected substructure, should generally be viewed as provisional because we will assume a well-mixed steady-state system which may be a poor description at large radii.

The discrepancy between the old and new data at intermediate radii ($\sim 25$--40 kpc) is puzzling. A strongly increasing dispersion at the radial limit of the old data was a key result relied on by ensuing dynamical analyses (which we will discuss further in \S\ref{sec:dyn}). We have seen in \S\ref{sec:merge} that some of the high relative velocity measurements that drove the high dispersion were due to catastrophic observational errors, which we suspect may also account for some of the remaining high-velocity measurements. However, even excluding the most extreme velocities, the older dispersion estimate in this region appears higher than ours.

There do not appear to be any obvious differences in the GC properties or azimuthal positions to explain this discrepancy (which we will examine in more detail in \S\ref{sec:oldnew}),
but intriguingly, there are four dEs located at  projected distances of $7'$--$9'$ from M87. It is possible that one or more of these galaxies is somehow ``stirring the pot'' or else shedding stripped GCs in such a way as to cause localized spikes in the velocity dispersion. Such features might also be related to the aforementioned substructures that begin around these radii. Systematic study of this region around M87 would clearly be of interest.

There are two low-velocity confirmed GCs (and one transition object with \rh~$\sim$~8~pc) around $\sim$~300--400~\kms, which as shown by the dotted curves in Figure~\ref{fig:velR} reside around the 3-$\sigma$ boundary of the overall GC velocity distribution. For $\sim$~460 velocities drawn from a Gaussian distribution, it is expected that one object on average is found in the 3-$\sigma$ tail, and more than one if the distribution is leptokurtotic, i.e., with extended tails as generally predicted for a radially-biased distribution of orbits. In addition, there are several high-velocity objects just inside the upper 3-$\sigma$ boundary, so we will by default consider the low-velocity duo to be part of the ``normal'' GC population bound to M87 (rather than, e.g., part of an unbound IGC population).

\subsection{Kinemetry}\label{sec:kinem}

We next begin to put the kinematics information into a simplified quantifiable format using the general approach of ``kinemetry'', which extends the standard techniques used in ellipse-based galaxy surface photometry to quantify spatial variations of the kinematics
\citep{2006MNRAS.366..787K,2008MNRAS.390...93K}. This approach is designed to compress two dimensions of kinematics information
into one dimension by assuming that the relevant quantities (rotation in particular) are approximately stratified on ellipses. In a circular system, the kinemetry reduces trivially to sine-curve fitting, which is a well worn technique in the context of GC kinematics. However, the strong ellipticity of the outer regions of M87 suggests the more general approach---even
though we will find in this case that the kinematics results are not strongly influenced
by the ellipticity modification, and we will end up using the circular model in order
to allow for strong kinematic twists.

Below we present the general techniques in \S\ref{sec:form}, which we then apply to the stellar kinematics in \S\ref{sec:stelkin} and to the overall GC system in \S\ref{sec:oldnew}.

\subsubsection{Methods}\label{sec:form}

Following the methodology introduced in \citet{2009MNRAS.398...91P} for application to sparsely sampled data, our basic kinemetric model for fitting a rotation field as a function of azimuth $\theta$ in an elliptical annulus is:
\begin{equation}\label{eqn:vmod}
v_{\rm mod}(\theta) = v_{\rm sys} \pm \frac{v_0}{\sqrt{1+\frac{\tan^2(\theta-\theta_0)}{q^2}}} , \newline
\end{equation}
where $v_0$ is the maximum rotation amplitude, $\theta_0$ is the direction of maximum rotation (the kinematical major axis)\footnote{This convention for $\theta_0$ differs from \citet{2009AJ....137.4956R}, where we used the direction of the angular momentum vector.}, $q$ is the ``axis ratio'' of the velocity field,
and the $+$ and $-$ signs are for a position angle inside and outside the range
$(\theta-\theta_0) = [-\pi/2$, $+\pi/2$], respectively.

In the case of integrated-light stellar kinematics, the mean velocity is directly measured at each data point, $\langle v_i \rangle$, which can be matched to the model $v_{\rm mod}(\theta)$ using $\chi^2$ fitting. A model for the dispersion $\sigma_{\rm p, mod}$ is then fitted separately. We also 
augment the velocity measurement uncertainties by $\sim$~5~\kms\ in quadrature, to keep outliers and systematic departures from the simple kinemetric model from unduly influencing the results.

For discrete velocities, the kinematical model involves both rotation and dispersion simultaneously, and uses a likelihood function to evaluate the probability of a measurement $(v_i \pm \Delta v_i)$ being drawn from a Gaussian distribution of model velocities. The equivalent $\chi^2$ function to be minimized is then:
\begin{equation}\label{eqn:chisq}
\chi^2 = \sum_i \frac{(v_i-v_{\rm mod})^2}{\sigma_{\rm p,mod}^2+(\Delta v_i)^2} 
+ \ln\left[\sigma_{\rm p,mod}^2 + (\Delta v_i)^2\right] 
\end{equation}
(cf~equation~A.2 of \citealt{2006A&A...448..155B} and equation~4 of \citealt{2010ApJ...720L.108G}). The same approach is also used for kinemetry with discrete velocities in \citet{2011MNRAS.415.3393F} and \citet{2011ApJ...736L..26A}\footnote{In the case of equal uncertainties $\Delta v_i$, equation~\ref{eqn:chisq} reduces to standard least-squares fitting. On the other hand, standard $\chi^2$ fitting as used in some previous studies  is mathematically incorrect and can yield very wrong answers when $\Delta v_i \ll \sigma_{\rm p}$ and when $v_{\rm rot} \ll \sigma_{\rm p}$.}.

With integrated light kinematics, a kinemetric model may be refined in great detail, including fitting for separate flattenings of both rotation and dispersion. With discrete velocities, more restrictive assumptions must usually be adopted, as at least $\sim$~1000 velocities are needed to constrain rotational flattening. We try out two model alternatives with the GC data: one that is circular and allows the position angle to change ($q_{\rm kin}=1, \theta_0$~free), and another where the kinematical ellipticity follows the stellar isophotes including their variations with radius ($q_{\rm kin}=q_{\rm phot}, \theta_{\rm 0, kin}=\theta_{\rm 0,phot}$). The galactocentric radii for the GC measurements are modified accordingly for each scheme.

The uncertainties in the fitted parameters are estimated by constructing a Monte Carlo series of 3000 mock data sets, using the best-fit kinemetric model as a starting point. Then at each GC location, a random velocity is drawn from an underlying Gaussian distribution defined by the intrinsic dispersion and the spectroscopic measurement uncertainties. We fit a kinemetric model to each mock data set and use the 68\% range of resulting model fits to define the 1~$\sigma$ uncertainties. The error bars estimated in this way are probably somewhat underestimated because the azimuthal sampling effects are difficult to fully take into account.

We also use the Monte Carlo simulations to correct for bias in the kinemetric parameters. In particular, when $\theta_0$ is a free parameter, $v_{\rm rot}$ will be biased high.  E.g., a non-rotating system that is sampled with a finite number of velocity measurements $N_i$ will always appear to have non-zero rotation if the direction of rotation is freely chosen. In general, the bias becomes important when the rotation is relatively low, with the rule of thumb that an observed free-P.A. rotation estimate is insecure if $(v_{\rm rot}/\sigma_{\rm p}) \lsim 0.55 \times \sqrt{20/N_i}$. 

As far as we know, this effect has not been explicitly considered in previous studies of rotation in GC and PN systems.

For the simple dynamical models that we will construct in this paper, we need only an azimuthally averaged second-order velocity moment, $v_{\rm rms}(R)$.
We could use the maximum likelihood techniques as in the kinemetry formalism, assuming Gaussian velocity distributions
and forcing the rotation to be zero. However, to avoid the Gaussian assumption as much as possible, we instead
use a classical velocity dispersion estimator:
\begin{equation}
v_{\rm rms}^2 = \frac{1}{N} \sum_{i=1}^{i=N} (v_i-v_{\rm sys})^2 - (\Delta v_i)^2 .
\end{equation}
The uncertainty in this quantity $\Delta v_{\rm rms}$ is then estimated through formulae provided by \citet{1980A&A....82..322D} which do have to assume Gaussian velocity distributions.

It is also important to estimate the deviations of the velocity distributions from Gaussianity. To this end, we adopt the classical fourth-order moment, the projected velocity kurtosis:
 \begin{equation}
             \kappa_{\rm p} = \left[ \frac{1}{N} \sum_{i=1}^{i=N} 
             \left( \frac{v_i-v_{\rm sys}}{v_{\rm rms}} \right)^{4} \right] - 3 \pm \sqrt{\frac{24}{N}}, 
             \label{eqn:kurt}
          \end{equation}
where $v_{\rm rms}$ is calculated {\it before} correction for the observational uncertainties $\Delta v_i$, and in practice we modify the simple expression
above for $\kappa_{\rm p}$ based on bias-corrected formulae from \citet{1998jg}.

\subsubsection{Stellar Kinematics}\label{sec:stelkin}

Before continuing to the GC kinemetry, we present results on the stellar kinematics, where the kinemetry is conceptually more straightforward.
The two main datasets we will consider are from  two-dimensional integral field spectrographs: SAURON \citep{2007MNRAS.379..418C} out to $\sim$~0.6\arcmin\ ($\sim$~3~kpc), and VIRUS-P \citep{2011ApJ...729..129M} out to $\sim$~4\arcmin\ ($\sim$~19~kpc). Note that the SAURON data are a revised version using an improved stellar template library, which does make a significant difference in the case of M87.

\begin{figure}
%\epsscale{0.85}
\epsscale{1.2}
\plotone{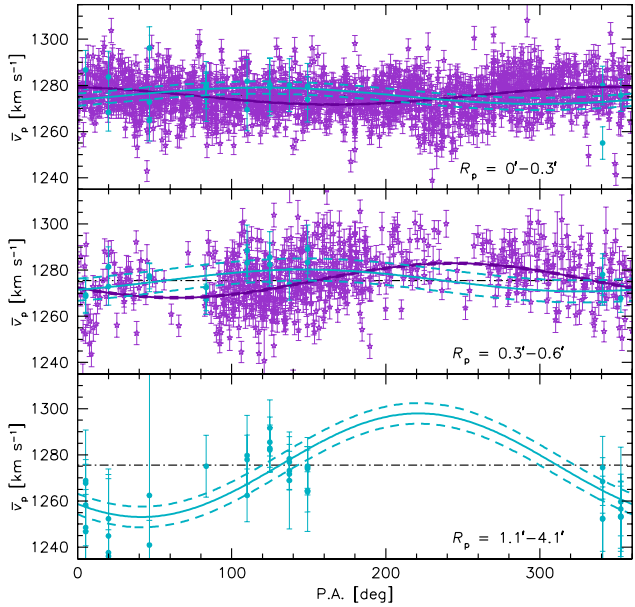}
\figcaption[velR8av.ps]{\label{fig:stelkin}
Mean velocity versus position angle in three radial bins (as labeled in the panels).
The purple stars and blue circles are SAURON and VIRUS-P data, respectively.
Kinemetric fits to the data are shown as solid curves with dashed curves
showing the uncertainties on the rotation amplitudes.
}
\end{figure}

The kinemetric results in three sample bins in radius are shown in Figure~\ref{fig:stelkin}. In the innermost regions (inside $\sim$~0.6\arcmin\ or $\sim$~3~kpc), both the SAURON and VIRUS-P datasets imply very low rotation, at the $\sim$~5~\kms\ level. Differences between these datasets in the direction of the rotation may not be significant owing to incomplete azimuthal sampling. At larger radii ($\sim$~1\arcmin--4\arcmin\ or $\sim$~5--19~kpc), the VIRUS-P data suggest the rotation increases but to merely $\sim$~20~\kms.

\begin{figure}
%\epsscale{0.85}
\epsscale{1.2}
\plotone{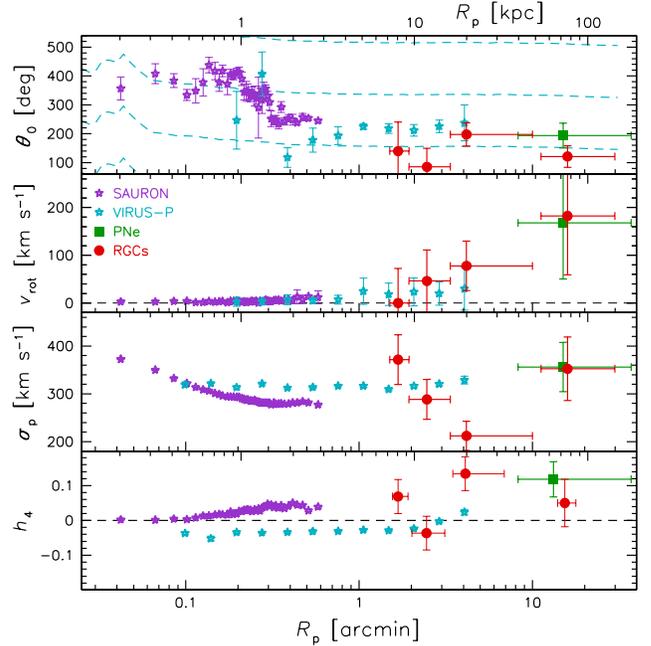}
\figcaption[M87GCkin4u.ps]{\label{fig:kinprof1}
Kinematics of M87 versus galactocentric radius.
The panels from top to bottom show rotation position angle and amplitude,
velocity dispersion, 
and fourth-order Gauss-Hermite moment $h_4$.
Different data sources are indicated by symbols as given in the legend of
the second panel.
The kurtoses from the discrete velocity data (PNe and RGCs) are transformed to
$h_4$ by the approximation $\kappa_{\rm p} \simeq 8\sqrt{6}h_4$
\citep{1993ApJ...407..525V}.
The dashed blue curves in the top panel show the $V$-band stellar photometric major-axis
\citep{2009ApJS..182..216K}.
}
\end{figure}

We next show a summary of kinemetric parameters plotted versus galactocentric radius in Figure~\ref{fig:kinprof1}. Some of the features of this plot will be discussed later, but first we highlight the rotational behavior.  It appears that the rotation is aligned with the major axis inside $\sim$~0.3\arcmin\ ($\sim$~1.5~kpc), and then twists fairly sharply by $90^\circ$ to alignment with the {\it minor axis}. Such a kinematical twist does not necessarily imply a radical dynamical transition but may reflect the rearrangement of a small fraction of the stellar orbits, as found through dynamical modeling of the famous ``kinematically decoupled core'' in NGC~4365 \citep{2008MNRAS.385..647V}.

The next thing we notice is that the SAURON and VIRUS-P data differ in their velocity dispersion and $h_4$ (fourth-order Gauss-Hermite moment) profiles, in some areas at very high confidence levels given the apparent error bars on the data. This is a demonstration of the sometimes severe systematic uncertainties in analyzing stellar kinematics data. \citet{2011ApJ...729..129M} et al. found for the dispersion results at least that the difference could be attributed to template mismatch involving the Mg~{\it b} region, which is a more dominant component in the SAURON data.

\subsubsection{Globular Cluster Kinematics: Old Versus New}\label{sec:oldnew}

Moving on to kinemetry of the M87 GCs, we begin with the simple circular model, combining all of the data irrespective of color (i.e., blues and reds together) but separating the analysis into the ``old'' and ``new'' measurements, where the year 2003 is used as the boundary of obsolescence. For the purposes of this comparison, we do {\it not} remove old velocity measurements with known catastrophic errors (e.g., S878, S1074), but we do reclassify the objects in some cases based on new multi-color photometry and {\it HST} size measurements. Some ``GCs'' are now identified as foreground stars or background galaxies (e.g., S1472, S7008, S7012), while some ``contaminants'' are reclassified as GCs (e.g., S5002, S7017). This is discussed in \S\ref{sec:class}. We also remove UCD candidates and transition objects (\rh~$>$~5.25~pc; \S\ref{sec:ucds}), as well as bright objects ($i_0 < 20$) because of their potentially disparate kinematics (to be examined in \S\ref{sec:subpop}).
These procedures allow us to focus our comparison of old and new data sets on the spectroscopic aspects.

Before showing our results, we will describe but not show explicitly the results from \citet[also using a circular model]{2001ApJ...559..828C}.
They found a rotation of $v_{\rm rot} \sim$~150~\kms\ along the {\it minor} axis inside 3.5\arcmin\ ($\sim$~18~kpc), particularly for the blue GCs. At larger distances they found {\it major}-axis rotation rising with radius, up to $v_{\rm rot} \sim$~400~\kms\ at $\sim$~7\arcmin\ ($\sim$~35~kpc). They also found velocity dispersions of $\sigma_{\rm p} \sim$~350~\kms\ in the inner regions, rising suddenly outside $\sim$~6\arcmin\ ($\sim$~27~kpc) to $\sigma_{\rm p} \sim$~500~\kms. The P.A. twist and high outer rotation and dispersion were also noted from earlier data by \citet{1997ApJ...486..230C} and \citet{1998AJ....116.2237K}.

\begin{figure}
%\epsscale{0.85}
\epsscale{1.2}
\plotone{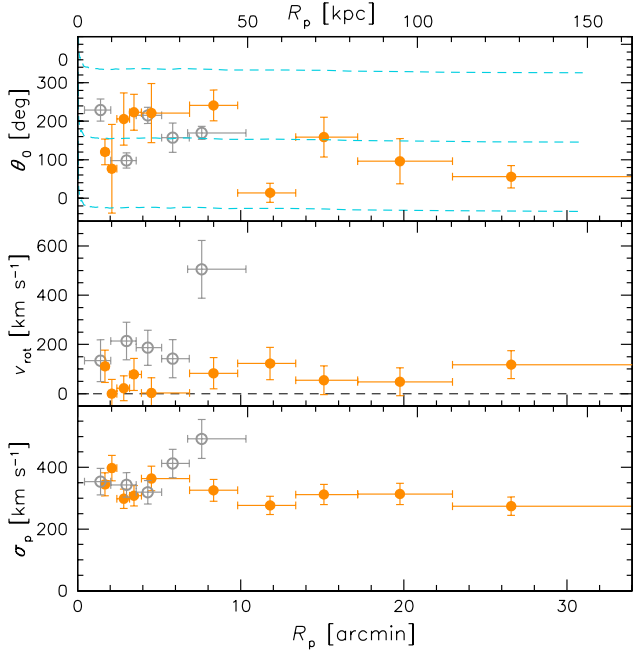}
\figcaption[M87GCkin3ab]{\label{fig:kincomp}
Kinemetric fits to M87 GC data, versus galactocentric radius.
The fitted parameters are the direction of maximum rotation (which wraps around
past 360$^\circ$),
the rotation amplitude, and the velocity dispersion
({\it panels from top to bottom}).
The filled orange circles show our new data, and the open gray circles
show the older literature data.
Dashed blue curves in the top panel show the photometric major axis of the
M87 stellar light from \citet[with $\pm 180^\circ$ ambiguity]{2009ApJS..182..216K}.
}
\end{figure}

Now binning the data by galactocentric radius, and adopting a fixed $v_{\rm sys}=1307$~\kms, we show our kinemetric outcomes in Figure~\ref{fig:kincomp}.
Using the old data, we generally reproduce the previous results except that we find 
even {\it stronger} rotation than \citet{2001ApJ...559..828C} did at the outermost radii 
($v_{\rm rot} \sim 500$~\kms).
These differences are presumably due to having a ``purer'' GC catalog.

The results from our {\it new} data show some further remarkable differences, as already suggested in \S\ref{sec:gen}. Inside $\sim$~5\arcmin\ ($\sim$~25~kpc) we do again see signs of kinematical twisting, but the rotation amplitude is lower, and the twist orientations are different from in the old data. We suspect these inner P.A. fluctuations are not significant, since with $v_{\rm rot}/\sigma_{\rm p} \sim 0.15$, it is difficult to determine the P.A. robustly. The new rotation is in most locations consistent with major axis rotation, although there is a hint of twisting to minor axis rotation outside $\sim$~11\arcmin\ ($\sim$~55~kpc).

In general, the rotation amplitude is lower than in the old data, with $v_{\rm rot} \sim$~50~\kms\ typically.
In particular, the drastic outer rotational increase has vanished. The old and new dispersion profiles are generally consistent, except outside $\sim$~5\arcmin\ ($\sim$~25~kpc): the new $\sigma_{\rm p}$ declines
where the old data showed a rise (as discussed above).

\begin{figure}
%\epsscale{0.85}
\epsscale{1.2}
\plotone{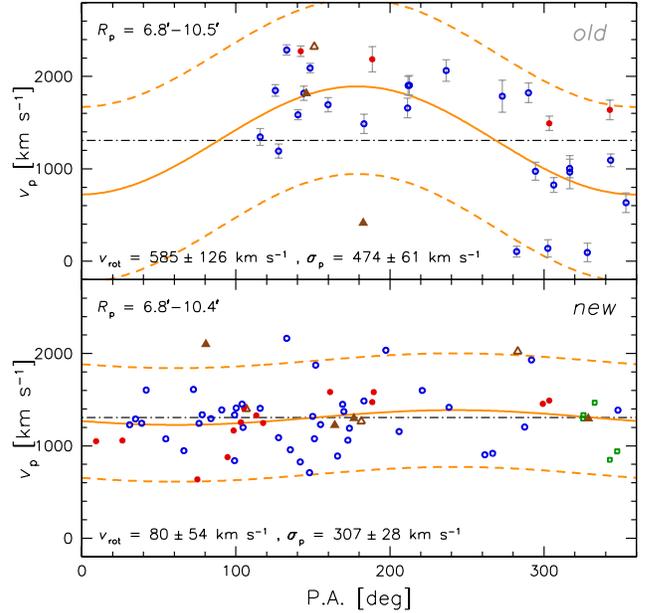}
\figcaption[velR8aw.ps]{\label{fig:PAcomp}
Radial velocities versus position angle in a radial bin at 
$R_{\rm p}$~$\sim$~8\arcmin\ ($\sim$~40 kpc),
for old ({\it top}) and new ({\it bottom}) data.
Symbol types are is in the previous two figures.
The kinemetric solutions for these bins are shown as solid orange sinusoidal curves with
dashed curves outlining the $\pm$~2~$\sigma$ envelope;
the solutions' rotation and dispersion parameters are listed in the panels.
The PNe, UCDs (including the peculiar object S7023), 
and transition objects are not used in the fits.
The black dot-dashed horizontal lines mark the systemic velocity.
}
\end{figure}

In order to understand the reasons for these differences, we consider the results at $\sim$~8\arcmin\ ($\sim$~40~kpc) in more detail.
In Figure~\ref{fig:PAcomp} we plot the individual GC velocities versus position angle for the old and new data in the same radial bin.
These two data sets share the same systemic velocity and general direction of rotation, but otherwise look as though they were drawn from a different galaxy. The clear high-rotation pattern in the old data is strongly inconsistent with the new data, and the new velocity dispersion is significantly lower.
Some of the highest old velocities in this bin have now been identified as catastrophic measurement errors, and it is plausible that some of the other high velocities are caused by similar errors. However, the differences are not simply due to this effect, as the velocity distribution at $\sim~180^\circ$ is {\it peaked} near 1700~\kms\ (i.e., 400~\kms\ higher than $v_{\rm sys}$) and some low velocities near $0^\circ$ are found as would be expected from real rotation.

\begin{figure*}
%\epsscale{0.85}
\epsscale{1.0}
\plottwo{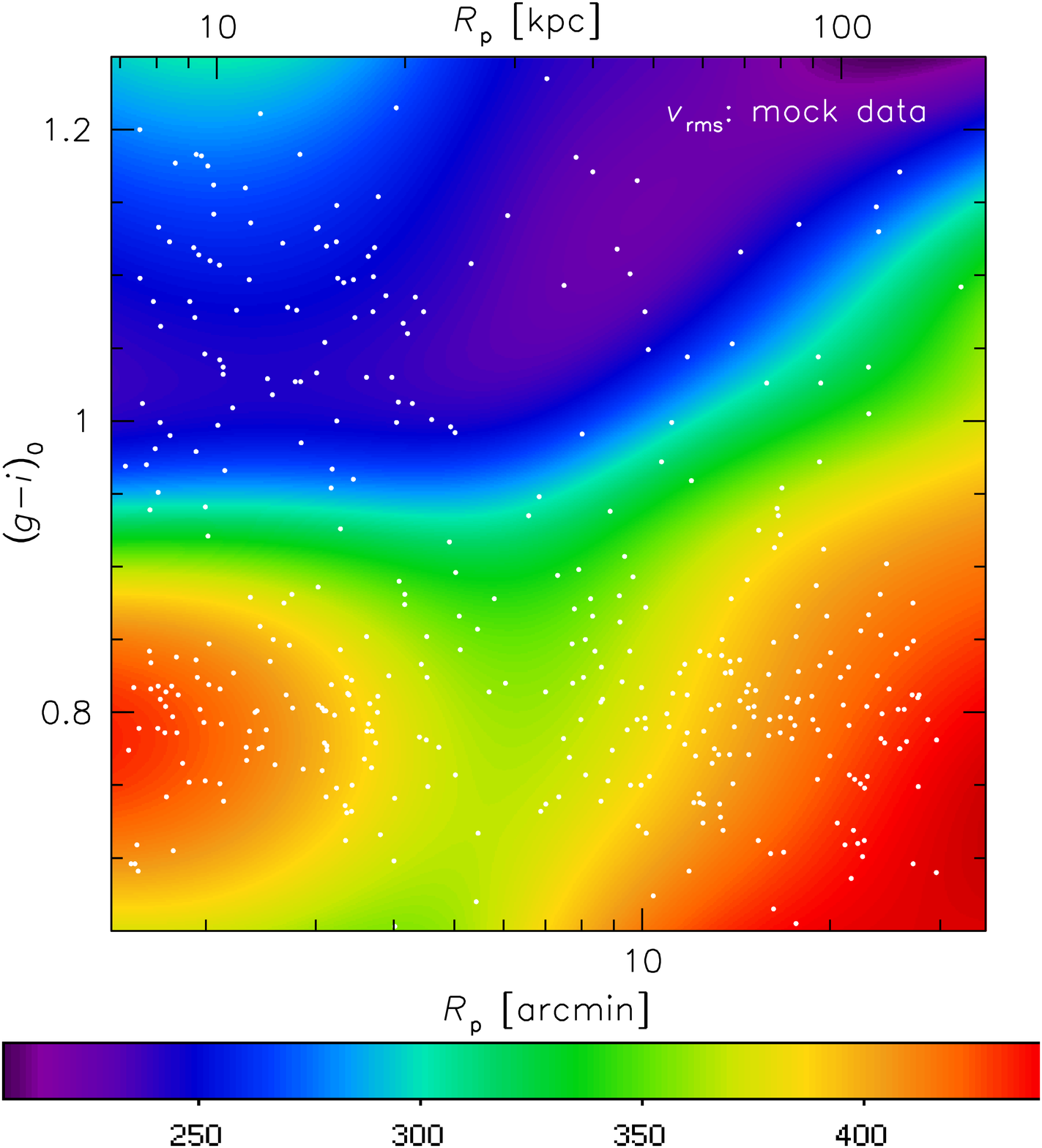}{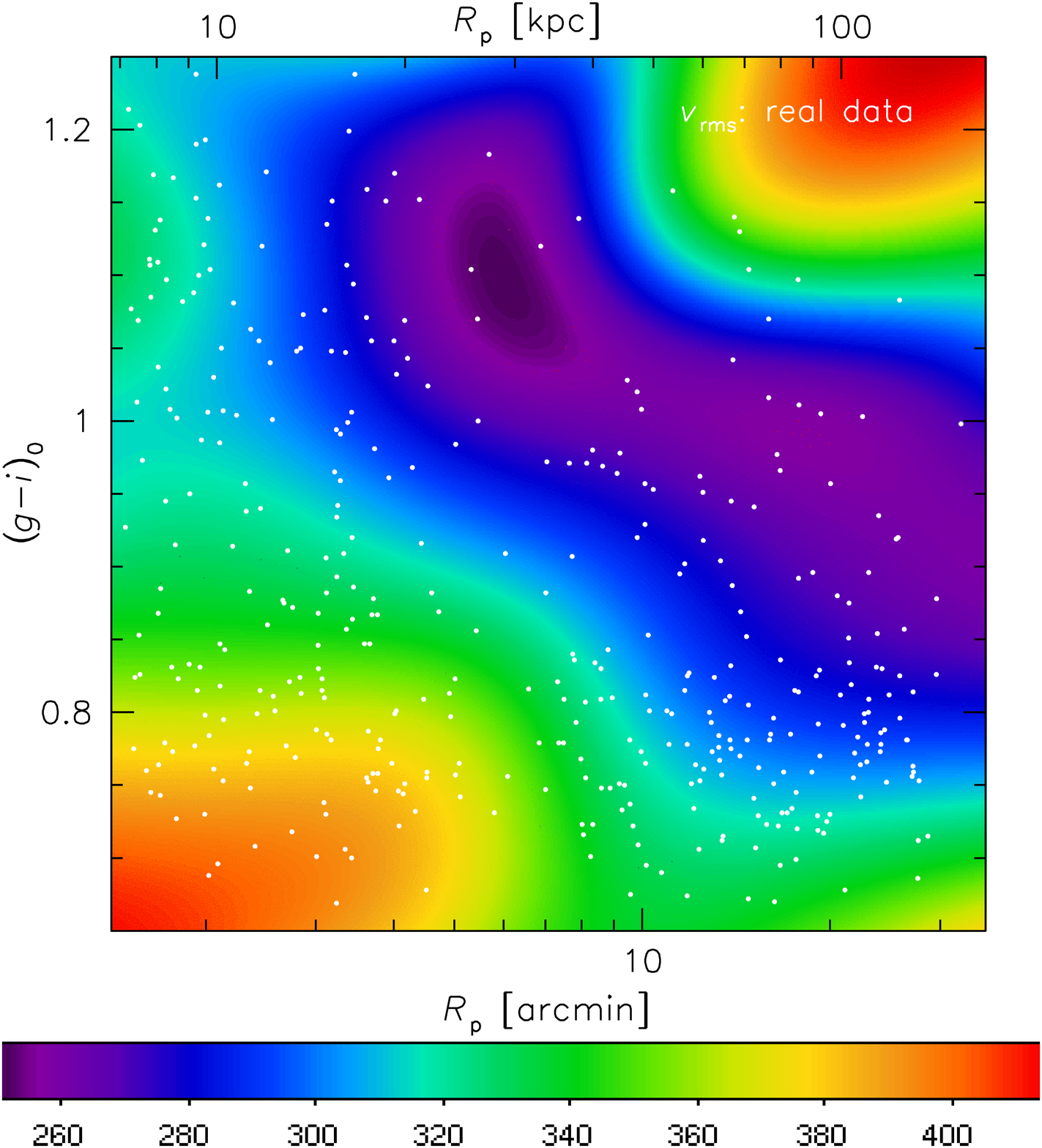}
\figcaption[M87GCgrid]{\label{fig:grid0}
Smoothed rms velocity fields of GCs in the plane of color versus log of the 
galactocentric distance.
White points show the individual velocity measurement locations, and the
color bars at the bottom indicate the $v_{\rm rms}$ scale (in \kms).
On the left is a mock data set, and on the right is the true data set of new M87 GCs.
A joint color-kinematical bimodality was input in the mock data, and correctly recovered
as seen by the strong $v_{\rm rms}$ transition at small radii.
The transition seen in the real data is more complicated, suggesting multiple subpopulations.
}
\end{figure*}

Both data sets do show strong fluctuations in the kinematics around this radial range, which may be part of the story, if there are very localized substructures with distinct kinematics whose detection depends on having complete or fortunate target selection.
In particular, the new data set has sparse coverage at $R \sim$~6.7\arcmin\ ($\sim$~32~kpc) where many of the old extreme velocities were, and we highlight the need to provide more coverage in this area and to repeat more of the old velocity measurements. Another factor that will be discussed in \S\ref{sec:subpop} is that the brighter GCs targeted in the old surveys are systematically different from the fainter ones that we can now reach.

At present we are unable to conclusively assess the reliability of the old data. The differences between the old and new data may be due both to sample selection and to problems with the old data. Since a small number of large outliers can significantly skew the results, for the rest of the paper we will rely only on the new data set unless otherwise stated.

\subsection{Subpopulations: Parameter Surveys}\label{sec:param}

We next consider the kinematics of various subpopulations, starting with trends with
color in \S\S\ref{sec:over}--\ref{sec:kinev}, 
and with luminosity in \S\ref{sec:trendlum}. 

\subsubsection{Overview}\label{sec:over}

This extensive data set, with high quality photometry and velocities, allows us
to consider the generic question of {\it bimodality} (``blue'' and ``red'' GCs) in a new way:
we investigate whether or not the GC {\it kinematics} show a strong transition or discontinuity with color that would imply two disconnected subpopulations.

We could attempt to define subpopulations based on the six potentially-relevant parameters in our data set (color, magnitude, size, velocity, 2-D position).
However, such an exercise is beyond the scope of this paper (cf \citealt{2009ApJ...705.1533C}), and even our extensive data set will provide only sparse coverage of a six-dimensional parameter space, so for now we will carry out a more cursory treatment.

We begin with a non-parametric study in the space of color, $(g-i)_0$, versus galactocentric distance, $R_{\rm p}$, since both of these parameters are likely to be important to the GC kinematics. The simplest kinematical parameter that does not require detailed modeling is the rms velocity $v_{\rm rms}$. We will therefore examine the dependence of $v_{\rm rms}$ on color and radius jointly.

\subsubsection{Simulations}\label{sec:sim}

Before analyzing the real data, we generate some mock data sets in order to optimize our techniques to pick up only real features in the data. We take the same positions $R_{\rm p}$ for the mock data as in our real data set, and randomly generate a color and velocity for each object. The color is drawn from either a blue or red subpopulation, with its probability of belonging to each subpopulation being a function of radius motivated by our surface density profile analysis (\S\ref{sec:raddist}). This probability ranges from 45\% blue at $R_{\rm p}$~$\sim$~1\arcmin\ to 85\% blue at $\sim$~40\arcmin. For simplicity, we choose constant rms velocity profiles with radius, with $v_{\rm rms}=400$~\kms\ for the blue GCs, and $250$~\kms\ for the red GCs. The $(g-i)_0$ color distributions are assumed to be Gaussian with peaks and 1-$\sigma$ widths of $(0.79,0.06)$ and $(1.05,0.09)$ for the blue and red GCs, respectively.

After generating the data set, we construct a grid of color versus log-radius, and at each grid point derive a smoothed $v_{\rm rms}$ value based on a weighted average of the surrounding data points.  The weighting uses a Gaussian kernel with ``distances'' of 0.22~dex in log-radius and 0.07~dex in color. 

The results for one mock data set are shown in the left panel of Figure~\ref{fig:grid0}. An obvious feature in the central regions is the strong transition in $v_{\rm rms}$ correctly recovered at $(g-i)_0 \sim 0.95$, reflecting the presence of two distinct underlying subpopulations. At larger radii where there are few red GCs, the color-location of the kinematical transition drifts away from the input value, presumably because of redward contamination for the blue subpopulation.
There are also localized fluctuations in $v_{\rm rms}$ caused by statistical fluctuations---particularly in regions with very few  nearby data points. 

\subsubsection{Kinematical Evidence for Subpopulations}\label{sec:kinev}

We next consider the real data in the right panel of Figure~\ref{fig:grid0}, where again, we are now using only the new data set, and are omitting the known ``UCDs'' (\rh~$>$~5.25) as well as the bright objects ($i_0 < 20$); there are 410 objects in this revised sample. Some ``hot'' and ``cold'' spots can be seen which are in regions with few data constraints and so can be dismissed as noise. As in the simulation, a kinematical ``boundary'' is indeed seen at $(g-i)_0 \sim 0.95$. However, there are a number of dissimilarities with the simple model that appear subtle but may be important.  

One difference is that the kinematical transition with color is less gradual in the data than expected, with $v_{\rm rms}$ reaches its maximum value at very blue colors of $(g-i)_0 \la 0.75$.  Another is that the transition color shifts blueward at larger radii, $R_{\rm p}\ga$~8\arcmin\ ($\ga$~40~kpc), while is seen in the simulation, we would expect a {\it redward} shift if anything.

\begin{figure*}
%\epsscale{0.85}
\epsscale{1.1}
\plotone{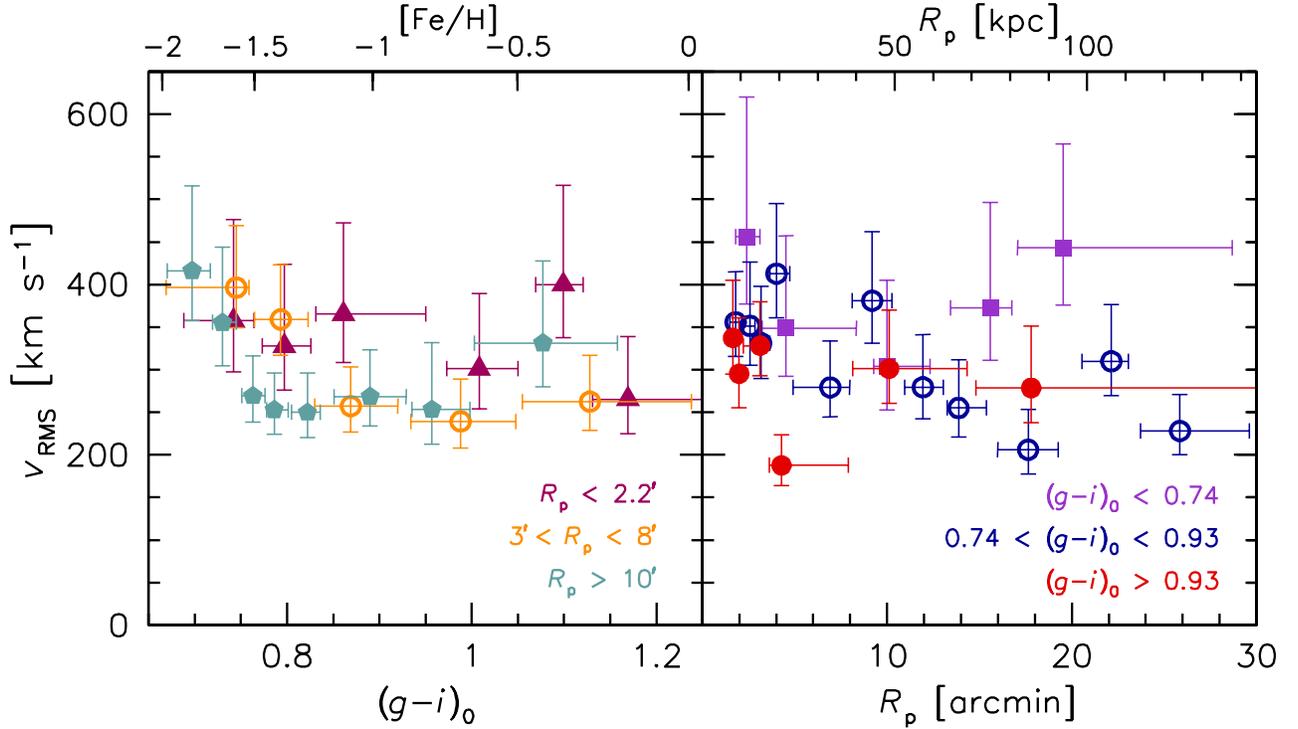}
\figcaption[M87GCvel5n]{\label{fig:vrms2}
Root-mean-square velocity profiles of GC subsamples.
The left panel shows $v_{\rm rms}$ versus GC color, in three radial bins
(see legend in panel).
The right panel shows $v_{\rm rms}$ versus galactocentric distance, in
three color bins (see legend).
}
\end{figure*}

To investigate these features in more detail, we make standard one-dimensional profiles of $v_{\rm rms}$ versus color in select radial bins, and versus radius in select color bins. Figure~\ref{fig:vrms2} shows the results. We first consider the radial bin $R_{\rm p}=$~3\arcmin--8\arcmin\ (15--40~kpc), which we may consider the main body of the GC system. We see that there may be a discontinuity at $(g-i)_0 \sim 0.83$, with GCs bluewards and redwards having $v_{\rm rms}=370\pm38$~\kms\ and $255\pm23$~\kms\, respectively. 

This result would seem initially to {\it not} confirm the simple bimodality model
with a color boundary at $(g-i)_0 \sim 0.93$, and may even be consistent with
a non-linear color-metallicity relation (e.g., \citealt{2006Sci...311.1129Y}),
since an assumed smooth $v_{\rm rms}$-metallicity anti-correlation might then
lead to a strong increase of observed $v_{\rm rms}$ at very blue colors.
However, such a scenario would need additional data and modeling to test,
and in any case we have already seen that the M87 GC color distribution is more 
complicated than in a classical bimodality picture.
We found evidence in \S\ref{sec:beyond} for at least a {\it third} GC subpopulation with intermediate colors of $(g-i)_0 \sim$~0.86--1.01.
It appears that these objects have relatively ``cold'' kinematics, which
is similar to the redder GCs, and complicates the kinematics tests for distinct subpopulations.

In Figure~\ref{fig:vrms2} we also plot $v_{\rm rms}$ versus color in other radial bins,
and versus radius in different color bins, and see that the picture becomes even
more complicated.  
For example,
at large radii, both the very bluest and the very reddest GCs have higher $v_{\rm rms}$
than the bulk of the GCs at the same radii (see also Figure~\ref{fig:grid0}).
These kinematical irregularities seem to be driven largely by a small number of extreme-velocity objects\footnote{Note that among the far-red GCs,
H20573 and H66419 have similar colors, luminosities, and phase-space locations, supporting a common origin. For the far-blues, the extreme-velocity objects are T16997, H23346, and H58443, which also lie in a similar region of the color-color diagram (see earlier discussion of the latter object). None of these objects has a measured size but we do not consider them as likely stars (in the case of those with low velocities), but instead as probably parts of unrelaxed substructure around M87 (see also two PNe close to H58443 in phase-space in Figure~\ref{fig:velR}).},
and could be due to recent, discrete accretion events resulting in families of
GCs that are clumped in chemo-dynamical phase-space.

We therefore conclude that the outer regions of M87 may host a significant fraction of unrelaxed material that makes kinematical analysis of subpopulations in these regions somewhat precarious.  The outer subpopulations may be considered still in the process of assembly, and might not even bear much connection to
the central subpopulations. These conclusions, based on basic color and $v_{\rm rms}$ trends, parallel those of \citet{Roman11}, who analyzed phase-space in more detail.

\begin{figure}
%\epsscale{0.85}
\epsscale{1.2}
\plotone{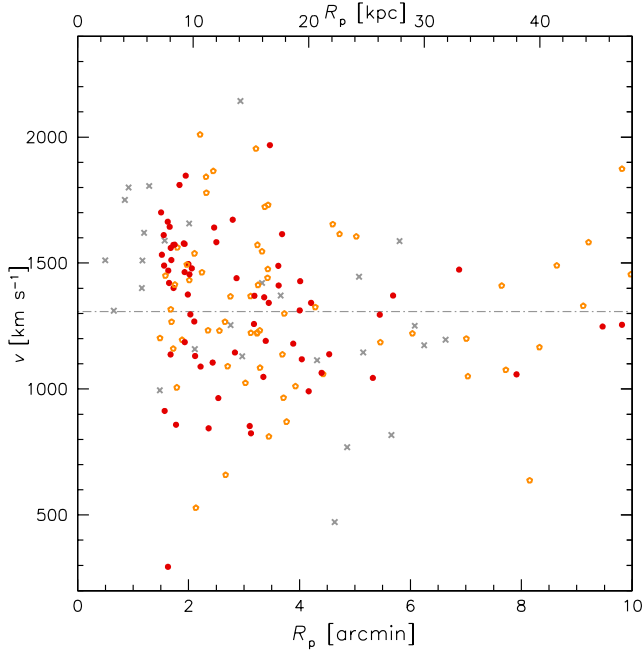}
\figcaption[velR6ei]{\label{fig:vsys}
Velocities versus galactocentric distance for redder GCs.
Old data with $(g-i)_0 > 1.01$ are shown with gray $\times$~symbols, 
while the new data are shown as orange open pentagons for $0.86 < (g-i)_0 < 1.01$,
and red filled circles for $(g-i)_0 > 1.01$. At small radii, the reddest GCs
show a strong velocity asymmetry relative to $v_{\rm sys}$ (the UCD candidates,
which are not plotted here, have the same behavior).}
\end{figure}

\begin{figure*}
%\epsscale{0.85}
%\epsscale{1.2}
\epsscale{1.16}
\plottwo{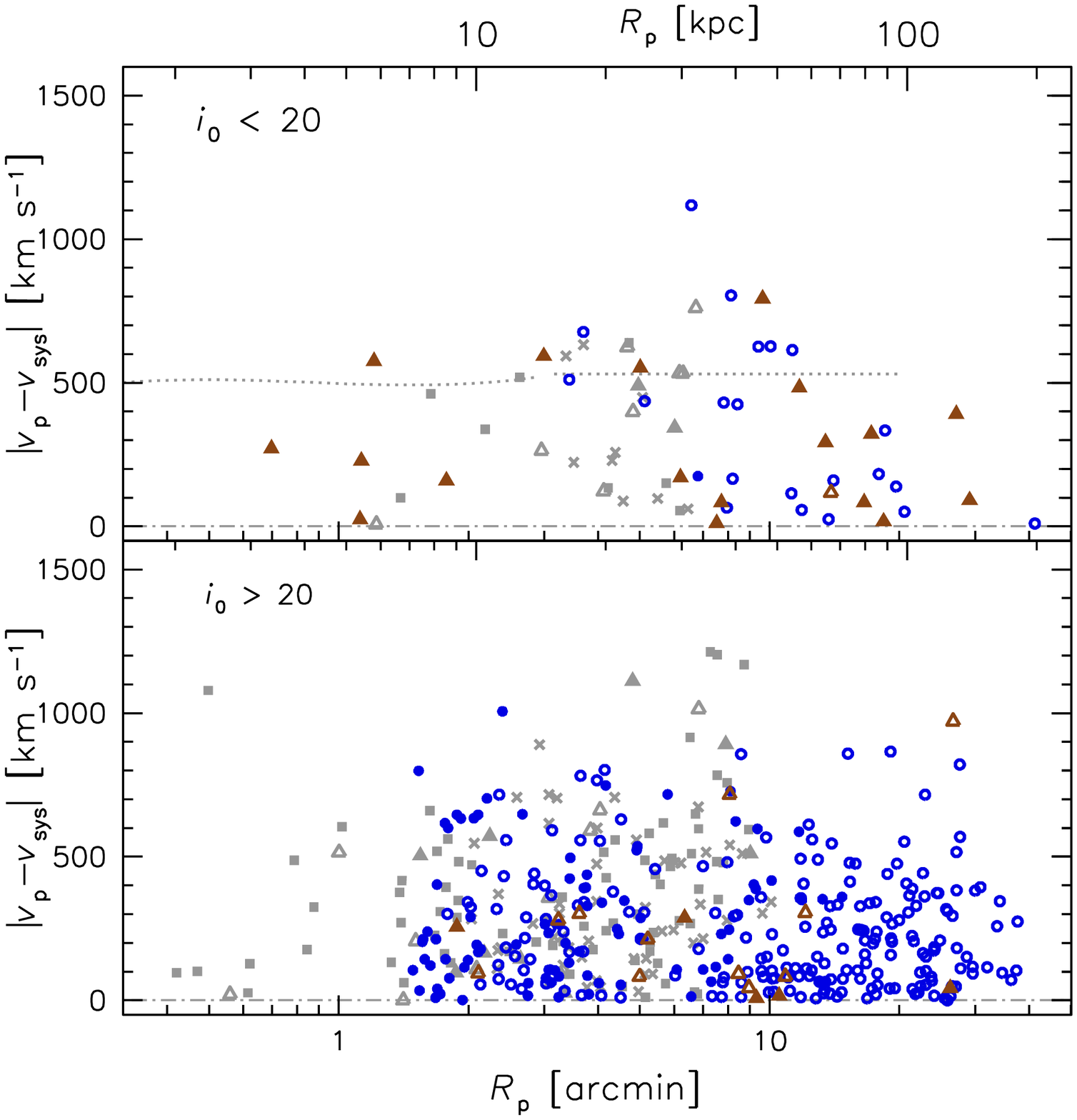}{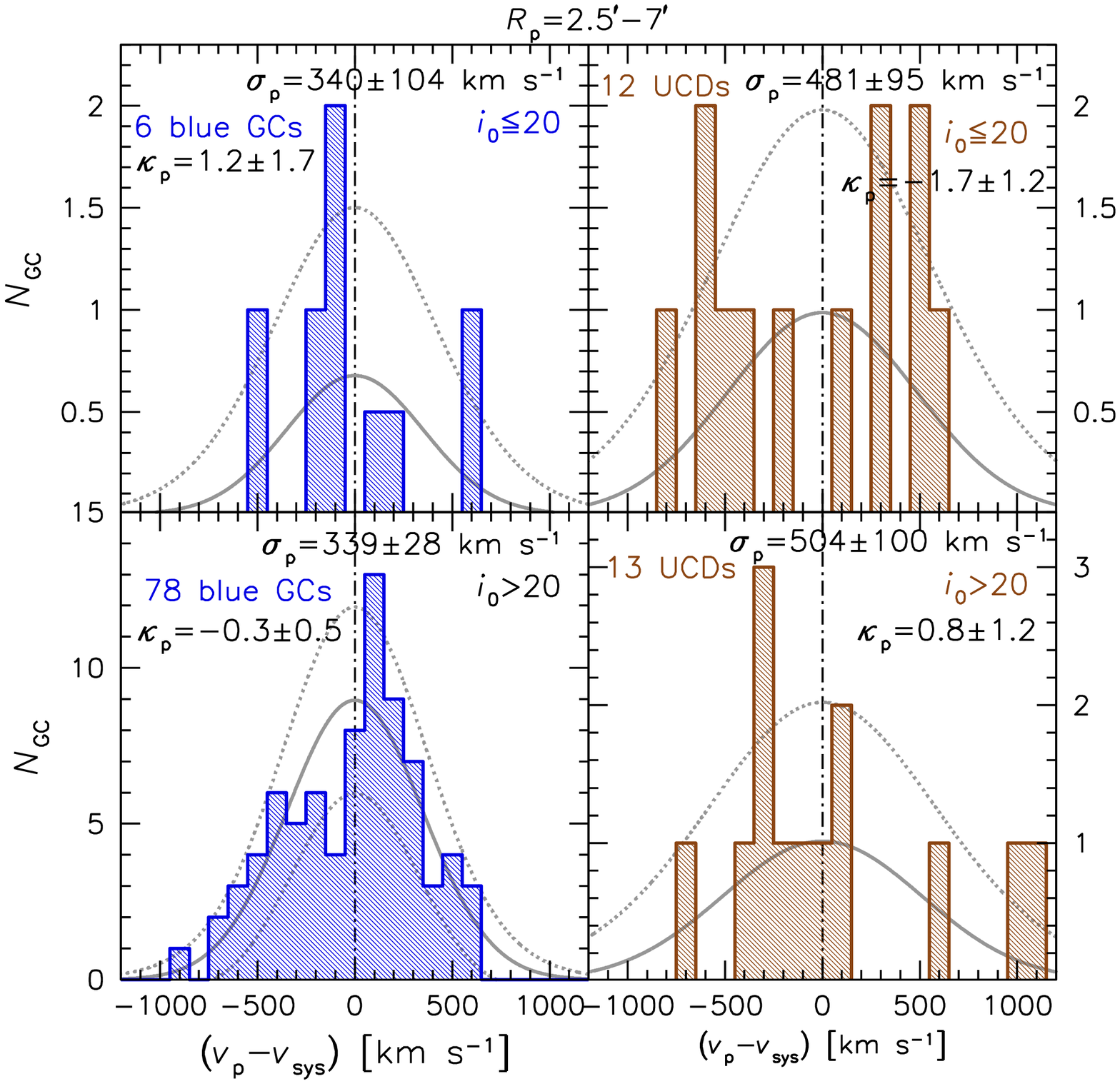}
\figcaption[velR6ejM87GCvel6ca]{\label{fig:mags}
Demonstration of magnitude dependence of GC kinematics, using 
$i_0=20$ as the boundary between ``bright'' and ``faint'' objects,
and \rh=5.25~pc as the boundary between compact and extended objects.
{\it Left panel:}
Absolute value of the difference between object velocity and systemic velocity,
as a function of galactocentric radius.
Blue GCs and UCDs (of all colors) are shown, in bright and faint subsamples
(top and bottom).
As in previous plots, colored and gray symbols show new and old data, respectively.
Symbol types denote sizes, with filled and open triangles the UCDs and transition objects,
filled squares and circles the ``normal'' compact GCs,
and crosses and open circles for unknown sizes.
An approximate circular velocity profile is shown in the top panel 
(see \S\ref{sec:dyn}).
{\it Right panel:} Velocity distributions, for data 
in the radial range $R_{\rm p}=$~2.5\arcmin--7\arcmin\
(in histograms with bins of 100~\kms),
and best-fit Gaussians (solid curves with dotted outlines of $\sqrt{N}$ scatter).
The top and bottom panels show the bright and faint objects, respectively.
The left and right panels show ``GCs'' and ``UCDs'', distinguished by being compact
or extended, respectively.
The best-fit values for the Gaussian dispersions and the kurtoses are
indicated in each panel, along with the 1~$\sigma$ uncertainties.
}
\end{figure*}

A similar explanation may apply to a kinematical peculiarity of the reddest GCs at 
{\it small} radii ($R_{\rm p}=$~1.6\arcmin--2.2\arcmin\ or 8--10~kpc).
These objects have a much higher $v_{\rm rms}$ than at intermediate radii,
which we trace to a peculiar offset in their overall velocities.
As shown in Figure~\ref{fig:vsys}, the red GCs inside 2.2\arcmin\ (10~kpc)
have a median velocity of $1464 \pm 49$~\kms, which is significantly offset from the overall $v_{\rm sys}=1307$~\kms, and is driven mostly by the far-red objects with $(g-i)_0 \gsim 1.01$. This effect does not seem to be driven by uneven sampling of a strongly rotating system, since the offset objects are distributed over a wide range of azimuths, and we cannot think of a plausible observational or instrumental effect to explain it.

Inspecting the older data set, the offset also appears in the red GCs, and can be seen to extend to even smaller radii (Figure~\ref{fig:vsys}).
As mentioned in \S\ref{sec:beyond}, there are also indications of peculiar shifts in the red GC colors at precisely these radii, supporting a picture of the inner regions of M87 as incompletely mixed.

\subsubsection{Trends with Luminosity}\label{sec:trendlum}

We next consider magnitude dependencies of the GC kinematics. Because of the strong variations in the magnitude selection with radius in our data (Figure~\ref{fig:magrad}), it is difficult to consider this theme in much detail. We adopt the simple approach of examining the velocity distribution  as a function of radius for ``bright'' and ``faint'' subsamples, separately for blue and red GCs.

For the red GCs we do not find any strong trends with magnitude, but for the blue GCs we find a possible transition at $i_0 \sim 20$ ($M_i \sim -11$) which we illustrate in Figure~\ref{fig:mags} (left panel). For the innermost regions where we have kinematics data on bright blue GCs, these objects appear to avoid the systemic velocity.
Although we have very low number statistics for these objects, intriguingly,
the bright UCDs (of all colors) appear to share in the same pattern.

A key question here is whether these kinematical peculiarities are primarily linked  to {\it luminosity} or to {\it size}, since four out of the five bright blue ``GCs'' have unmeasured sizes and might be UCDs\footnote{These four GCs are S279, S348, S501, and VUCD10. The fifth, S1265, has a normal GC size, and is the closest of the five objects to $v_{\rm sys}$, but intriguingly has a very similar distance, color, magnitude, and velocity to the bona fide UCD S1629. Also, two bright transition objects that fit the same pattern are S77 and S137, which have similar positions, velocities (from \citealt{2000AJ....119..162C} and \citealt{2001ApJ...559..812H}), colors, magnitudes, and sizes to each other. These seem likely to share a common origin.}. 
To investigate further,
we break down the sample in the radial range $R_{\rm p} \sim $~2.5\arcmin--7\arcmin\ ($\sim$~10--30~kpc) into four subsamples of size and magnitude, and show
their velocity histograms in the right panel of Figure~\ref{fig:mags}.
Because some of the subsamples include very few objects, we have boosted the
statistics by calling objects with intermediate sizes (\rh~$\sim$~5--10 pc) ``UCDs'',
and by including all of the old data in the hope that the occasional rogue
velocity measurement will not muddy the waters.

For each of the four subsamples, the velocity dispersion (from a Gaussian fit) and
kurtosis are indicated in the Figure panels. We see that the bright and faint compact blue
GCs have consistent dispersions and kurtoses, suggesting that these belong to the
same population.  Both UCD subsamples, on the other hand, have higher dispersions
than the GCs, which provides another line of evidence that UCDs (as defined
by large sizes but not necessarily high luminosities) are a distinct population from 
normal GCs.

This picture becomes more complicated when comparing the UCD subsamples, which although
having similar velocity dispersions, have discrepant kurtoses.  The bright UCDs have 
a negative kurtosis, driving the peculiar ``double-peaked'' velocity distribution previously
mentioned, while the faint UCDs have a higher kurtosis.
There are a number of potential explanations for this difference:
multiple populations of UCDs; additional cases of catastrophic measurement errors;
or a statistical fluctuation (the kurtoses are different at only 1.5~$\sigma$).
Firmer conclusions about UCD kinematics
will require an enlarged set of new data.

\begin{figure*}[t]
%\epsscale{0.85}
%\epsscale{1.0}
\epsscale{1.1}
\plotone{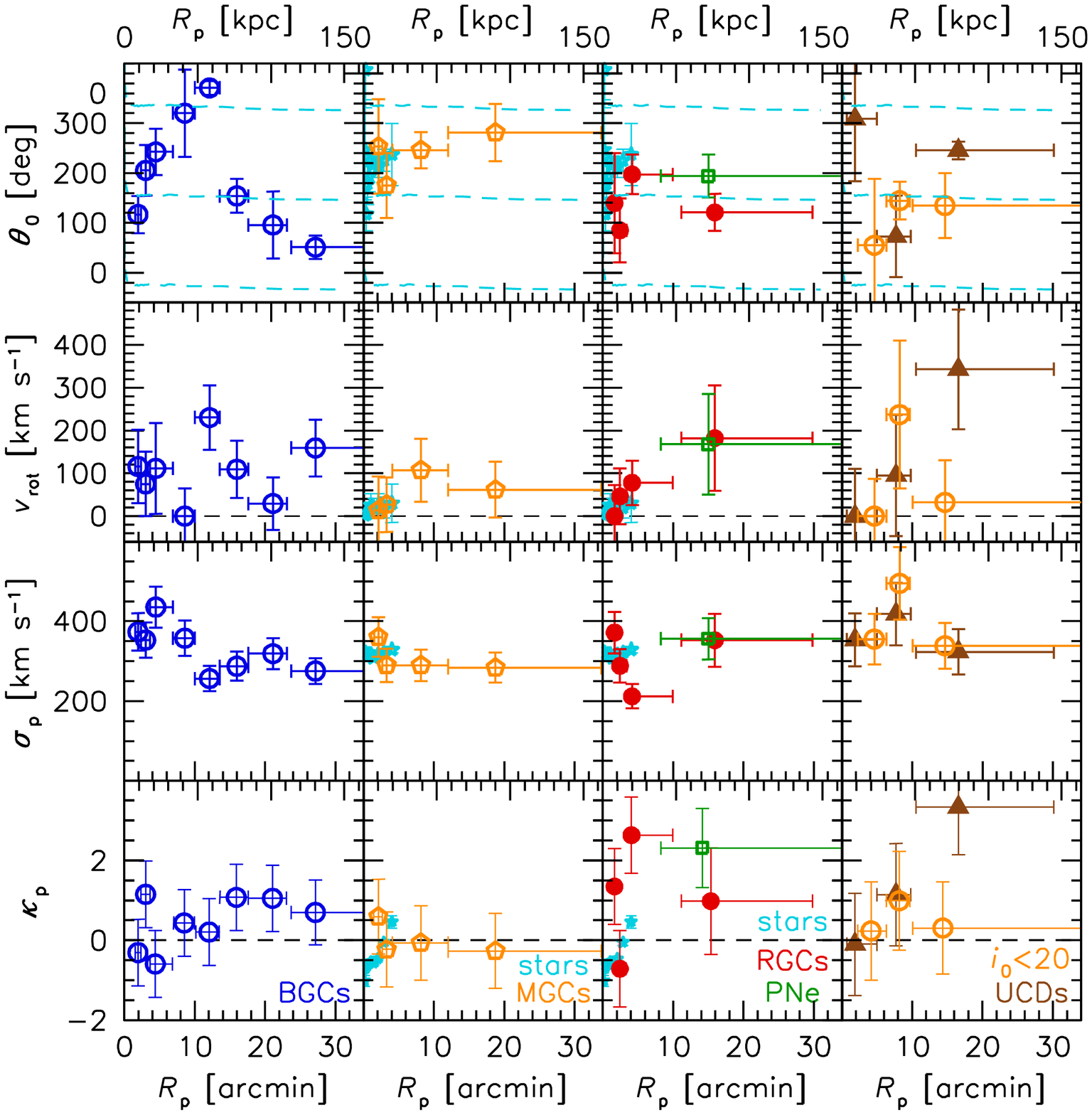}
\figcaption[M87GCkin3ac.ps]{\label{fig:grid}
Summary plot of kinematical profiles for M87 subpopulations,
as functions of galactocentric radius.
The first column shows blue GCs; the second shows intermediate-color GCs;
the third shows red GCs and PNe (red and green color-schemes, respectively);
and the fourth shows bright and large objects (``UCDs'' with \rh~$>$~5.25~pc).
The second and third columns also include the VIRUS-P stellar kinematics data.
The top three rows show the kinemetric fit parameters
$\theta_0$, $v_{\rm rot}$, and $\sigma_{\rm p}$ 
(position angle wrapped around past 360$^\circ$, 
rotation amplitude, and projected velocity dispersion).
The bottom row shows the velocity kurtosis.
Points with error bars represent best fits in independent radial bins,
along with the 1~$\sigma$ uncertainties.
The dashed light blue curves in the top row show the photometric major axis
as in Figure~\ref{fig:kincomp}.
}
\end{figure*}

We will return to additional luminosity trends in \S\ref{sec:lum} and discuss some general implications in \S\ref{sec:bright}.
To be safe, we have in general simply omitted all objects with $i_0 < 20$ (whether blue or red)
from our kinematical analyses in this paper.

\subsection{Subpopulations: Kinemetry}\label{sec:subpop}

\begin{figure*}
%\epsscale{0.85}
\epsscale{1.0}
\plotone{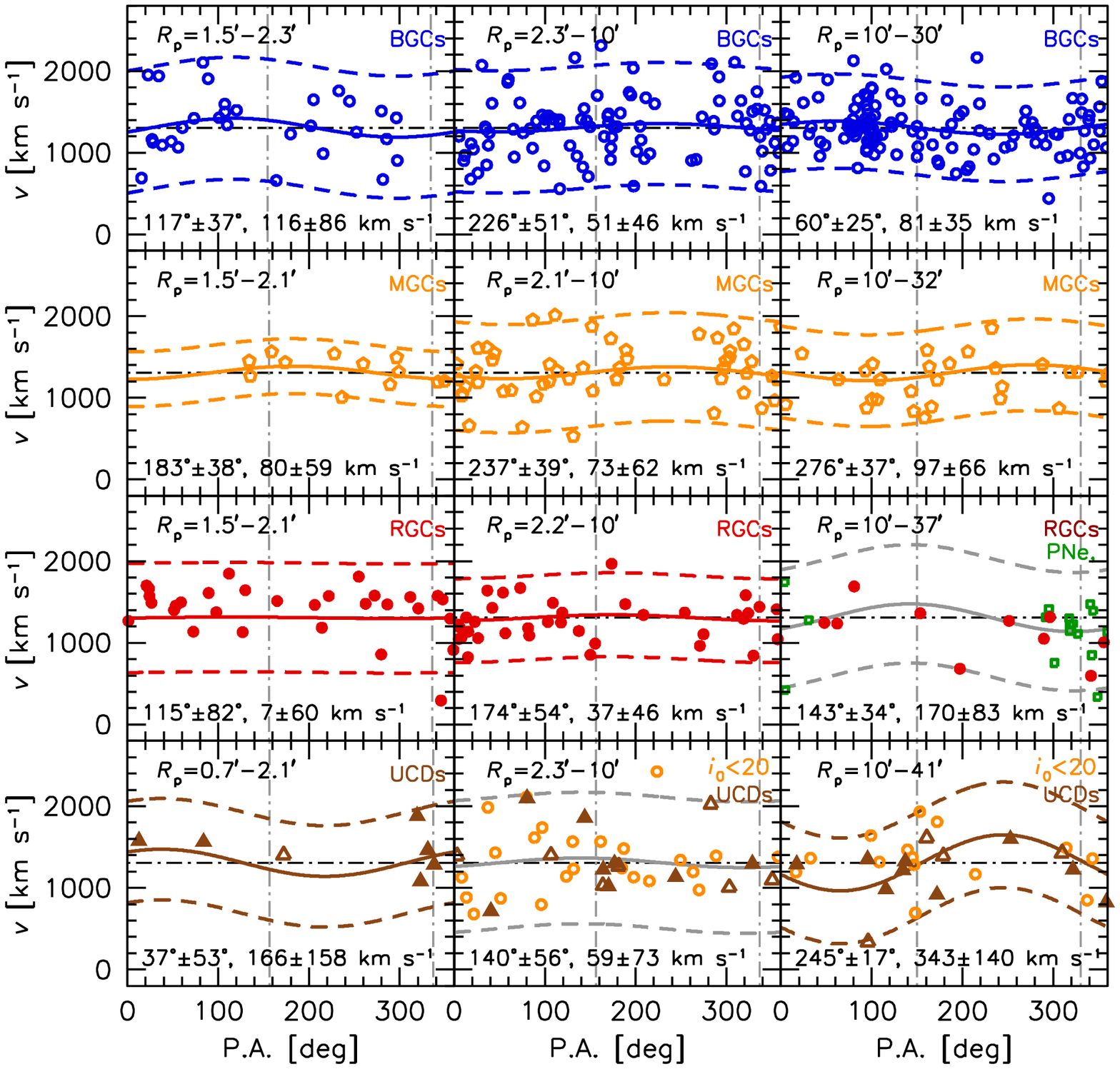}
\figcaption[velR8ax.ps]{\label{fig:PAcomp3}
Velocities versus position angle in bins of radius and subpopulation
(cf Figure~\ref{fig:PAcomp}).
The subpopulations (blue GCs, intermediate-color GCs, red GCs, UCDs, and
bright GCs) are as labeled in each row.
The columns show inner, intermediate, and outer radial bins, as
labeled in the panels.
The best-fit rotation position angle and amplitude values are reported in each panel.
There are additionally PNe included along with the red GCs at large radii, and the
kinemetric model there is a joint fit to both subpopulations.
The lower panels plot the UCDs and bright GCs together where possible, but the
kinemetric models shown are generally fits to the UCDs, except for the middle panel
where a joint fit is applied.
Vertical lines mark the semi-major axes of the stellar isophotes.
}
\end{figure*}

We now move on to kinemetric models of the GC subpopulations (see methods in \S\ref{sec:form} and overall analysis in \S\ref{sec:oldnew}).
Given our discussion of multiple GC subpopulations in preceding sections, we adopt a {\it trimodal} analysis as our new default. The three subpopulations are the {\it blue} GCs or ``BGCs'' (analyzed in \S\ref{sec:BGC}), the {\it intermediate color} GCs or ``MGCs''(\S\ref{sec:MGC}), and the {\it red} GCs or ``RGCs''(\ref{sec:RGC}). The color boundaries are $(g-i)_0=$~0.86 and 1.01 (derived in \S\ref{sec:colsub}). We also exclude the large $(r_{\rm h} > 5.25$~pc) objects (or ``UCDs'') as well as all bright objects $(i_0 < 20)$ because of the strong suspicion that they represent different subpopulations (\S\ref{sec:lum}).
Our kinemetry results will be summarized in \S\ref{sec:sum}.

An overview of the kinemetric results is shown in Figure~\ref{fig:grid}. For each subpopulation, radial profiles are shown for rotation position angle and amplitude ($\theta_0$ and $v_{\rm rot}$),  projected velocity dispersion ($\sigma_{\rm p}$),
and projected velocity kurtosis ($\kappa_{\rm p}$).
For clarity, the radial binning is the same for all kinematic parameters of a given subpopulation, and so represents a compromise between optimal binning for, e.g., dispersion and kurtosis.

We continue to use circular kinemetric models ($q=1$) for the results and figures discussed here, in order to allow for kinematic position angle variations (see \S 6.2.1). We have also carried out parallel kinemetric analyses using flattened models (based on the photometric distributions of the stars or the GCs), but, at least for the present data, these models turn out to yield similar results.

Although we have used Monte Carlo methods to estimate the uncertainties in our kinemetry fits, we suspect that these uncertainties are underestimated, particularly for cases where azimuthal coverage is sparse. To help evaluate the results in Figure~\ref{fig:grid}, we therefore provide Figure~\ref{fig:PAcomp3} to show some details of the fits. We also plot some velocity distributions in Figure~\ref{fig:LOSVD} as a counterpart to the kurtosis metrics,
and in Table~\ref{tab:kinem} we report some numerical values for the kinematics in bins of subpopulation and radius.

\subsubsection{Blue GCs}\label{sec:BGC}

Considering first the BGCs, we find some indications that they do not obey a smooth, simple rotation field. 
In the central regions (inside $\sim$~6.6\arcmin\ or $\sim$~30~kpc), there appears to be significant rotation $(v_{\rm rot}/\sigma_{\rm p}\sim0.3$), but with a nearly $180^\circ$ twist near $\sim$~2.3\arcmin\ ($\sim$~11~kpc) from minor-axis rotation to the east to minor-axis rotation to the west (in this region, the VIRUS-P data suggest the stars have minor-axis rotation toward the west). However, the azimuthal coverage along the direction of the maximum rotation is poor, and we cannot yet be confident in this result.

There are additional suggestions of BGC kinematical twists at larger radii. 
At two locations (10\arcmin--13\arcmin\ or $\sim$~50--65~kpc, and 23\arcmin--30\arcmin\ or $\sim$~110--140~kpc), high rotation along the minor axis to the {\it east} is inferred ($v_{\rm rot}/\sigma_{\rm p} \sim$~0.5--0.9). However, the azimuthal coverage and number of velocities measured in these bins are rather sparse, and the rotation does not seem to persist at radii in between, so we will regard this outer-rotation finding as provisional.
Considering all of the BGCs inside 27\arcmin\ together, they have $v_{\rm rot}/\sigma_{\rm p} < 0.22$.

The velocity dispersion and rms velocity profiles of the BGCs decline mildly with radius, with an overall power-law exponent for the latter of $ \sim -0.12 \pm 0.05$.  However, rather than declining smoothly, these profiles may have a sharp transition at $\sim$~10\arcmin\ ($\sim$~50~kpc), with constant profiles
inside and outside this radius (at $\sim$~370~\kms\ and $\sim$~300~\kms, respectively; see also top panels of Figure~\ref{fig:PAcomp3}). This feature coincides with the possible onset of a kinematical twist previously mentioned.

\begin{figure}
%\epsscale{0.85}
\epsscale{1.15}
\plotone{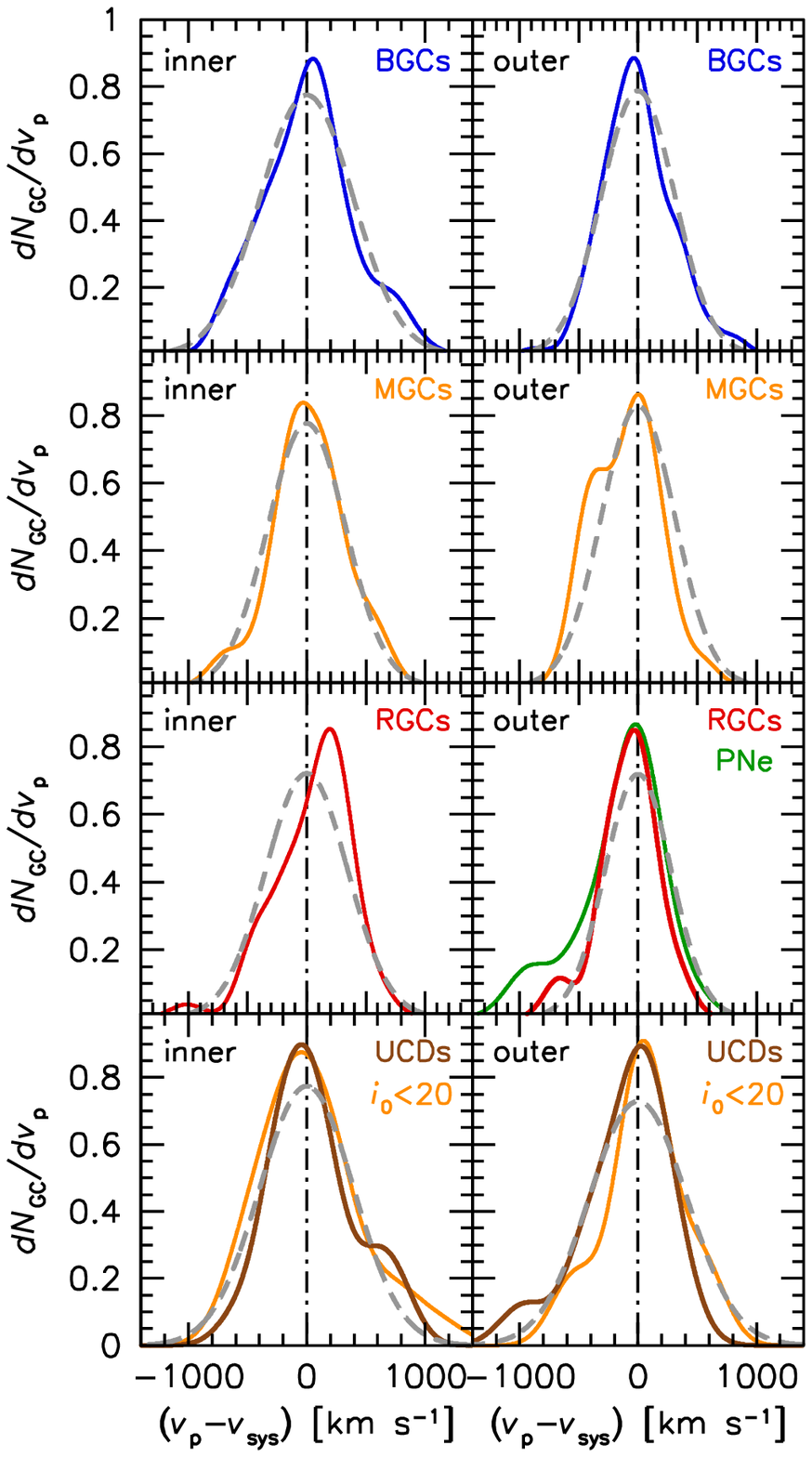}
\figcaption[M87GCvel6bz.eps]{\label{fig:LOSVD}
Velocity distributions of GCs in bins of subpopulation 
and galactocentric radius.
The solid curves show the observed results, which use optimal Gaussian-kernel smoothing.
The dashed curves show the best-fitting Gaussians, convolved with the smoothing kernel.
Visible differences between these curves are generally significant.
The rows from top to bottom show blue, intermediate color, and red GCs, and UCDs.
The left column shows the inner radii, and the right column shows the outer radii,
with the boundary radius $R_{\rm p}=10$\arcmin\ in all cases except 3.5\arcmin\
for the red GCs.
The sixth panel includes PNe at the same radii for reference,
and the lowest panels show the bright GCs.
Optimal Gaussian smoothing has been used for the velocity distributions of the data.
The normalizations are arbitrary.
}
\end{figure}

The kurtosis of the BGCs may also change at $\sim$~10\arcmin, from $\kappa_{\rm p}=-0.1\pm0.4$ in the inner regions to $0.6\pm0.5$ in the outer regions (a 1~$\sigma$ difference).
We will discuss this theme in more detail later, but the implication would be for a shift from isotropic to more radial orbits. As an alternative check on this result, we show the reconstructed velocity distributions of these two GC subsamples in the top panels of Figure~\ref{fig:LOSVD}, with best-fit Gaussians shown for comparison. Here it can be seen that the two velocity distributions show similar non-Gaussian deviations (peaky profiles in the center, and excess objects at high relative velocities), even though their kurtoses are formally inconsistent. These deviations are suggestive of radially-biased orbits in both bins and are a reminder that the kurtosis metric can be a blunt tool.

\subsubsection{Intermediate-color GCs}\label{sec:MGC}

Turning next to the MGCs, we found earlier that their $v_{\rm rms}$ values were closer to those of the RGCs than those of the BGCs. Figure~\ref{fig:grid} suggests that the MGCs have a different rotation profile from the RGCs, particularly at large radii, where the two subpopulations may be counter-rotating. This would support the idea (already suggested by photometry) that there are more than the two standard GC subpopulations. 

The MGCs at small radii ($R_{\rm p} \leq$~2.1~arcmin or $10$~kpc) appear to be kinematically very cold ($\sigma_{\rm p}= 168\pm31$~\kms; see Figures~\ref{fig:vsys} and \ref{fig:PAcomp3}). However, with only 12 objects in this bin, this result might be a chance  sampling fluctuation.

The MGC kurtosis is generally consistent with zero. Figure~\ref{fig:LOSVD} shows that ther is also an asymmetric excess of lower GC velocities at large radii, which may be part of a substructure in the MGCs
\citep{Roman11}.

\subsubsection{Red GCs and PNe}\label{sec:RGC}

The RGCs also have generally weak rotation close to zero at all radii, except for a rise suggested outside $\sim$~10~\arcmin\ ($\sim$~50~kpc). Figures~\ref{fig:grid} also shows results from the PN sample of \citet{2009A&A...502..771D}, adopting $R_{\rm p}=40$\arcmin\ (190~kpc) as a cut-off radius, beyond which there may be severe contamination from intracluster PNe (see Figure~\ref{fig:velR}).

There is remarkable agreement between the PNe and the outer RGCs in all four kinematical parameters, which is not shared by the other subpopulations, or even by the red GCs when using the standard bimodal color division. Unfortunately, this conclusion is not certain because as shown in Figure~\ref{fig:PAcomp3}, there are very few RGC data points at large radii, and the PN azimuthal coverage is very uneven (affecting the rotation conclusions in particular).

Also, it is possible that at these large radii, the very low velocities are not from objects bound to M87 but from free-floating intergalactic populations that are near M87 only in projection. \citet{2009A&A...502..771D} adopted this interpretation for their three low-velocity PNe, but as discussed when introducing Figure~\ref{fig:velR}, we consider it also plausible that these are {\it bona fide} M87 objects on very radial orbits. 

\citet{2009A&A...502..771D} commented that their inferred halo rotation from the PNe was low, suggesting that the fast rotation of the GCs from \citet{2001ApJ...559..828C} was driven by contaminants or else meant that the GCs do not trace the stars. We have found that there were apparent issues with the older data that produced the very high rotation signal, but also that some rotation is still possible in both the PNe and the RGCs, depending on how outliers are handled. If we did exclude the 5 lowest PN and RGC velocities, we would find an dramatically lower overall dispersion for these objects at large radius ($\sim$~200~\kms\ rather than $\sim$~400~\kms). One important avenue for future investigation would be to increase the number of PNe and RGCs with measured velocities at large radii.

The velocity distributions of the outer RGCs and PNe do support the kurtosis agreement, with peaky profiles and an excess of very low velocities (Figure~\ref{fig:LOSVD}).  Such behavior suggests fairly radial orbits (again, assuming that these are not intergalactic contaminants). Note that the velocity distribution of the inner RGCs is asymmetric, reflecting the velocity offset we previously identified for these objects (see Figures~\ref{fig:vsys} and \ref{fig:PAcomp3}).

There is some overlap between the GCs and the stellar kinematics from VIRUS-P \citep{2011ApJ...729..129M}. Both the MGC and RGC subsamples show some kinematic similarities to the stars (see also Figure~\ref{fig:kinprof1}), although the stars do not show the large mean velocity offset found in the RGCs in particular.

\subsubsection{Luminous GCs and UCDs}\label{sec:lum}

We also attempt to examine the kinemetry of the bright GCs and the UCDs, which is difficult because of the fairly small data sets for these objects. First of all, Figure~\ref{fig:grid} suggests that the bright GCs and UCDs have consistent kinematics in general, except for remarkably high rotation at large radii for the UCDs, which is not shared by the bright GCs.  However, as shown by Figure~\ref{fig:PAcomp3}, there is one low-velocity UCD (H38554) that is driving {\it half} of the rotation, and without this object the two subpopulations would be more consistent. The velocity distributions further show some similar asymmetries (Figure~\ref{fig:LOSVD}; see also Figure~\ref{fig:mags}). 

More generally surveying the different color subpopulations' kinemetry, the UCDs seem to be generally consistent with the MGCs as well. 
Because of the lower GC velocity dispersion that we have now found, 
we do not reproduce the finding of \citet{2008MNRAS.389.1539F} that these have a higher
dispersion than the UCDs around M87.

One interesting result here is that not only do the bright GCs have a higher velocity dispersion overall than the faint ones (\S\ref{sec:param}), but their rotation and dispersion profiles both show spikes (of marginal significance) to relatively high values at $\sim$~8\arcmin\ (Figure~\ref{fig:grid}).
This is the same radial region where older work found elevated rotation and dispersion values (see \S\ref{sec:oldnew}), suggesting that those results were driven by behavior in the bright GCs that does not reflect the bulk of the GC system, and may trace primarily UCD kinematics
(see also \S\ref{sec:trendlum}).

\subsubsection{Kinemetry Summary}\label{sec:sum}

To summarize the kinematic results, overall there is little dynamically significant rotation found in M87, with some localized suggestions of more substantial rotation. The rotational ``blips'' may be due to unmixed substructures, and in general will require more complete spectroscopic coverage to ascertain definitively. We see various differences between the BGCs, MGCs, and RGCs that tentatively support the distinct nature of these subpopulations. The UCDs and bright GCs also appear to have distinct kinematics which are most closely related to the MGCs. The central stellar kinematics has some similarities to both the MGCs and RGCs, while the PN kinematics is most similar to the RGCs.
See also Table~\ref{tab:kinem} for a summary.

\section{Dynamics}\label{sec:dyn}

M87 has been one of the most intensively modeled galaxies in the Universe, and it is beyond the scope of this paper to carry out another detailed dynamical analysis or even to adequately review the previous work. This will be the subject of future papers, and for now we will derive simple scale-free estimates of the mass profile and compare them to previous results.
Our dynamical analysis is presented in \S\ref{sec:scale}, comparisons to other studies in
\S\ref{sec:comp}, and implications for the dark matter halo in \S\ref{sec:DM}.

\subsection{Scale-free Analyses}\label{sec:scale}

Our dynamical analysis begins with a method descriptions in \S\ref{sec:methods},
our results in \S\ref{sec:results},
and checks of the methods with other data sets in \S\ref{sec:sanity}.

\subsubsection{Methods}\label{sec:methods}

The first key approximation we will make is of dynamical equilibrium, which is certainly questionable in the light of the indications of substructure found in M87, and will be an issue for even the more advanced dynamical models. Nevertheless, we are restricted at this point to simple assumptions, which will also include spherical symmetry.

The basic Jeans equation describing the relation between the gravitating mass profile and the kinematics and density of a tracer population can then be expressed in the simplified form:
  % eqn:jeans
  %____________________________________________________________________
  \begin{equation}
    v_{\rm c}^2(r) = \left[\alpha(r) - 2\beta(r) + \gamma(r)\right] \sigma_r^2(r) ,
\\
    \label{eqn:jeans}
  \end{equation}
where $r$ is the three-dimensional galactocentric radius, $\alpha(r)\equiv -d\ln \nu/d\ln r$ is the slope of the tracer density profile $\nu(r)$, $\beta(r)\equiv 1-\sigma_\theta^2/\sigma_r^2$ is the anisotropy parameter that describes the balance of the tangential and radial components of the velocity dispersion $\sigma_\theta(r)$ and $\sigma_r(r)$, $\gamma(r)\equiv -d\ln \sigma_r^2/d\ln r$ is the slope of the internal velocity dispersion, and the circular velocity is related to the cumulative mass profile by $v^2_{\rm c}(r)\equiv G M(r)/r$ (cf Eq. 4.215 of \citealt{2008gady.book.....B}).

One advantage of the circular velocity expressed as a function of angular radius is that once derived, this quantity is independent of the distance to the galaxy and of any luminosity model and filter bandpass characterizing the stellar distribution. Note also that unlike the case of a thin-disk galaxy, the luminosity and basic velocity profiles are not enough to specify the mass profile because of the additional factor $\beta$.  This is the ``mass--anisotropy degeneracy'' that is the bugaboo of studying elliptical galaxy dynamics and which may be alleviated through several techniques, some of which we will incorporate here.

We next adopt the simplification that the galaxy's dynamics are scale-free, i.e. the quantities $\alpha$, $\beta$, $\gamma$, and $v_{\rm c}$ in equation~\ref{eqn:jeans} are all independent of the radius, which also implies that the projected velocity dispersion and kurtosis profiles $\sigma_{\rm p}(R_{\rm p})$ and $\kappa_{\rm p}(R_{\rm p})$ are scale-free. The conversion between observed dispersion and circular velocity can then be expressed as a constant:
\begin{eqnarray*}
 v_{\rm c}(r=R_{\rm p}) = k \sigma_{\rm p}(R_{\rm p}), \, \, {\rm where} 
 \end{eqnarray*}
 \begin{equation}
 k^2 \equiv \frac{(\alpha+\gamma)(\alpha+\gamma-2\beta)}{\alpha+\gamma-(\alpha+\gamma-1)\beta}
  \frac{\Gamma[(\alpha+\gamma)/2] \Gamma[(\alpha-1)/2]}{\Gamma[(\alpha+\gamma-1)/2] \Gamma[\alpha/2]}
 \label{eqn:k}
 \end{equation}
 (\citealt{2005Natur.437..707D}; see also \citealt{1980MNRAS.193..931E,1993MNRAS.265..213G,2010MNRAS.406..264W}). 
 Note that the observed dispersion $\sigma_{\rm p}$ we will use here actually has the (dynamically weak) rotation folded in as $v_{\rm rms}$. Also, real galaxies and GC systems are not exactly power-law,  but we will adopt the approximation that the slope of $\sigma_{\rm p}$ at $R_{\rm p}$ provides the line-of-sight averaged slope $\gamma$ for $\sigma_r$ at $r=R_{\rm p}$, and similarly, the slope of the surface density profile yields the effective $\alpha$ after adding $1$.

Some interesting cases are an isotropic system ($\beta=0$) where the pre-factor in $k^2$ reduces to $(\alpha+\gamma$); a constant-dispersion system ($\gamma=0$) where $k^2=\alpha(\alpha-2\beta)/(\alpha+\beta-\alpha\beta)$; and an isotropic constant-dispersion system where $k^2=\alpha$. Note also the complicated dependence of $k$ on the parameters. For example, the classic situation where radial anisotropy suppresses $\sigma_{\rm p}$ in galaxy halos (i.e., $k$ increases with $\beta$) applies only when $(\alpha+\gamma) > 3$; the reverse is true for shallower density profiles with flat velocity dispersions
such as the case of the observed M87 GC system, where radial anisotropy should actually {\it boost} the dispersion ($k$ decreases with $\beta$). This also means that in regions where $(\alpha+\gamma) \sim 3$, the inferred mass will be fairly insensitive to the (often uncertain) value of $\beta$;
for M87, this should happen at $\sim$~90~kpc for the total or blue GC system, and at $\sim$~25~kpc for the red GC system.

An additional modeling ingredient is the use of the observed velocity kurtosis $\kappa_{\rm p}$ to estimate the anisotropy $\beta$. \citet{2009MNRAS.393..329N} found that for a certain class of spherical systems with constant profiles of $\beta$ and velocity dispersion ($\gamma=0$), $\beta$ can be derived geometrically by a set of projection integrals (see their equations B10--B12). While all of these assumptions will be at some level inaccurate for M87, we can use this approach to derive a first plausible guess for $\beta$.

\subsubsection{Results}\label{sec:results}

\begin{figure}
%\epsscale{0.85}
\epsscale{1.2}
\plotone{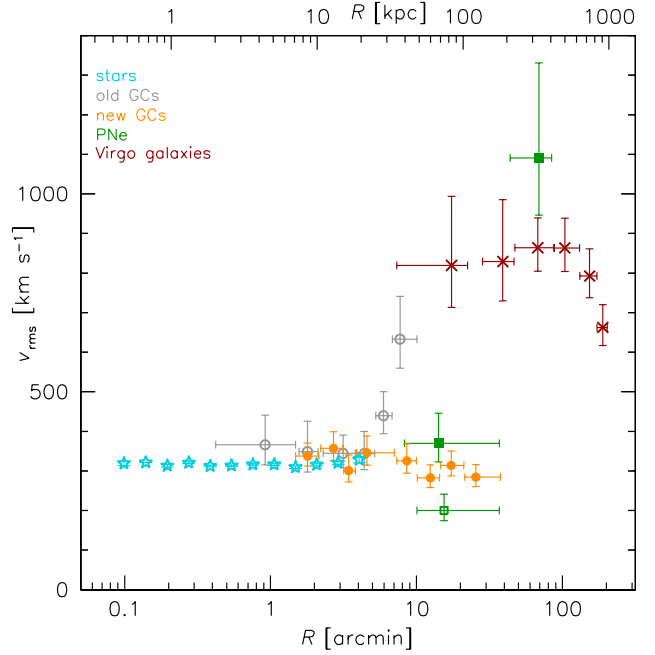}
\figcaption[M87GCkin4v.ps]{\label{fig:vrms}
Root-mean-square velocity profiles for subpopulations in Virgo, relative to the systemic velocity of M87. The different subpopulations are indicated in the panel legend; the stellar kinematics are from VIRUS-P, using true second-moment estimates provided by J. Murphy \& K. Gebhardt. The differences between GC subpopulations are relatively minor on this scale, and we lump all of the GCs together.
The open and filled green squares show the PNe with high-velocity ``outliers'' alternatively excluded or included.
}
\end{figure}

Now we carry out the dynamical analysis using all of the M87 GCs from the ``new'' data set\footnote{401 objects, where we have omitted the IGCs, UCDs, bright objects, the possibly dwarf-bound GC H35970, and the weird, high-velocity object S923.}. We show the corresponding rms velocity profile in Figure~\ref{fig:vrms}  (which includes other details that will be discussed later). Before binning the data with radius, we illustrate the method with the data combined. The GCS dispersion is fairly constant overall, with $\gamma=0.13\pm0.07$ and $\sigma_{\rm p}=$~$320\pm11$~\kms\ at a median radius of 6.9\arcmin\ (33~kpc), while the kurtosis is $\kappa_{\rm p} = 0.32 \pm 0.24$. We then infer $\beta \simeq 0.3\pm0.2$ and after substitution in equation~\ref{eqn:k} we find $k\simeq1.64\pm0.09$ and therefore $v_{\rm c} \simeq 525\pm28$~\kms\ (allowing for the statistical uncertainties in $\alpha$, $\beta$, $\gamma$, and $\sigma_{\rm p}$).

The inclusion of S923 would have boosted both the dispersion and kurtosis, and implied a higher anisotropy $\beta \simeq 0.7\pm0.1$: a large change induced by just one object out of $\sim$~400\footnote{Investigating the dynamics of the subpopulations in more detail is beyond the scope of this paper, but if we included the UCDs and bright objects in our analysis, then the results $\beta$ and $v_{\rm c}$ results above would not be affected significantly. We also infer the blue and red GCs to have overall $\beta=0.2\pm0.3$ and $0.6\pm0.2$, respectively---although these values should not be compared directly since they are  measured at different characteristic radii.}. The net effect on the $v_{\rm c}$ estimate would be very small because the dispersion and kurtosis work in opposite directions in the central regions of the GCS.  However, the same would not be true if such an object were found in the outer regions, highlighting the critical importance of contamination rejection in discrete velocity samples.

\begin{figure}
%\epsscale{1.17}
\epsscale{1.2}
\plotone{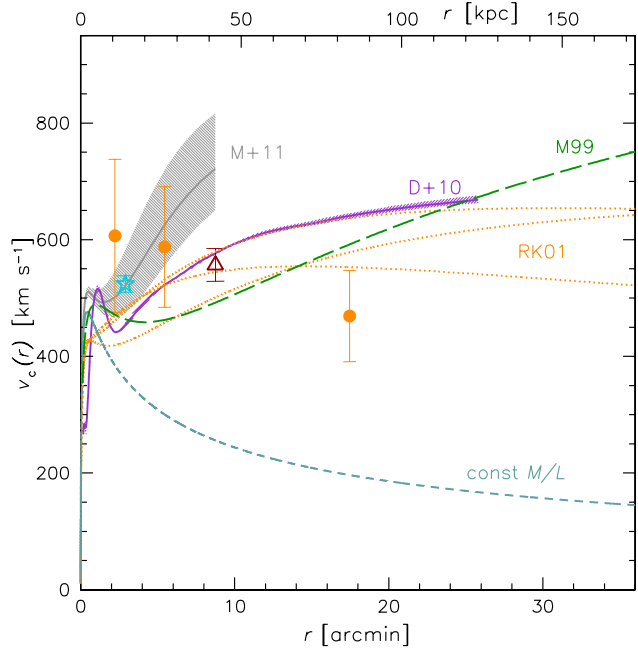}
\figcaption[vc14q.ps]{\label{fig:vc}
Mass profile of M87, expressed as
circular velocity versus radius.
Orange filled points with error bars show results from this paper based on simple 
dynamical modeling of the new GC data;
the equivalent modeling using the outer stellar kinematics data of VIRUS-P
is shown as a blue star symbol.
Curves with shaded regions show models from the literature based
on X-ray data (\citealt{2010MNRAS.409.1362D}; very narrow uncertainty band)
and old GC data \citep{2011ApJ...729..129M}.
A constant mass-to-light ratio model is shown as a dashed curve for comparison.
Green dashed and orange dotted curves show mass models from
\citet{1999ApJ...512L...9M} and \citet{2001ApJ...553..722R}, respectively.
The red triangle with error bars shows another GC-based result, from
\citet{2011arXiv1110.0833D}.
}
\end{figure}

We next follow the same technique, while breaking the dataset down into three radial bins with $\sim$~100--150 GCs each. The results are shown in Figure~\ref{fig:vc}: the circular velocity shows some indication of a decline with radius but is consistent with being constant  at $v_{\rm c}\sim$~530~\kms\ over the range of radii probed by the GCs. The large uncertainties in each radial bin are driven by the uncertain local dispersion slope $\gamma$.

We leave the next steps to future work, including separate examination of the dynamics of GC subpopulations in color, luminosity, and size, and the use of more detailed models (e.g., \citealt{2011ApJ...729..129M}).

\subsubsection{Sanity Checks}\label{sec:sanity}

Our final step here is to carry out two checks on the reliability of our methods. The first is to apply the same simple modeling techniques to the {\it stellar} kinematics data from VIRUS-P (introduced in \S\ref{sec:stelkin}). Using the data from the last three points (where $h_4$ is approximately converted to $\kappa_{\rm p}$), we derive an estimate of $v_{\rm c}\simeq 524\pm14$~\kms\ at $\sim$~14~kpc, as shown in Figure~\ref{fig:vc}.

This $v_{\rm c}$ value is almost identical to the full \citet{2011ApJ...729..129M} modeling results at the same radius (where the dispersions of the stars and the old and new GC data are all very similar).  Our inferred $\beta \sim -0.5 \pm 0.1$ for the stars can be compared to the Murphy~\etal\ finding of $\beta = -0.1^{+0.3}_{-0.5}$, which should be fairly insensitive to any errors in the mass profile caused by the old GC data, given the slope $\alpha\sim-2.8$ of the luminosity profile in this region, which should mean that the anisotropy is driven by the velocity distribution shape. These anisotropy results are compatible but the formal error bars in our method are clearly missing some sizable systematic uncertainties.

The second check is to apply our simple model to the old GC data, to verify that our new finding of decreased mass is caused by the data, not the dynamical methods. However, the steep outer dispersion rise in the old data turns out to be problematic for our scale-free model,  apparently allowing no physical solution for $k$ with near-isotropic  orbits, while slightly more anisotropic assumptions formally permit {\it any} value for $k$. Put another way, the ``observation'' $\gamma \sim -3$ implies a mass density profile that impossibly {\it increases} with radius, $\rho(r) \propto r$. This is a warning of the limitations of scale-free models, and also suggests that sharp increases in galaxy velocity dispersion may signify problems in the data (note though that the rising stellar dispersion profiles found in some cD galaxies are not as steep as in the old M87 GC data).

Some benchmark values to report from our models are a dynamical mass of
$M=(9.2\pm4.0)\times10^{11} M_\odot$ and 
$(4.3\pm1.1)\times10^{12} M_\odot$ at $r=11$ and $84$~kpc, respectively.
The cumulative mass-to-light ratios are then
$\Upsilon_B = (20 \pm 9)\Upsilon_{B,\odot}$ and 
$\Upsilon_B = (46 \pm 15)\Upsilon_{B,\odot}$, respectively.

\subsection{Comparisons}\label{sec:comp}

In Figure~\ref{fig:vc} we also show for comparison some mass profiles derived from the literature.  The first ones we will discuss come from state-of-the-art analysis of the latest M87 data till now: X-ray gas emission \citep{2010MNRAS.409.1362D} and combined stellar and GC kinematics \citep{2011ApJ...729..129M}. Both techniques found a circular velocity that increases outside $\sim$~10~kpc, but with inconsistent amplitudes.

Our new results at large radii are incompatible with any reasonable extrapolation outwards of the Murphy~\etal\ models, which would have three times as much dynamical mass within $\sim$~85~kpc as our new estimate. This is not surprising since those models were based on the old GC kinematics data set with a high outer dispersion which we no longer find (Figures~\ref{fig:kincomp}
and \ref{fig:vrms}).
There may also be inconsistency of our mass results with the outer X-ray profile, but we do not regard this as a firm result given the systematics in our dynamical analysis.

Three other sets of models are also shown in Figure~\ref{fig:vc}. One is a mass model constructed for Virgo by \citet{1999ApJ...512L...9M} that was later found by \citet{2001ApJ...559..828C} to reproduce the GC dynamics if roughly isotropic orbits were adopted. The three dotted curves are sample models from the joint dynamical analysis of stars and GCs in \citet{2001ApJ...553..722R}. The red open triangle is from a constant-anisotropy power-law analysis of the GC dynamics
from \citet{2011arXiv1110.0833D}. It can be seen that all these models, which used the old GC data set, are generally close to our new results using the new GC data, and dissimilar to the Murphy~\etal\ model based on the old data.

These comparisons are puzzling.  It is possible that some of the models got the ``right'' answer for the wrong reasons, but we did notice in a broader inventory of previous models of M87 that a surprisingly wide range of mass results were obtained from different studies that used basically the same data sets. This suggests that some methods may be less sensitive to errors and outliers in the data than others, which is a possibility that should be tested further through modeling of simulated data sets.

\subsection{Dark Matter Halo}\label{sec:DM}

We next consider some general implications that can be drawn for the dark matter (DM) surrounding M87---without doing any additional dynamical modeling. We begin with the assumption that the Murphy~\etal\ results are robust within the region probed by the stellar dynamics ($r \sim$~14~kpc) and that our new mass constraint at $\sim$~85~kpc is accurate. We then explore a range of cosmologically-motivated mass models that may be compared to these constraints.

Our basic model consists of a stellar mass distribution and a $\Lambda$CDM halo.  For the former, we use the triple power-law model of \citet{2001ApJ...553..722R} based on the $B$-band surface brightness profile: although this does not use the most modern photometric results (e.g., \citealt{2009ApJS..182..216K}), it is good enough for our purposes. For the latter, we adopt the classic \citet{1997ApJ...490..493N} profile: 
\begin{equation}
v^2_{\rm c,NFW}(r) = \frac{4 \pi G \rho_{\rm s} r_{\rm s}^3}{r}\left[\ln\left(1+\frac{r}{r_{\rm s}}\right)-\frac{r}{r_{\rm s}+r}\right] ,
\end{equation}
where $\rho_{\rm s}$ and $r_{\rm s}$ are the characteristic density and scale radius. The overall model is specified by the latter parameters along with  the (constant) stellar mass-to-light ratio $\Upsilon_{*,B}$.

Given three free parameters, it is fairly easy to fit a wide range of constraints. However, there are strong prior probabilities on all of these parameters. In particular, there is a statistical relation expected between $\rho_{\rm s}$ and $r_{\rm s}$ (or alternatively, between halo virial mass $M_{\rm vir}$ and concentration $c_{\rm vir}\equiv r_{\rm vir}/r_{\rm s}$; e.g., \citealt{2011arXiv1104.5130P}).

As a first rough guess for $M_{\rm vir}$, we may use a simple relation between velocity dispersion and virial radius, based on the approximation of a singular isothermal sphere (after \citealt{1997ApJ...478..462C}):
\begin{equation}
r_{\rm vir} \simeq \frac{\sigma_{\rm p}}{4 H_0} .
\end{equation}
We consider two alternatives: that the DM halo of M87 is smoothly contiguous with a relaxed Virgo cluster halo; or that M87 hosts its own group-scale halo that is still decoupled from the larger Virgo environment. In the first case, we use the cluster galaxies' overall  $\sigma_{\rm p}=803\pm29$~\kms\ to find $r_{\rm vir} \simeq$~2.8~Mpc, with a corresponding $M_{\rm vir}\simeq1.2\times10^{15} M_\odot$. For comparison, more detailed analyses of the Virgo kinematics by \citet{2001A&A...375..770F} and by \citet{2005ApJ...618..214T} imply $M_{\rm vir}\sim10^{15} M_\odot$, and by \citet{2006AJ....132.1275R} implies $r_{\rm vir} \simeq$~1.9~Mpc and $M_{\rm vir}\sim4\times10^{14} M_\odot$.

For the second case, we use the GC velocity dispersion, which will provide a lower limit to the mass, since the GCs are probably colder than the DM halo itself. We find $r_{\rm vir} \gsim$~1.1~Mpc and $M_{\rm vir} \gsim 8\times10^{13} M_\odot$. The $\sim$~2.3~keV temperature of the X-ray gas around M87 also suggests  $r_{\rm vir} \sim$~1.4~Mpc and $M_{\rm vir}\sim1.7\times10^{14} M_\odot$ from standard scaling relations (e.g., \citealt{2011MNRAS.414.2101U}).

For comparison to these dynamical metrics, classical number-density surveys of the Virgo cluster at optical and X-ray wavelengths characterize it as a region of $\sim$~2~Mpc projected radius with distinct substructures (e.g., \citealt{1987AJ.....94..251B,1994Natur.368..828B,1999A&A...343..420S}). For example the ``BCG" M49 (NGC~4472) hosts a subcluster ``B"
at 1.3~Mpc projected distance from M87, while the overdensity ``A" associated with M87 itself may be characterized as a region of $\sim$~0.6--1.0~Mpc projected radius. Given the overall constraints, we consider $r_{\rm vir}\sim$~1.2~Mpc and $M_{\rm vir} \sim 10^{14} M_\odot$ to be a reasonable model for the subhalo around M87.

The \citet{1999ApJ...512L...9M} model already shown in Figure~\ref{fig:vc} has $r_{\rm vir}\simeq$~2.2~Mpc, $M_{\rm vir} \simeq 6\times10^{14} M_\odot$, and $c_{\rm vir}\simeq$~4.  This would be a fairly reasonable model for a relaxed, cluster-wide halo, but as the Figure shows, it does not accommodate as much mass in the $\sim$~20~kpc region as currently inferred.  In fact, it turns out that this is a generic problem, and {\it no} standard halo (of either group or cluster mass) fits our constraints\footnote{In \citet{Spitler11} we focused on modeling the ``blue'' GC subpopulation and found that if the central DM density cusp has a log slope of $-1.2$ rather than $-1$ as in the NFW profile above, then a halo with $M_{\rm vir}\sim5\times10^{13}M_\odot$ can fit the data, using a plausible $\beta(r)$ anisotropy profile.}.

There are several possibilities for a DM halo with a higher central density. One is a halo with a much higher concentration than average; this is shown conceptually by the \citet{2001ApJ...553..722R} model in Figure~\ref{fig:vc} that comes close to matching the ``data''. The implication would be that M87 is at the center of a group halo that collapsed at an earlier time than average for its mass.

Another possibility is that a process like adiabatic contraction has raised the central halo density (e.g., \citealt{1986ApJ...301...27B}). Alternatively, a ``cored'' isothermal halo (e.g., \citealt{2011MNRAS.411.2035N}) would work, but this is not surprising given the extra free parameter with no prior constraint, not to mention the lack of theoretical motivation for the profile itself. Finally, there is the possibility that our simple dynamical models are wrong, and that the mass at $\sim$~85~kpc is somewhat higher, allowing for a normal $M_{\rm vir} \sim 10^{14} M_\odot$ halo.

Unfortunately we cannot resolve these possibilities without further extensive modeling.  For now, we conclude that the properties of the DM halo(s) of M87 and Virgo are poorly known---an unfortunate circumstance for the nearest galaxy cluster which should be remedied as soon as possible through more detailed modeling.
In particular, the DM density is of great interest in potentially providing 
observable gamma-ray signatures from the DM particles
(e.g., \citealt{2010PhRvD..82h3514G,2011ApJ...726L...6C,2011arXiv1104.3530S,2011arXiv1109.3810S}).

\section{Discussion}\label{sec:disc}

We now bring together the preceding compendium of new observational results on M87 into discussions of several thematic areas. In \S\ref{sec:revisit} we address the discrepancies between our new GC rotation and velocity dispersion profiles with those from the literature. \S\ref{sec:subs} discusses the evidence for multiple GC subpopulations and their relationships to individual field star components. In \S\ref{sec:bright} we delve further into the kinematical properties of candidate UCDs and weigh the evidence for different formation scenarios. \S\ref{sec:transit} is an analysis of the reality of proposed structural transitions in M87, especially in the halo, and the related issue of intergalactic GCs. In \S\ref{sec:merger} we consider various possibilities for recent mergers. In \S\ref{sec:n1399} we present a detailed comparison of M87 with the cluster-central Fornax elliptical NGC 1399. Finally, in \S\ref{sec:imp} we discuss the larger implications of this work for understanding the formation of massive galaxies. 

\subsection{Rotation and Dispersion Revisited}\label{sec:revisit}

One of our key results is that we do not reproduce the higher outer rotation and dispersion previously found for GCs in M87 (e.g., \citealt{2001ApJ...559..828C}).
Neither do we find any evidence for a steeply falling dispersion profile indicative of a truncation of the stellar halo at a radius of $\sim 150$ kpc, as suggested by kinematics of a small sample of PNe \citep{2009A&A...502..771D}. Instead, we find a fairly constant velocity dispersion profile out to a projected radius of nearly 200 kpc.

In previous sections we have briefly discussed the origin of the discrepancies with previous GC kinematics findings.
Part of the problem appears to be the presence of what we term catastrophic outliers; e.g., two GCs (S878 and S1074) had published radial velocities of $\sim$~2200 and $\sim$~2500~\kms\ that were erroneous by $\sim$~900 to 1000~\kms. Both are located in the radial range 30--35 kpc, and thus contributed to the high dispersion inferred (another relevant case is S7023, discussed in  \S\ref{sec:clean}). However, other GCs with extreme velocities have been confirmed by multiple measurements (e.g., S66 and S176). Therefore, we cannot conclude that such outliers are the sole cause of the disagreement; additional duplicate measurements of older radial velocities are needed.

Another factor that may be relevant is the luminosity difference between the samples. A subset of the more massive GCs may be ``contaminating" the sample as UCDs with distinct kinematics (see below), and the median magnitude of the C{\^o}t{\'e} \etal\ GC sample is $\sim 0.8$ mag brighter than in our sample.

Improved photometry is also important for more accurate kinematics because of the often underappreciated complication of high-velocity Galactic star contamination in GC data sets. We have used both high-precision color--color diagnostics, and {\it HST}-measured sizes, to help weed out stars and (to a lesser extent) galaxies.

\citet{Roman11} found that there is substantial kinematical substructure among GCs in the halo of M87, especially at radii $\ga 50$ kpc. Therefore we consider it plausible that another important reason for our discrepancy with previous work (see especially Figure~\ref{fig:PAcomp}) is that each study is sampling different regions of substructure. If true, this could indicate the dominance of substructure starting at even smaller radii (perhaps 30 kpc) than currently demonstrated. Additional velocities in this ``transition" region, with good azimuthal coverage, are needed to assess the importance of substructure.

\subsection{Subpopulations: Bimodality and Beyond}\label{sec:subs}

The prototype of GC bimodality is the Milky Way, where the disk/bulge and halo subpopulations are distinguished by strongly different distributions in metallicity, rotation, and positions relative to the center of the galaxy. Similar studies of external galaxies 
are generally confined to GC color, which is used as a metallicity proxy because of the generally large ages found for the GCs (e.g., \citealt{2006ARA&A..44..193B}).

Bimodal GC color distributions have been established for many galaxies including M87 (e.g., \citealt{2006ApJ...639...95P}), but there are lingering controversies over whether color bimodality reflects a true underlying metallicity bimodality, or a nonlinearity of the color-metallicity relation (e.g., \citealt{2006BASI...34...83R,2006Sci...311.1129Y,2007ApJ...669..982C,2007ApJ...660L.109K,2007AJ....133.2015S,2008MNRAS.389.1150S}).
A more general analysis of GC chemo-dynamical phase-space would thus be helpful in resolving this matter definitively.

In NGC 1399, \citet{2010A&A...513A..52S} showed that there is a distinct offset in the GC velocity dispersion at the same color as the bimodal separation between the blue and red GC subpopulations. We have attempted a similar analysis in the case of M87 (\S\ref{sec:param}), but our findings are less clear. Certainly we have found strong kinematical differences between different color subpopulations (e.g., Figures~\ref{fig:grid} and \ref{fig:PAcomp3}), but the picture is complicated by the evidence that we have uncovered for at least a {\it third} distinct GC subpopulation (cf \citealt{2011arXiv1110.1378B}).

The third population has intermediate spatial distribution and colors with respect to the blue and red subpopulations, a stronger tail to high luminosities, and somewhat different kinematics. Curiously, this population appears to provide the best kinematical match with the VIRUS-P stellar kinematics, while the far-red GCs match well with the PNe. The radial density profile of the full subpopulation of red GCs is also a very good match to the $V$-band stellar surface brightness of M87.

Such comparisons are important for understanding the origins and interrelations of different subpopulations within galaxies.  The conventional wisdom about GCs is that the metal-poor subpopulation corresponds to the faint metal-poor stellar halo,
while the metal-rich subpopulation is strongly associated with the formation of the hot spheroid (``bulge"). Our results in M87 generally support this picture.

The parent stellar populations of the bright PNe in early-type galaxies are somewhat mysterious (e.g., \citealt{2005ApJ...629..499C}). Empirically, the PN number densities and kinematics seem to generally agree well with the properties of the surrounding field stars  (e.g., \citealt{2009MNRAS.394.1249C}). In M87 we find a preliminary indication that the PNe are associated with the most metal-rich GCs. This can be tested in the future with a kinematical sample of PNe at small radii, where the far-red GCs show a peculiar velocity offset relative to systemic.

Detailed, orbit-based dynamical models will be needed to understand more clearly the relations between these different subpopulations.

\subsection{Bright GCs and UCDs}\label{sec:bright}

When using GCs as discrete kinematical tracers, a rarely-considered point is the luminosity of the clusters. Brighter GCs are more frequently targeted, for obvious reasons, and this has probably caused significant biases in many of the GC kinematics studies to date. Not only can tide-driven evolution produce correlations between GC luminosity and kinematics (e.g., \citealt{2003ApJ...593..760V}), but there is now abundant photometric evidence for a transition between normal GCs and extended UCDs at the bright end of the ``GC'' luminosity function.

Understanding the nature of UCDs and the degree to which they ``contaminate'' the normal GC population is a ongoing area of exploration, and here we add just a few lessons learned from our analysis of M87. The first is that a joint examination of size, luminosity, and color (\S\ref{sec:ucds}) suggests that there is a genuinely distinct population of extended objects that are not simply a continuation of the bright GC population. We develop this theme further in \citet{2011arXiv1109.5696B}.

Next, we have found that the most luminous objects in M87 ($i_0 \lsim 20$)
have kinematics distinct from fainter GCs (\S\ref{sec:param} and \S\ref{sec:subpop}), including elevated velocity dispersions and peculiar radial velocity distributions that avoid the systemic velocity. Most of the bright objects are also extended, and there is some evidence that {\it size} rather than
luminosity may be the key parameter in these peculiar kinematics trends (although this conclusion hinges on the reliability of the older velocity data).
This reinforces our conclusions from photometry that the UCDs and GCs are distinct classes of objects.

These kinematical differences are part of an emerging pattern discussed by \citet{2009AJ....137.4956R} where the brighter ``GCs'' (with sizes unknown) in massive ellipticals show peculiar kinematics. This is typically manifested as a higher dispersion and/or double-peaked velocity distribution, with the transition at $M_i \sim$~$-10$ to $-11$. Additional, recent support for this picture has come from NGC~5128 \citep{2010AJ....139.1871W} and NGC~1399 \citep{2010A&A...513A..52S}. In the latter study, the blue GCs seem more strongly affected, which also appears to be the case in M87.
On the other hand, \citet{2011A&A...531A...4M} found for NGC~3311 that velocity dispersion {\it decreases} for the brighter objects. 

To interpret these findings, we consider two simple 
scenarios. One is that the UCDs are the remnant cores of larger galaxies which have been stripped through a process such as tidal threshing (e.g., \citealt{1994ApJ...431..634B,2001ApJ...552L.105B}).  The other is that they are bona fide star clusters with large sizes from a range of possible causes, e.g., from birth in a dwarf galaxy environment or from mergers of star clusters
(e.g., \citealt{2002MNRAS.330..642F,2005ApJ...628..231B,2008ApJ...672.1006E,2009AJ....137.4361D,2009A&A...498L..37P,2010MNRAS.401.1832B,2010MNRAS.408.2353H,2011A&A...529A.138B,2011MNRAS.413.2606A}).

These two scenarios may be difficult to distinguish if they both lead to final sizes that are established by tidal limitations from the surrounding M87 gravity well (see further discussion in \citealt{2011arXiv1109.5696B}). However, there should be residual kinematical and dynamical signatures of the origins of UCDs.  If they began as normal nucleated dwarf galaxies, then they have become stripped down to UCD size by making close passages to the center of M87. In this scenario, one would naturally expect them to be on preferentially
{\it radial} orbits (\citealt{1994ApJ...431..634B,2003MNRAS.344..399B}, although \citealt{2008MNRAS.385.2136G} found more complicated orbital trends). The UCDs as a population should then reside in a centrally concentrated distribution, with radial velocities that decline steeply with increasing galactocentric distance \citep{2007MNRAS.380.1177B}.

In the second scenario, the current UCD sizes are similar to their original sizes when formed as extended star clusters, and they represent the surviving objects that have avoided plunging close to center of M87.  The UCDs would then be expected to reside
on more {\it tangential} orbits, showing dispersions that are low in the center and increase outwards. The number density profile should also have a shallow core, with large objects found near the center only in projection.

The implications for velocity distribution shape are less clear. Although radial and tangential orbits are classically expected to produce ``peaky'' and ``double-horned'' velocity distributions, respectively, this is for objects following a smooth power-law density distribution with radius (e.g., \citealt{1993ApJ...407..525V}).  The velocity distributions might differ in a situation with a cored density law caused by ongoing tidal disruption. For example, the shell of objects on radial infall around M87 has a velocity distribution that changes from peaky near its edge to double-horned at smaller radii (a ``chevron'' pattern in phase-space; \citealt{Roman11}).

Very qualitatively, we can outline some simple geometrically-based expectations for a well-mixed, quasi-equilibrium population following a cored density law and with radially-biased orbits (the stripped dwarf scenario). This population can be approximated as a superposition of many shell structures, with near-zero projected velocities for objects near apocenter, and high velocities exceeding the circular velocity $v_{\rm c}$ for objects near pericenter (cf~Figure~3 of \citealt{1998MNRAS.297.1292M}). A broad range of velocities is expected, except in the core region---here there are few objects near apocenter, and thus only the high-velocity objects near pericenter are seen. Therefore at small radii, one might expect to see a shell-like diverging chevron pattern of velocities.

For near-circular orbits (the star cluster scenario), we would again expect a broad distribution of projected velocities, but with the range extending no higher than $\sim v_{\rm c}$. At small projected radii, the circular orbits would generally be moving in the plane of  the sky, and so the observed velocities would decrease towards zero.

In summary, the origins of UCDs could be revealed by consideration of several kinematic aspects: the slope of their projected velocity dispersion profile, the detailed shape of their radius--velocity phase-space distribution at small radii, and their maximum observed velocities. The final aspect may be difficult to test in practice, since a sufficiently accurate $v_{\rm c}(r)$ profile may not generally be available for comparison.

Now considering the overall properties of the M87 system of UCDs, we find a fair fraction of these to be at large radii ($\sim$~20\% at $\sim$~100~kpc), which may argue against the galaxy-threshing scenario (as \citealt{2008MNRAS.389..102T} concluded in Fornax). However, we have not yet attempted to construct the M87 UCD density profile, which would require a careful accounting of selection effects on both size and velocity measurements. The challenge of estimating sizes over a wide field of view is such that no existing wide-field UCD study of any system has yet achieved this goal.

Turning to the M87 UCD kinematics, their projected velocity dispersion profile is high compared to the normal GCs, but is nearly constant with radius (Figure~\ref{fig:grid}), which does not seem consistent with either UCD scenario. As discussed in \S\ref{sec:trendlum}, the bright UCDs show a remarkable double-peaked velocity distribution at small radii, but their phase-space distribution shows both diverging and converging behavior with radius (top left panel of Figure~\ref{fig:mags}), which suggests a combination of the behaviors predicted above for the two UCD scenarios (cf.~the top and bottom panels of Figure~1 in \citealt{1997ApJ...488..702R}). There is one UCD with a velocity well in excess of $v_{\rm c}$, but otherwise the highest velocities are close enough to $v_{\rm c}$ to require more detailed modeling for interpretation.

The overall kinematic picture of the M87 UCDs may argue for a mixture of objects with different origins, with additional data and orbital analyses needed for clarification. In NGC~1399, the UCD velocity dispersion was found to {\it increase} with radius (once the central UCDs are included in the analysis; \citealt{2008MNRAS.389..102T,2009AJ....137..498G}), which could imply that the Fornax UCDs are predominantly star clusters.

\subsection{Structural Transitions and Intergalactic GCs}\label{sec:transit}

As discussed in \S\ref{sec:intro}, a key goal of this study was to search for transitions in the halo of M87 between the galaxy and the surrounding cluster.  One would naturally look at the stellar surface photometry, which in many elliptical galaxies does show sharp changes in isophote position angle, ellipticity, or higher-order shape that surely reflect internal transitions in the orbital structure.

Despite the extensive photometric work on M87 over the years (e.g., \citealt{1978ApJ...220..449D,2009ApJS..182..216K,2010ApJ...715..972J}), it has been difficult even to establish whether or not it hosts a cD envelope.
We have made an inventory of the M87 data and found no obvious, agreed-upon photometric features, other than the strong increase of ellipticity with radius. This behavior has previously been interpreted as rotational flattening \citep{1998AJ....116.2237K}, but, given our revised rotational results, we suggest instead that it may be due to preferential accretion of material along a dark matter filament.

We consider next the possibility of transitions in kinematics and GC metallicities, reviewing our findings so far (recall that 1\arcmin\ corresponds to 4.8~kpc). There is a twist in the stellar kinematics at $\sim$~1.5~kpc, and changes in the colors and mean velocities of the red GCs inside $\sim$~10~kpc, extending down to at least $\sim$~2~kpc. The blue GCs show a peak-color transition somewhere around $\sim$~15--25~kpc. The velocity distribution of the UCDs and bright blue GCs has an unusual ``double-peaked'' shape over the range $\sim$~10--30~kpc. The far-red and blue GCs have dispersion drops at $\sim$~17~kpc and $\sim$~40--50~kpc, respectively. The 50~kpc radius seems to be a zone of multiple transitions, including the  blue GC, far-red GC, and PN rotation and dispersion, and the UCD rotation and velocity distribution shape.

\begin{figure*}
\epsscale{0.9}
\plotone{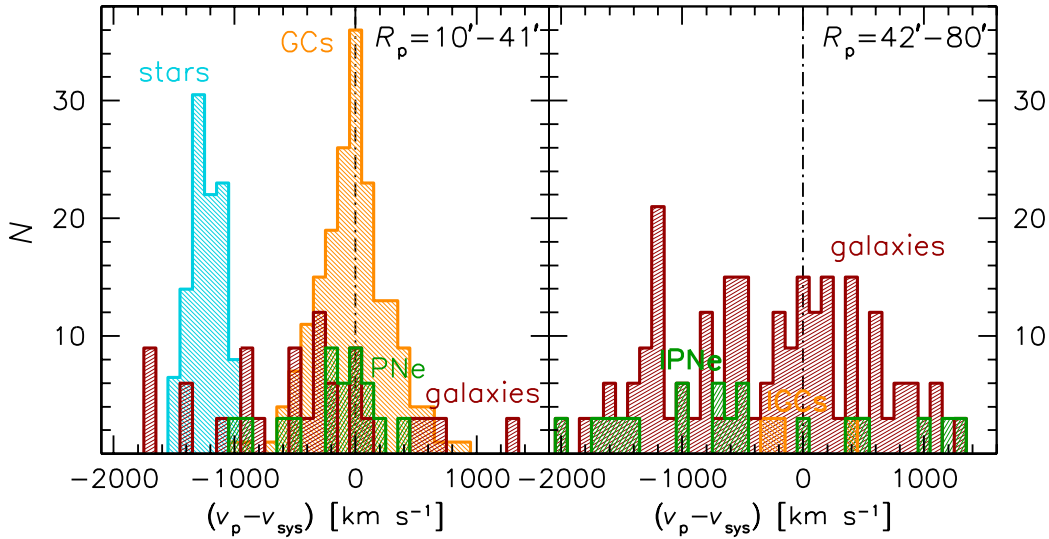}
\figcaption[M87GCvel6br.eps]{\label{fig:igc}
Distribution of radial velocities of objects around M87, in two radial bins (as labeled in the panels). Light blue, orange, red, and green histograms show data for stars, GCs, Virgo galaxies, and PNe, respectively; all of the histograms, except the stars and the GCs in the left panel, are rescaled by a factor of three for visibility. The PNe in the right panel include a few that extend to $R_{\rm p}=84$\arcmin; also shown are three IGCs at larger radii.}
\end{figure*}

This is a portrait of complex behavior that does not match up clearly with any transitions in the galaxy photometry. We do not reproduce the result of \citet{2001ApJ...559..828C} that a twist in the BGC rotation coincides with an onset of the cD envelope at a radius of $\sim$~21~kpc. However the $\sim$~50~kpc kinematical transition zone does correspond with the onset of a halo substructure identified by \citet{Roman11} with different techniques.

As a final exercise we look for transitions in the far outer halo of M87. Given that (i) \citet{2001ApJ...559..828C} found a strongly increasing outer GC dispersion that extrapolated smoothly to the high dispersion of cluster galaxies at larger radii;  and (ii) \citet{2009A&A...502..771D} found a low outer dispersion in the PNe, along with a paucity of PN detections that they interpreted as a truncation of the stellar halo at $\sim$~160~kpc, it is of great interest to revisit this issue with our new data.

As shown with the rms velocity profiles in Figure~\ref{fig:vrms}, we do not confirm either of the above claims. All of the subpopulations that we have studied around M87 stay at $v_{\rm rms}\sim$~300--500~\kms, out to the largest radii we have measured---in contrast to the cluster galaxies \citep{2008AJ....135.1837R} which have $v_{\rm rms}\sim$~700--1000~\kms, even at radii where they overlap with the GCs and PNe. 

We have already seen that the high GC dispersion previously reported may have been caused by problems with the older data, and we have discussed how the low PN dispersion was predicated on ``outlier'' removal. We find no evidence for a steeply declining dispersion in the GCs near 160~kpc. Similarly, the photometric number density of metal-poor GCs shows no notable features beyond 100~kpc (Figure~\ref{fig:sd}); a single S\'ersic profile is a good fit all the way to the edge of the photometric sample (see also similar constraints from \citealt{2006MNRAS.373..601T}). We can therefore categorically state that there is no edge to the stellar halo of M87 around these radii.

As discussed in \S\ref{sec:dyn}, there is already ample evidence that the Virgo cluster consists of multiple subsystems which have yet to merge and relax. Thus it is plausible that the dark matter halo and GC system surrounding M87 are basically decoupled from the greater Virgo environment, which would explain the relative coldness of the GC kinematics. However, kinematical measurements at larger radii should pick up a 
transition eventually, either due to the dynamical interface between M87 and Virgo, or because of an intergalactic (IGC) population seen in projection.

IGCs are known to exist in Virgo from photometry \citep{2007ApJ...654..835W,2011AAS...21715203D}, becoming dominant at perhaps $\sim$~190~kpc from M87 \citep{2010Sci...328..334L}. Although we have not probed this far out yet with kinematics, projection effects should make many IGCs appear at smaller radii, with some of them standing out by their extreme velocities.

To investigate this issue quantitatively, we estimate the number of IGCs expected among our ``new'' spectroscopic dataset of 485 objects.  For each of these objects, we calculate its {\it a priori} probability for being an IGC via the local ratio of IGC surface density (estimated to be 0.1--0.5~arcmin$^{-2}$ in \citealt{2006MNRAS.373..601T}), and the IGC$+$GC density using our S\'ersic model for the GCS density profile from \S\ref{sec:raddist}. Summing over all objects and allowing for statistical fluctuations, we find that we should have obtained between 17 and 87 IGC spectra, with 90\% of these occurring at $R_{\rm p} \gsim$~8\arcmin\ ($\gsim$~40~kpc).

To estimate how many of these IGCs should have ``Virgo-like'' velocities that clearly distinguish them from GCs bound to M87, in Figure~\ref{fig:igc} we plot histograms of observed velocities for various populations. First we show GCs, stars, PNe, and Virgo galaxies in the outer regions of M87 where we have GC spectroscopy (left panel, $R_{\rm p}=$10\arcmin--41\arcmin). For a larger sample of galaxies, we also consider their velocity distribution from a larger-radius bin (right panel), assuming this is similar to the region at smaller (projected) radius. We also show the PN velocities at larger radii, which as mentioned previously, show a transition to Virgo-like velocities outside $\sim$~40\arcmin.

If we define extreme velocities to be more than 1000~\kms\ relative to $v_{\rm sys}$ for M87 (i.e., $v_{\rm p} < 307$~\kms\ or $v_{\rm p} > 2307$~\kms), then the various subsamples of galaxy and intergalactic PNe imply an ``extreme'' fraction of between 16\% and 67\%. Therefore our spectroscopic data set should include between 3 and 58 IGCs with extreme velocities.

We do find 4 objects (including the weird S923) with extreme velocities (see Figure~\ref{fig:velR}), but these are all at small radii (8--30~kpc), and except for S923, are only slightly in the ``extreme'' range. Therefore we deem them very likely to be objects bound to M87, seen around the pericenters of near-radial orbits. There are also several objects (GCs and PNe) at large radii (90--130~kpc) with relative velocities close to $-1000$~\kms.  As previously discussed, these are somewhat ambiguous: whether or not they are bound to M87 is highly dependent on the uncertain distribution of mass outside these radii.

There could be additional lower-velocity GCs lurking in our sample of foreground ``stars'', many of which do not have size measurements to verify their classifications. Indeed, there are three confirmed GCs in the older velocity data with reported relative velocities of around $-1200$~\kms\ (see \S\ref{sec:stars}). However, we also expect a significant fraction of IGCs to be found with relative velocities below $\sim -1500$~\kms\ and above $\sim$~1000~\kms, which are not yet observed.

Therefore we conclude that the IGCs around the core of Virgo have either a surface density of $\sim$~0.1~arcmin$^{-2}$ or lower, or a surprisingly cold radial velocity dispersion. The latter possibility is supported by the three bona fide IGCs with spectroscopy (at $\sim$~800~kpc distances from M87; \citealt{2008MNRAS.389.1539F}). Their velocities are close to $v_{\rm sys}$ for M87 (see Figures~\ref{fig:velR} and \ref{fig:igc}), so we speculate that much of the intergalactic material in the core of the Virgo cluster may be relatively cold and flowing in roughly the plane of the sky. One might suppose such an arrangement to be a consequence of filamentary accretion into the cluster (e.g., \citealt{2004ApJ...603....7K}), but the Virgo filament is thought to run along rather than across the line-of-sight (\citealt{2007ApJ...655..144M}), i.e., probably not coincident with the major axis of M87.

\subsection{Recent Merger(s)?}\label{sec:merger}

The large substructure outside $\sim$~50~kpc is discussed in detail in \citet{Roman11}, where the overall conclusion is that a massive galaxy was accreted less than $\sim$~1~Gyr ago. Here we have also found examples of small pairs and groups of objects (\S\ref{sec:param}) that could be relics of disrupted dwarf galaxies.  Similar features have been found in other systems \citep{2003ApJ...591..850C,2009AJ....137.4956R,2011AJ....141...27W}, and it should be kept in mind that localized peculiarities in the GC kinematics might be caused by small substructures with coherent kinematics.

We now review additional signs of more significant interactions or mergers in M87,  starting with the central regions. There is a known velocity offset in the M87 stellar kinematics at a radius of $\sim$~0.2~kpc \citep{1990ApJ...348..120D,1991A&A...244L...1J,1991A&A...247..315J,1992MNRAS.257P...7C}. This feature has been suggested as the nuclear remnant of a smaller accreted galaxy that settled down to the center of M87 by dynamical friction. We have also found (\S\ref{sec:param}) that the far-red GCs show peculiar color and velocity shifts inside $\sim$~10~kpc, suggesting this region is unrelaxed.

There is a well known central jet that is estimated to have been active for $\sim$~0.1~Gyr \citep{2000ApJ...543..611O}, and which could have been fed by gas in a merger. There is furthermore a complex of dusty and filamentary warm and hot gas extending out to radii of $\sim$~15~kpc that has been suggested as a possible byproduct of a merger (e.g., \citealt{2004ApJ...607..294S}).  Additional ``cold fronts'' are found in the hot gas at $\sim$~30 and 90~kpc which have been explained as gas sloshing provoked by a ``fly-by'' of a massive galaxy group $\sim$~1~Gyr ago \citep{2011MNRAS.413.2057R}. 

It was also recently discovered that the supermassive black hole in M87 is slightly off-center, with one possible explanation being a galaxy merger within the past $\sim$~1~Gyr \citep{2010ApJ...717L...6B}.

Previous work on GCs in M87 derived high outer rotation, with a major merger as one possible explanation.  However, we have found that this rotation detection was probably spurious. We find no strong signatures of rotation.

\citet{1971ApJ...163..195A} and \citet{1997ApJ...490..664W} identified a broad fan of material at a radius of $\sim$~50~kpc. Weil et al. modeled this as the product of a smaller galaxy accretion within the past $\sim$~0.5~Gyr.  \citet{2010ApJ...715..972J} cast doubt on the existence of this structure, but did identify other small stellar substructures in the outer halo of M87 (see also \citealt{2005ApJ...631L..41M,2010ApJ...720..569R}). \citet{Roman11} identified a small moving group of GCs at $\sim$~150~kpc that may be associated with one of these stellar features, and could in principle be part of a substructure extending inwards.

As previously mentioned, several dE galaxies are found at $\sim$~40~kpc distance from M87 (NGC~4476, NGC~4478, NGC~4486A, NGC~4486B, and IC~3443), most of them with signs of severe disturbance as might be expected in these regions. These include peculiarities in the stellar isophotes, colors, and kinematics
\citep{1965AJ.....70T.689R,1973ApJ...179..423F,1984AJ.....89..919S,1987A&A...173...49P,1994AJ....108.1579V,2001MNRAS.326..473H,2005AJ....129..647L,2006ApJS..164..334F}.
Any or all of these galaxies could have disturbed the kinematics of the GCs in this $\sim$~40~kpc region, both from gravitational scattering of M87's pre-existing GCs, and (more likely) from depositing their own stripped GCs on unmixed, coherent orbits (see, e.g., discussion in \citealt{1998AJ....116.2237K}).

The overall picture in M87 to date provides various indications for interactions with galaxies from the surrounding environment, probably including multiple independent events. Until now, these indicators have all been relatively subtle, with the overall visible picture of M87 suggesting a placid galaxy where any recent interactions were presumably weak, involving accretion of low-mass galaxies.

Focusing on the inner substructure that we have detected at radii of $\sim$~4--9~kpc, the crossing time here is $\sim$~10--20~Myr, so assuming that the merger happened a few crossing times ago in order to still be very visible in phase space (Figure~\ref{fig:vsys}) but not in real space, we infer an event time no more than $\sim$~0.1~Gyr ago. This agrees with the timescales from the black-hole offset and central jet studies, and it may be that all of these features were caused by a significant gas-rich galaxy accretion $\sim$~0.1~Gyr ago that is still in the process of settling to the  center of M87.

There is a snag with this interpretation. The far-red, probably metal-rich GCs that show the velocity ``sloshing'' are unlikely to have been brought in by the accreted galaxy (which should typically host bluer clusters), but would instead be central GCs of M87 itself.  Even if the center of M87 were somehow moved around enough by the merger to show a velocity offset, the problem is that the same offset is not seen in the other central subpopulations of M87: the blue GCs and (particularly) the field stars. More data and simulations for this central region should help clarify the situation.

\subsection{Comparisons to NGC~1399 and Beyond}\label{sec:n1399}

\begin{figure*}
\epsscale{1.17}
\plottwo{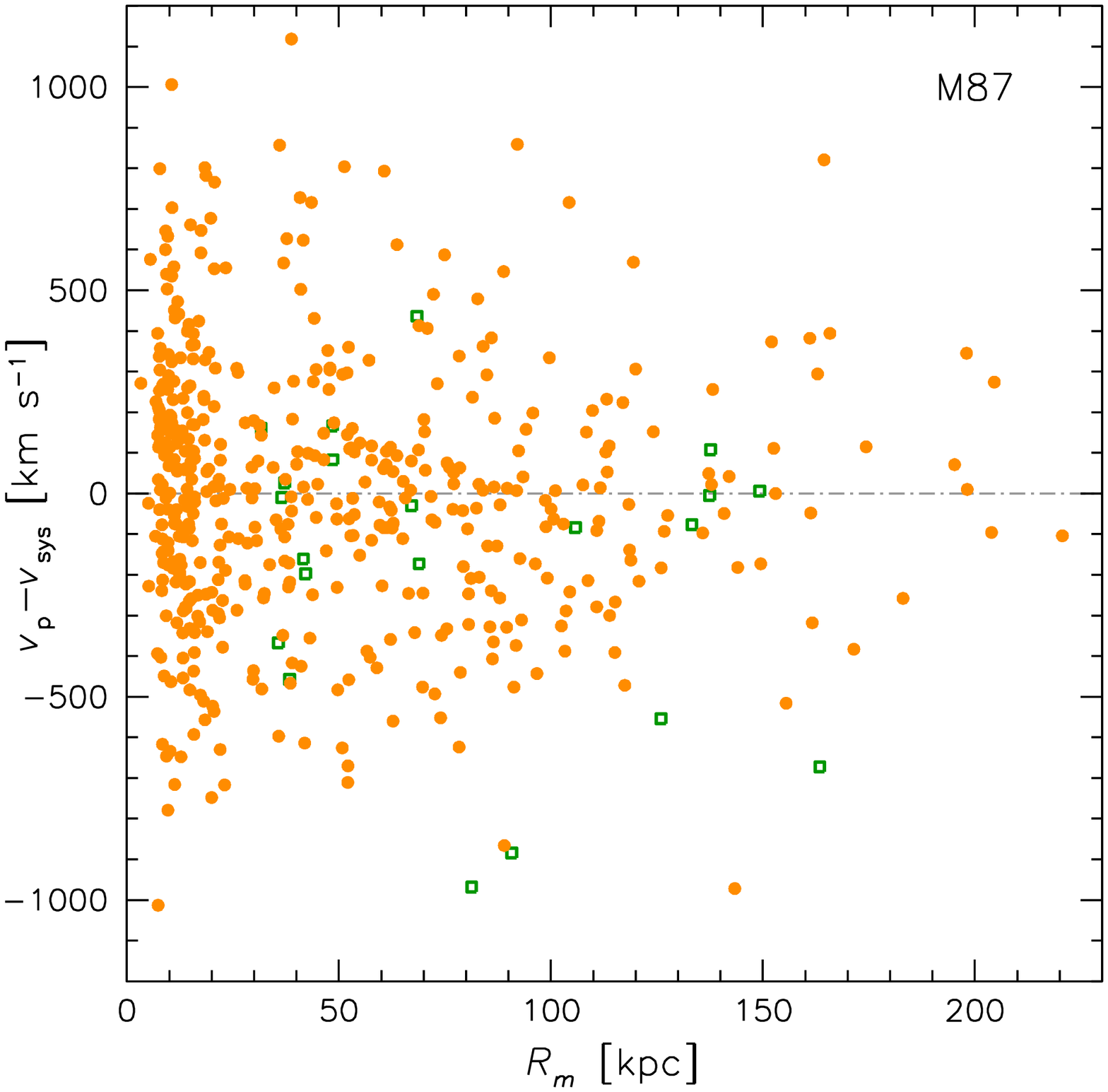}{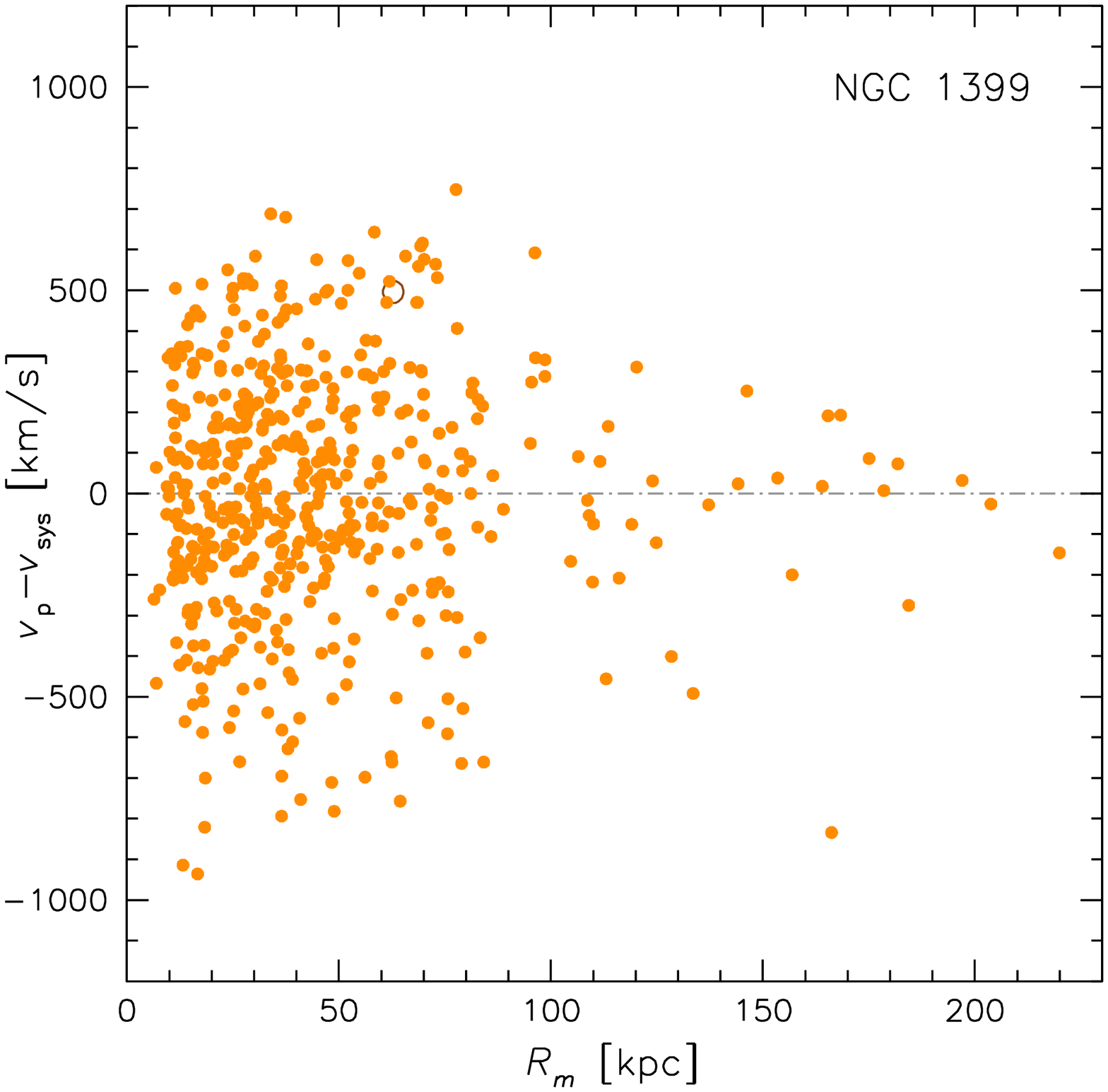}
\figcaption[velR6ci52.ps]{\label{fig:n1399}
Halo kinematics data in M87 (left) and NGC~1399 (right). 
Orange circles show GCs and green open squares show PNe (in the case of M87).
The known ``UCDs'' are included for M87 but not for NGC~1399.
A quality criterion has been used for both datasets: 
post-2003 measurements for M87, and ``class A'' for NGC~1399.
A brown open circle in the NGC~1399 panel shows the approximate phase-space
position of the satellite galaxy NGC~1404, whose accompanying GCs we have
made no attempt to exclude.
The equivalent radius $R_m$ for the NGC~1399 data was calculated approximately
by using an overall position angle of $90^\circ$ and ellipticity of $\epsilon=0.2$
\citep{2003AJ....125.1908D}.
}
\end{figure*}

We now have the opportunity to draw illuminating comparisons and contrasts between the halo kinematics of M87 and NGC~1399, the central galaxy in the second nearest cluster (Fornax).
An overview of the two systems is shown in Figure~\ref{fig:n1399}. Our new combined data set of GC radial velocities around M87 has 737 objects including 487 with high-resolution measurements (18~\kms\ median uncertainty), out to galactocentric radii of nearly 200~kpc. The combined NGC~1399 dataset is comparable, with 729 objects total and 527 ``class A'' measurements (30~\kms\ median uncertainty), also out to $\sim$~200~kpc (\citealt{2004AJ....127.2094R,2007A&A...464L..21B,2010A&A...513A..52S}).
The median luminosities for the high-quality GC measurements are also similar. NGC~1399 has a much larger PN dataset than M87 \citep{2010A&A...518A..44M}.

One of the first striking differences between these studies is that in M87 we had to eliminate only one GC from our sample as potentially bound to an adjacent dwarf galaxy, while NGC~1399 saw a great deal of ``interloper'' removal with dramatic ramifications for the kinematical and dynamical results. As we have already commented, overjudicious outlier trimming can be treacherous in the halos of galaxies, which may very well contain radially-biased orbits with velocities extending well past the normal wings of a Gaussian velocity distribution.

Some of the differences between the galaxies that we discuss below might be caused just by disparate treatments of outliers. Also, as emphasized by \citet{2008A&A...477L...9S}, a very small handful of erroneous velocity measurements (with presumably marginal $S/N$) can dramatically change the kinematics results, and so we consider only conclusions that are based on class-A measurements around NGC~1399. Even so, there were two cases of high-quality repeat velocity measurements with catastrophic ($\sim$~4~$\sigma$) differences.

Like M87, NGC~1399 had earlier reports of very high outer rotation and velocity dispersion (in its PN kinematics) which were overturned by the later higher-quality data. Both galaxies are now known to have extremely low rotation amplitudes overall, $\sim$~20~\kms.

NGC~1399 has a peak central stellar velocity dispersion of $\sim$~370~\kms, which declines steadily with radius and by $\sim$~50~kpc is $\sim$~200~\kms\ (as traced by the PNe and RGCs). This is very similar to the behavior in M87, where we have also found that  the $\sim$~50~kpc coldness may be related to a massive substructure, with the dispersions rising somewhat at larger radii. The NGC~1399 BGCs have an initially higher dispersion that also drops to $\sim$~200~\kms\ at $\sim$~150~kpc.

Thus NGC~1399, like M87, shows no transition to hotter cluster kinematics (which would be $\sim$~300--400~\kms; e.g., \citealt{2001ApJ...548L.139D}).
In particular, there is little sign of a transition at $\sim$~60~kpc where \citet{1996Natur.379..427I} claimed to see a galaxy-cluster interface from X-ray observations.

The amplitudes of the GCS projected velocity dispersions in M87 and NGC~1399 are not straightforward to interpret without understanding the mass, number density, and orbital anisotropy profiles in detail. However, it seems initially remarkable that the dispersions of the two systems are so similar while the overall mass of Virgo may be larger than Fornax by a factor of $\sim$~10 (e.g., \citealt{2005ApJ...618..214T}).
This situation further bolsters our suggestion that the M87 GCs are associated with a group-mass subhalo rather than with the entire Virgo cluster.

The NGC~1399 data were previously interpreted as showing a dynamically relaxed central galaxy accompanied by a subpopulation of vagrant' GCs (see references above), which are some combination of IGCs (i.e., associated with the entire Fornax cluster) and GCs on tidal streams stripped from other galaxies by  NGC~1399.  Such conclusions were motivated both by fluctuations in the velocity dispersion and by extreme radial velocities. The latter point is apparent in Figure~\ref{fig:n1399}, where there is one low velocity at $\sim$~170~kpc, and an asymmetric low-velocity tail at $\sim$~10--80~kpc (which is also seen in the PNe).

M87 is generally more symmetric than NGC~1399 in phase-space but harbors signatures of cold substructure and numerous extreme velocity objects that could mark highly eccentric orbits. Both systems also show strong fluctuations in the mean velocities of their RGCs at small radii that imply they are out of equilibrium. The outer kinematics in both cases also suggest significant infall and accretion, even though no major disturbances are apparent in optical imaging (e.g., \citealt{2009AJ....138.1417T}). More work is needed to characterize these processes quantitatively.

One area where the systems may differ is in their velocity distribution shapes (considered after removing the bright GCs with their potentially different kinematics). NGC~1399 was found to have a near-zero kurtosis for the GCs overall, becoming slightly negative for the BGCs.  In M87, we found generally positive kurtosis for both BGCs and RGCs. These differences are preliminary, pending uniform analysis of both datasets with the same outlier treatments, and considering also the variations of kurtosis with radius.

Keeping these caveats in mind, the kurtosis implication would be that the GC orbits are radially-biased in M87, but isotropic or tangential in NGC~1399, perhaps implying formational differences between the two galaxies. Radially-biased orbits are a fairly generic expectation for galaxy halos, whether because of cosmological infall or fall-back of tidal tails from major mergers, but a transition to isotropy may occur within the central regions of massive groups or clusters during their initial ``fast accretion'' phase (e.g., \citealt{2009A&A...501..419B}). Therefore NGC~1399 may be in a more advanced state of assembly than M87, consistent with the general idea that the Fornax cluster is more dynamically evolved than the Virgo (see also below).

Other comparisons between the systems, concerning their luminosity-dependent GC kinematics, their UCDs, and kinematical evidence for bimodality, were discussed in earlier sections. \citet{2010A&A...513A..52S} also found in NGC~1399 that the canonical connection between the RGCs and the visible starlight is supported by their similar density profiles, which in M87 we were able to establish further through ellipticity comparisons (NGC~1399 is nearly round so this test would be weaker).

The Fornax cluster is optically more dense and symmetric than Virgo, which is thought to reflect a more advanced stage of dynamical evolution (e.g., \citealt{2007ApJS..169..213J}).  Two other nearby massive systems have also been recently studied via the dynamics of both their group galaxies and their GCs: the Eridanus A group (with central galaxy NGC~1407; e.g., \citealt{2009AJ....137.4956R}) and the Hydra I cluster (with central galaxy NGC~3311; e.g., \citealt{2011A&A...531A.119R}). The GC dispersion profile around NGC~1407 remains constant at $\sim$~250~\kms\ out to at least 60~kpc, while in NGC~3311 there is a sharp dispersion increase (in both stars and GCs) that reaches $\sim$~700~\kms\ by 50~kpc.

We propose that these four systems represent a qualitative evolutionary sequence for clusters and their central galaxies.  Eridanus~A is a massive galaxy group that may be the progenitor of a cluster like Virgo which is in an active assembly phase. Fornax is a more evolved system, but like Virgo is still not internally relaxed enough to heat the central GC population or to accumulate a substantial cD envelope. Hydra~I is closer to a  ``fully realized'' cluster with a massive cD envelope in place that bridges the central galaxy and its host halo (although this region still does not appear well mixed in phase-space; \citealt{2011A&A...528A..24V}).

\subsection{Implications for BCG and GCS Formation}\label{sec:imp}

M87 is likely to remain a touchstone for formational tests of both BCGs and GCs, with the multi-dimensional properties of its field stars and GCs providing invaluable constraints. Here we will focus on a few specific aspects of the GCs:  rotation, velocity dispersion, spatial distribution, and orbital anisotropy.

\subsubsection{Rotation}

The high outer rotation measured in the past for the GCs in M87, as well as in early studies with GCs and PNe in other galaxies such as NGC~1316, NGC~1399, and NGC~5128, has been interpreted as evidence for a major merger that transferred initial orbital angular momentum to the outer parts of the remnant \citep{1995ApJ...449..592H,1998ApJ...507..759A,1998AJ....116.2237K,2001ApJ...559..828C,2002ApJ...581..799V}.  At one point, this appeared to be shaping up as a new trend: {\it ``rotation may in fact be a common by-product of the formation of GC systems''} \citep{2003ApJ...591..850C}. However, as we have seen in the case of M87, those earlier studies seem to have been prone to systematically overestimating the outer rotation, which with newer data on some of the same galaxies has turned out to be considerably lower \citep{2010AJ....139.1871W,2010A&A...513A..52S,2010A&A...518A..44M}. Other, more recent rotation studies could be affected by bias in the analysis techniques as we have discussed here  (e.g., \citealt{2009AJ....137.4956R,2010ApJ...709.1083L}).

Recent work on extended kinematic tracers in larger samples of early-type galaxies suggests that high outer rotation may be the exception rather than the rule (Proctor et al. 2009; \citealt{2009MNRAS.394.1249C}), and even in this context, M87 has gone from being one of the most dramatic outer rotators to one of the weakest. Although more theoretical work is needed on the halo rotation of BCGs in different formational scenarios, low rotation as in M87 and NGC~1399 is qualitatively suggestive of the accretion of multiple, small systems with uncorrelated angular momenta rather than of a single massive merger.

\subsubsection{Velocity Dispersion and Assembly Constraints}

We have already discussed in \S\ref{sec:transit} and \S\ref{sec:n1399} the remarkable lack of a velocity dispersion increase with radius in M87, and the possible implication of a decoupling with the overall Virgo cluster. More generally, the velocity dispersion of a population of objects in a cluster is thought to reflect their time of infall, assuming they were born outside of the cluster (e.g., \citealt{2005MNRAS.364..367D,2006MNRAS.368..563M}). For example, spiral galaxies are generally found with more extended radial distributions and correspondingly hotter velocity dispersions than ellipticals in clusters, probably reflecting their more recent infall (e.g., \citealt{1984ApJ...281...31T}).

Qualitatively, if the GCs were accreted by M87 at an ``early'' stage, this idea could explain the disparity observed between the velocity dispersions of GCs and cluster galaxies even if there is no decoupling between M87 and Virgo. An intriguing similarity can be seen between the GCs and the relatively cold kinematics of the subset of nucleated dEs around M87 with round isophotes, which has been interpreted as a signature of an early formational epoch \citep{2009ApJ...706L.124L}.

If we assume first an extreme scenario where the GC system of M87 is entirely accreted, and that to a first approximation it is built up monotonically from the inside-out like a series of tree rings (\citealt{2011MNRAS.413.1373W,2011arXiv1104.2334S}), then we can adopt a simple model to put limits on its assembly age.  The basic idea is that the GCs observed now at some radius $r$ cannot have arrived at a time when $r \gsim 2 r_{\rm vir}$ (twice the virial radius of M87; \citealt{2008ApJ...680L..25D}). Then using our mass model for M87 from \S\ref{sec:dyn} (i.e., $v_{\rm circ} \sim$~500~\kms) and cosmological formulae for virial quantities (e.g., \citealt{1998ApJ...495...80B}), we can map between $r$ and some upper-limit on the assembly redshift $z_{\rm f}(r)$. This is a conservative upper-limit because it assumes the mass within $r$ did not grow with time (\citealt{2008MNRAS.389..385C}).

Using this schematic, we find that the GC population (of all metallicities) at the $\sim$~100 and 200~kpc radii around M87 must have been accreted sometime after $z_{\rm f} \sim$~8 and 5, respectively. These are not very stringent constraints, but they do illustrate conceptually the kind of inferences that one might make about the accretion history with more detailed models, e.g., based on the ellipticity, kinematics, and radial density profile of the GC system \citep{2005MNRAS.364..367D,2006MNRAS.368..563M,2010MNRAS.405..375G}.  We pursue an approach like this in \citet{Spitler11}.

Another inference that one could make from standard $\Lambda$CDM cosmology is the characteristic radius of accreting material at the present day. Material on its first infall typically penetrates to $r \sim$~0.3~$r_{\rm vir}$ \citep{2007ApJ...667..859D}, so given our estimate in \S\ref{sec:DM} of $r_{\rm vir}\sim$~1.2~Mpc, we expect accretion around M87 to now be occurring typically at radii of $r \sim$ 350 kpc to 2.5~Mpc (where the infalling galaxies should not be disrupted but might lose some of their outermost GCs to tidal stripping).

This expectation, along with the lack of a velocity dispersion increase at large radii (suggesting little recent accretion from the cluster environment), would appear to create some tension with the discovery of a massive accretion event at $r \sim$~50--100~kpc \citep{Roman11}.  The nominal dynamical friction timescales at 350~kpc are too long for even a $10^{12} M_\odot$ subhalo to sink far inward,  but it is plausible that an entire group of galaxies fell in and broke apart (e.g., \citealt{2006ApJ...648..936R}), delivering the shell progenitor to small radii.

We may apply similar arguments to the scenario at the other extreme, where the GCs are all formed {\it in situ}. Considering that the associated starbursts would probably occur inside the virial radius, we can conclude that the formation of the M87 GC system as a whole, out to $\sim$~200~kpc, continued until after $z \sim$~2 (otherwise at higher $z$ the GCs would have been forming {\it outside} the galaxy). Such timing would not sit comfortably with direct estimates of the GC ages, and would contradict the current observational picture of BCGs experiencing the bulk of their in situ starburst activity at $z \gsim 4$, with only fairly quiescent merging since then.  

This argument thus appears to rule out a wholly in situ formation scenario for the M87 GC system (and possibly for the cD envelope by association), with scope remaining for {\it in situ} formation within the inner $\sim$~100~kpc radius.  There is in principle a loophole where the in situ GCs could have formed early on in a compact configuration, and been propelled to larger radii much later by bombardment from infalling substructure. However, it is likely that the energy requirement of this ``puffing up'' would involve a large amount of infalling mass which would add its own considerable population of accreted GCs.

\subsubsection{Orbital Anisotropy}

A last clue to the formation of BCG halos is orbital anisotropy (also discussed in \S\ref{sec:n1399}). As summarized by \citet{2008ApJ...674..869H} and \citet{2009AJ....137.4956R}, M87 and other high-mass ellipticals were previously found to host near-isotropic or even tangentially-biased orbits that contradict general expectations from theory for radially-biased orbits (see also \citealt{2011MNRAS.411.2035N,2011arXiv1110.0833D}). An implication  of these results might have been the existence of a process that drives the orbits toward isotropy in the central regions of groups or clusters (e.g., \citealt{2009A&A...501..419B}). 

However, our new observational results overturn the previous findings of isotropy in M87, and we instead find indications of radially-biased orbits overall (\S\ref{sec:dyn}). We have not tracked down the exact reason for this difference, but it may be caused in large part by ``contamination'' of the previous sample by UCDs and bright GCs with their distinct kinematics (\S\ref{sec:param}). The kurtosis trends for the blue GCs also suggest that the system may be near-isotropic inside $\sim$~50~kpc and more radial at larger distances, which might reflect the two-phase assembly discussed earlier, with the inner and outer regions forming in fast and slow accretion phases, respectively.

\subsubsection{Metallicity Gradients and General Conclusions}

Returning to the original questions raised in \S\ref{sec:intro} about BCG assembly, our overall conclusion is that the outer regions of M87 do not show the hot kinematics expected if they were formed from infalling cluster material.  It is possible that here the bona fide cD envelope is still in a very early stage of development, with the halo around M87 only now beginning to merge with the larger cluster environment. The multiple transitions at $\sim$~50~kpc (including the appearance of strong substructure) may mark the region now experiencing cluster infall. 

Another important feature in this context is the metallicity increase of the M87 BGCs that we have found inside $\sim$~20~kpc (\S\ref{sec:colgrad}). Similar transitions have been found in the Milky Way and NGC~1407, at $\sim$~10 and $\sim$~60~kpc, respectively \citep{2001stcl.conf..223H,2011MNRAS.413.2943F}.
A metallicity transition with radius may be seen as a sign of a two-phase assembly scenario, with the inner component having formed {\it in situ}
\citep{2009ApJ...702.1058Z,2010MNRAS.407L..26C,2011MNRAS.416.2802F}. However, this scenario generally describes the overall stellar population
(which is mostly metal-rich in the inner regions), while here we are considering only the metal-poor component.

Understanding the gradient goes to heart of the GC bimodality mystery: when and where were the metallicity subpopulations formed? In each giant galaxy, an interruption in the early, intense star formation  period may have produced a break in GC metallicity (e.g., \citealt{2002MNRAS.333..383B}), with a radial gradient in each subpopulation reflecting the dissipative nature of this period. Alternatively, virtually all of the metal-poor GCs might have been acquired through the accretion of smaller galaxies, with the inner gradient caused by a small number of massive accretors (with more metal-rich stars) preferentially settling at smaller radii because of dynamical friction. However, this interpretation may simply shuffle the bimodality puzzle off to the smaller galaxies, and indeed less massive galaxies also show evidence for GC subpopulations \citep{2006ApJ...639...95P,2006AJ....132.2333S}.

There is a critical need for theoretical work to help interpret the wealth of kinematical and dynamical data now being obtained on M87 and other BCGs. Although recent years have seen a few high-resolution simulations of BCGs in a cosmological context \citep{2006ApJ...648..936R,2009ApJ...696.1094R,2010MNRAS.405.1544D}, as far as we know, there has been no study of their predicted kinematics since \citet{1998ApJ...502..141D}. Neither has there been any study of the long-term dynamical evolution of GCs within galaxies assembling in a cosmological context, except for the case of the Milky Way \citep{2008ApJ...689..919P}.

\section{Summary}\label{sec:summ}

We have presented a new photometric and spectroscopic catalog of 737 confirmed GCs around M87, the giant elliptical at the center of the Virgo cluster. For 451 of these GCs, we have new, precise radial velocities from observations with the MMT instrument Hectospec and the Keck instruments DEIMOS and LRIS. Our catalog also includes half-light radii, measured from high-resolution {\it HST} images, for nearly half of the GCs. The majority of these sizes are new measurements from archival data.

The high precision of our new radial velocities, with a median uncertainty of 18~\kms, has allowed us to identify phase-space structures with low velocity dispersions. These features are discussed in detail in a companion paper \citep{Roman11} and are likely due to a recent accretion event.

The construction of our catalog required a critical assessment of literature data. A large portion of this effort was the classification of individual objects as GCs or foreground stars. This task is difficult using radial velocities alone, partially because of the existence of the substructures mentioned above. We used a combination of multi-color photometry, velocity, and {\it HST} imaging to reliably identify nearly all of our objects as stars or GCs; few ambiguous objects remain. Conversely, the firm identification of many such foreground stars and background galaxies has allowed us to accurately delineate the $gri$ colors of GCs, enabling better selection for future photometric and spectroscopic studies.

We have found that a subset of published radial velocities, including some of the most extreme values, are erroneous, although it is unclear whether or not similar issues affect a substantial fraction of literature data. We generally (see below) come to kinematical conclusions that differ from those of previous work, even using observations taken at the same galactocentric radius. The causes of these discrepancies are not entirely clear; the erroneous literature measurements may play a role. Other factors include the possibility of pervasive substructure and the confounding effects of sample selection, such as a correlation between GC luminosity and kinematics. 

The half-light radii of confirmed M87 objects in our catalog allow us to help clarify the relationship between true GCs and UCDs with a more unbiased sample than before (since UCDs were not specifically targeted in our spectroscopic observations). Defining the \emph{certain} UCDs as luminous objects with sizes $\ga$~10~pc, we have discovered ~18 new ones, and found that overall the UCDs follow a narrow track in color-magnitude space. We infer that the UCDs are a population distinct from a mere extension of the bright GCs. Among luminous objects, we estimate that few true GCs have sizes $\ga 5$ pc, while some UCDs range to smaller sizes that overlap with the GCs. The UCDs are discussed further in another companion paper \citep{2011arXiv1109.5696B}.

We have presented new S\'ersic profile fits to the GC surface densities of both individual metallicity subpopulations (metal-poor and metal-rich) and the full set of M87 GCs. These fits are more accurate than single power-laws and can be readily deprojected to yield the three-dimensional number densities required for mass modeling. Blind fitting for the surface density of background objects in the S\'ersic profiles gives similar results to the spectroscopic fraction of contaminants, lending credence to S\'ersic fits for GC systems in general. 

The surface density profile of the metal-rich GCs is a good match to that of the stellar light of M87 itself over an enormous radial range, from $\sim 8$--100 kpc.
Separately, we have also estimated the radial profiles of ellipticity $\epsilon$ for the GC subpopulations. The metal-poor GCs have a flat radial profile with $\epsilon \sim 0.3$. The metal-rich GCs have $\epsilon$ that increases with radius and is generally consistent with the integrated light of M87. The latter finding, along with the radial profile comparison, supports an association between the metal-rich GCs and the bulk of the field starlight.

We have critically examined the $gri$ color distribution of GCs as a function of luminosity and galactocentric radius, using spectroscopically confirmed GCs to guide the analysis. There is evidence for a third old subpopulation of objects, in addition to the classic blue and red groups, with intermediate colors and brighter than typical luminosities. Only a fraction of these objects appear to be UCDs. This additional subpopulation may be partially responsible for the identification of a significant, extended color gradient among the metal-poor GCs in previous work. Our photometric analysis yielded only marginal evidence for such a gradient outside of $\sim$~3\arcmin. By contrast, while we have found no evidence for a monotonic color gradient among metal-rich GCs, the mean color of this subpopulation within 5\arcmin\ ($\sim$~25 kpc) has much larger radial variations than expected from counting statistics. This suggests the presence of accreted metal-rich GCs  that are still radially unmixed.

The kinematics of the GC system of M87 have been analyzed, while remaining cognizant of the presence of substructure in the distribution of radial velocities. Previous work found high amounts of rotation ($v_{\rm rot}$ up to $\sim$~400~\kms); our new analysis show little evidence for rotation ($v_{\rm rot} <$~150~\kms\ at all radii, and typically $\sim$~20~\kms). 

We find a much lower value for the outer velocity dispersion than in the literature (with $v_{\rm rms} \sim$~300~\kms), and no indication of a previously claimed transition between the outer halo of M87 and the potential of the Virgo cluster (which has $v_{\rm rms} \sim$~800~\kms). In particular, the current data suggest that M87, to radii $\ga 100$ kpc, may be dynamically decoupled from Virgo itself. We similarly cannot confirm the assertion that M87 has an ``edge" at $\sim$~150 kpc. These apparently mistaken conclusions seem to principally be due to the confluence of three separate effects: (i) the presence of radial orbits and phase-space substructure, (ii) small number statistics, and (iii) a few catastrophic errors in determining GC radial velocities.
We observe fewer IGCs than expected; the surface density of these objects may be lower than reported in the literature, or their kinematics may be cold along our line of sight so that they are difficult to identify.

Considering the blue and red GCs separately, we have attempted to find evidence for a clear kinematical dichotomy between these two classical subpopulations, but have been stymied by the complicating effects of an additional, intermediate-color subpopulation. These subpopulations have complex kinematics and are not well-described by a single value of rotation or position angle. There are indications of unusual kinematics at even finer gradations---for example, the reddest GCs inside of 10 kpc (projected) have a mean velocity that is offset from systemic by nearly 200~\kms. Conclusions based on smaller samples of GCs are necessarily less certain than those discussed thus far, and additional data are needed for confirmation of tentative features.

Using a scale-free analysis we have estimated the mass enclosed within $\sim$~85~kpc of M87 to be $(5\pm2)\times10^{12}M_\odot$. For the GC system as a whole, we infer mildly radial orbits with $\beta \sim 0.4$, with the derived mass only weakly dependent on the anisotropy. The mass is lower than the most recent X-ray based determination at the same radii, although the difference may not be significant given the systematic uncertainties in our analysis.  Our mass results are also {\it much} lower than the previous estimate using the old GC data set. This suggests that some of the discrepancies found between optical and X-ray based mass analysis in various galaxies (e.g., \citealt{2010ApJ...711..484S}) might be driven by problems with the optical data.

Looking toward the future, we identify the following priorities for future improvements. The first is a need for complete radial and azimuthal velocity coverage, with special attention given to regions with unusual features. This would require data in the very center ($\la 8$ kpc) where the far-red GCs have an offset $v_{\rm sys}$; the region of maximal disagreement between new and literature data (25--50 kpc); and in the distant outer halo ($\ga 150$ kpc), hopefully extending at least to 250 kpc where a population of intergalactic PNe has been found, and where the transition to the Virgo Cluster may finally be revealed. Improvements in the supporting data are equally important, including photometry and {\it HST} sizes, to enable the identification of subpopulations, exclude foreground stars, and continue to clarify the relationship between GCs and UCDs.

The extensive chemo-dynamical dataset of GCs around M87 provides a unique opportunity for insight into the assembly of dark matter and stellar halos around galaxies. The results so far, while bedeviled with complexities at all scales, are broadly consistent with the overall conclusion that M87 is in active assembly, but still decoupled dynamically from the Virgo cluster. Despite our wish list above, in this new era of ultra wide-field, high-precision kinematics, {\it data} have ceased to be the main limiting factor. As near-field cosmology {\it theory} starts to reach beyond the Local Group, M87 and Virgo would be a worthy next step.

\acknowledgments

This paper is dedicated to the memory of John P.~Huchra, a fearless pioneer in the spectroscopic study of M87 GCs. Special thanks to Karl Gebhardt and Jeremy Murphy for extensive discussions and for providing their results in advance of publications. We also thank Magda Arnaboldi, Michele Cappellari, Payel Das, Juan Madrid, and Soeren Larsen for providing their results in electronic form, and J\"urg Diemand and Chris Mihos for helpful comments. We thank Nelson Caldwell for his assistance with the Hectospec observations, and Jason X.~Prochaska and Kate Rubin for software help. J.~S.~was supported by NASA through a Hubble Fellowship, administered by the Space Telescope Science Institute, which is operated by the Association of Universities for Research in Astronomy, Incorporated, under NASA contract NAS5-26555. J.~B. and A.~J.~R. acknowledge support from the NSF through grants AST-0808099, AST-0909237, AST-1101733, and AST-1109878.. This research used the facilities of the Canadian Astronomy Data Centre operated by the National Research Council of Canada with the support of the Canadian Space Agency. Much of the data presented herein were obtained at the W.~M.~Keck Observatory, which is operated as a scientific partnership among the California Institute of Technology, the University of California and the National Aeronautics and Space Administration. The Observatory was made possible by the generous financial support of the W.~M.~Keck Foundation. Observations reported here were obtained at the MMT Observatory, a joint facility of the Smithsonian Institution and the University of Arizona. This paper was based in part on data collected at Subaru Telescope, which is operated by the National Astronomical Observatory of Japan. This paper uses data products produced by the OIR Telescope Data Center, supported by the Smithsonian Astrophysical Observatory. This research was supported in part by the National Science Foundation under Grant No. NSF PHY05-51164 while at the Kavli Institute for Theoretical Physics.

%\end{document}

%\LongTables

\begin{deluxetable}{lll}
\tablewidth{0pt}
\tablecaption{Spectroscopic Observing Log
        \label{tab:obsrun}}
\tablehead{Instrument & Date & Exposure Time \\
 & &  }
\startdata
Keck/DEIMOS	    &  2007 Mar 20/21 & 3--3.5 hr \\
MMT/Hectospec   &  2010 Feb 17 & 2 hr \\
Keck/DEIMOS      & 2010 Mar 12 & 0.6 hr \\
Keck/LRIS             & 2010 Apr 8--11 & 2.5--3.5 hr \\
\enddata
\end{deluxetable}

\begin{deluxetable}{lccccccccc}
\tablewidth{0pt}
\tabletypesize{\footnotesize}
\tablecaption{Data for Keck/DEIMOS Spectroscopic Candidates
       \label{tab:allcand}}
\startdata
 & & & & & & & & & \\
\enddata
\end{deluxetable}

\begin{deluxetable}{lcccccccc}
\tablewidth{0pt}
\tabletypesize{\footnotesize}
\tablecaption{Data for Supplementary Keck/DEIMOS Spectroscopy
        \label{tab:keck2}}
\startdata
 & & & & & & & & \\
\enddata

\end{deluxetable}

\begin{deluxetable}{lcccccccc}
\tablewidth{0pt}
\tabletypesize{\footnotesize}
\tablecaption{Data for MMT/Hectospec Spectroscopic Candidates
        \label{tab:allcand2}}
\startdata
 & & & & & & & & \\
\enddata

\end{deluxetable}

\begin{deluxetable}{lccccccccc}
\tablewidth{0pt}
\tabletypesize{\footnotesize}
\tablecaption{Data for Keck/LRIS Spectroscopic Candidates
        \label{tab:lriscand}}
\startdata
  & & & & & & & & & \\
\enddata
\end{deluxetable}

\begin{deluxetable}{lcccccccccccccc}
\tablewidth{0pt}
\setlength{\tabcolsep}{0.01in} 
\tabletypesize{\tiny}
\tablecaption{Globular Cluster/UCD Radial Velocities with Multiple Measurements
        \label{tab:glob_match}}
\startdata
  & & & & & & & & & & & & & & \\
\enddata

\end{deluxetable}

\begin{deluxetable}{lcll}
\tablewidth{0pt}
\tabletypesize{\footnotesize}
\tablecaption{New Half-Light Radii
        \label{tab:sizes}}
\startdata
 & & & \\
\enddata
\end{deluxetable}

\begin{deluxetable}{lccccccccll}
\tablewidth{0pt}
\tabletypesize{\tiny}
\tablecaption{Spectroscopic Data for All Foreground Stars
        \label{tab:stars}}
\startdata
& & &  & & & & & & &  \\
\enddata
\end{deluxetable}

\begin{deluxetable}{llccccccclcccc}
\tablewidth{0pt}
\tabletypesize{\tiny}
\tablecaption{Spectroscopic and Photometric Data for All M87 Globular Clusters and UCDs
        \label{tab:allcand3}}
\startdata
& & &  & & & & & & &  \\
\enddata
\end{deluxetable}

\begin{deluxetable}{lccccccc}
\tablewidth{0pt}
\tabletypesize{\scriptsize}
\tablecaption{Supplementary Photometric Data for M87 Globular Clusters and UCDs
        \label{tab:photsupp}}
\startdata
& & &  & & & &  \\
\enddata
\end{deluxetable}

\begin{deluxetable}{lcccc}
\tablewidth{0pt}
\tabletypesize{\tiny}
\tablecaption{S\'ersic Profile Fits
       \label{tab:fits}}
\tablehead{Sample & Population & $N_0$\tablenotemark{a} & $R_{s}$ &  $m$  \\
                                     & & (arcmin$^{-2}$)  & (arcmin) &    }
\startdata
Phot. & Full &    $2.72 \times 10^{4}$  &  $5.11 \times 10^{-5}$  & 5.50  \\
Phot. & Blue &   $6.10 \times 10^{2}$  &  $1.62 \times 10^{-2}$  & 3.69    \\
Phot. & Red  &  $5.08 \times 10^{4}$  &  $2.66 \times 10^{-5}$  & 5.33   \\
\hline
Spec. & Full &    $7.36 \times 10^{4}$  &  $4.96 \times 10^{-6}$  & 6.27 \\
Spec. & Blue &  $3.22 \times 10^{2}$  &  $4.99 \times 10^{-2}$  & 3.22 \\
Spec. & Red  &  $3.49 \times 10^{4}$  &  $5.51 \times 10^{-5}$  & 5.09 \\
\enddata
\tablenotetext{a}{These normalizations only include GCs with $19 < i_0 < 22.5$.}

\end{deluxetable}

\begin{deluxetable}{lcc}
\tablewidth{0pt}
\tabletypesize{\footnotesize}
\tablecaption{Interloper Surface Density Estimates
        \label{tab:interlop}}
\tablehead{Sample & Photometric & Spectroscopic  \\
           & (arcmin$^{-2}$)  & (arcmin$^{-2}$)  }
 
\startdata

Full    &   $0.51\pm0.16$    &  $0.44\pm0.15$     \\
Blue   &   $0.18\pm0.17$   &   $0.25\pm0.11$    \\
Red   &   $0.36\pm0.09$    &   $0.37\pm0.19$      \\

\enddata
\end{deluxetable}

\begin{deluxetable}{lccccc}
\tablewidth{0pt}
\tabletypesize{\footnotesize}
\tablecaption{Ellipticity and Position Angle Estimates
        \label{tab:elliptic}}
\tablehead{  &    & 1\arcmin $< R <$ 3\arcmin & 3 \arcmin $< R <$ 5\arcmin & 5\arcmin $< R <$ 8.5\arcmin & 8.5\arcmin $< R <$ 12\arcmin\\
                       &    &						& 						&						&  }
 
\startdata

Full    & $\epsilon$ & $0.24\pm0.05$ & $0.25\pm0.06$ & $0.31\pm0.07$ & $0.32\pm0.07$ \\
Full    & P.A.            & $-39\pm7$          & $-21\pm8$        & $-41\pm8$         & $-29\pm8$         \\
Blue  & $\epsilon$ & $0.27\pm0.07$ & $0.28\pm0.07$ & $0.30\pm0.07$ & $0.31\pm0.08$ \\
Blue    & P.A.          & $-38\pm8$         & $-19\pm8$         & $-41\pm8$        & $-25\pm9$         \\
Red   & $\epsilon$ & $0.12\pm0.03$ & $0.19\pm0.06$ & $0.34\pm0.10$ & $0.31\pm0.12$ \\
Red    & P.A.           & $-30\pm7$          & $-32\pm8$        & $-40\pm9$        & $-38\pm9$          \\

\enddata
\end{deluxetable}

\begin{deluxetable}{lccccccr}
\tablewidth{0pt}
\tabletypesize{\footnotesize}
\tablecaption{Kinematical Results
        \label{tab:kinem}}
\tablehead{ Subpopulation & Radius range ($R_{\rm p}$) & $N$ & $\theta_0$  & $v_{\rm rot}$ & $\sigma_{\rm p}$ & $v_{\rm rms}$& $\kappa_{\rm p}$ \\
& & & (\kms) & (\kms) & (\kms) & \\
}
 
\startdata

All GCs    & $1.5^\prime$--$38^\prime$ & 410 & $135^\circ\pm11^\circ$ & $21\pm19$ & $320\pm11$ & $320\pm11$ & $0.32\pm0.24$ \\
UCDs/bright GCs    & $0.7^\prime$--$41^\prime$ & 73 & $185^\circ\pm66^\circ$ & $17\pm47$ & $368\pm30$ & $368\pm16$ & $0.83\pm0.56$ \\
BGCs    & $1.5^\prime$--$38^\prime$ & 242 & $100^\circ\pm40^\circ$ & $32\pm26$ & $335\pm15$ & $335\pm15$ & $0.25\pm0.31$ \\
MGCs    & $1.5^\prime$--$32^\prime$ & 92 & $243^\circ\pm29^\circ$ & $79\pm44$ & $297\pm22$ & $298\pm22$ & $0.04\pm0.50$ \\
RGCs    & $1.5^\prime$--$26^\prime$ & 76 & $134^\circ\pm38^\circ$ & $62\pm42$ & $295\pm23$ & $295\pm24$ & $0.97\pm0.55$ \\
\hline
All GCs    & $1.5^\prime$--$10^\prime$ & 250 & $186^\circ\pm40^\circ$ & $38\pm26$ & $335\pm15$ & $335\pm15$ & $0.26\pm0.31$ \\
UCDs/bright GCs    & $0.7^\prime$--$10^\prime$ & 46 & $123^\circ\pm72^\circ$ & $17\pm61$ & $385\pm41$ & $385\pm41$ & $0.73\pm0.69$ \\
BGCs    & $1.5^\prime$--$10^\prime$ & 120 & $178^\circ\pm49^\circ$ & $43\pm41$ & $373\pm25$ & $373\pm24$ & $-0.08\pm0.44$ \\
MGCs    & $1.5^\prime$--$10^\prime$ & 64 & $225^\circ\pm37^\circ$ & $71\pm53$ & $306\pm26$ & $306\pm28$ & $0.25\pm0.59$ \\
RGCs    & $1.5^\prime$--$10^\prime$ & 66 & $126^\circ\pm54^\circ$ & $33\pm41$ & $286\pm25$ & $286\pm25$ & $1.16\pm0.58$ \\
\hline
All GCs    & $10^\prime$--$38^\prime$ & 160 & $62^\circ\pm31^\circ$ & $54\pm30$ & $293\pm16$ & $294\pm17$ & $0.36\pm0.38$ \\
UCDs/bright GCs    & $10^\prime$--$41^\prime$ & 27 & $230^\circ\pm27^\circ$ & $178\pm104$ & $338\pm44$ & $338\pm48$ & $1.46\pm0.87$ \\
BGCs    & $10^\prime$--$38^\prime$ & 122 & $60^\circ\pm25^\circ$ & $81\pm35$ & $289\pm18$ & $292\pm19$ & $0.57\pm0.44$ \\
MGCs    & $10^\prime$--$32^\prime$ & 28 & $276^\circ\pm37^\circ$ & $97\pm66$ & $280\pm36$ & $280\pm39$ & $-0.53\pm0.86$ \\
RGCs    & $11^\prime$--$26^\prime$ & 10 & $121^\circ\pm37^\circ$ & $182\pm123$ & $348\pm66$ & $348\pm87$ & $0.98\pm1.33$ \\
PNe    & $10^\prime$--$37^\prime$ & 16 & $182^\circ\pm46^\circ$ & $166\pm124$ & $387\pm63$ & $401\pm76$ & $1.91\pm1.09$ \\

\enddata
\tablecomments{The UCDs/bright GCs sample incorporates all objects with $r_{\rm h} >$~5.25~pc,
or $i_0 < 20$. These objects are excluded from all the other subsamples,
along with the IGCs, H35970, and S923.
The blue GCs (BGCs) have $(g-i)_0 < 0.86$, the red GCs (RGCs) have $(g-i)_0 > 1.01$,
and the MGCs have intermediate colors.}
\end{deluxetable}

\end{document}